\documentclass[prd,aps,onecolumn,floatfix,superscriptaddress,preprintnumbers,bibnotes]{revtex4-2}

\usepackage{amsmath, amssymb, amsfonts, mathtools, cases, slashed, bm, bbm}
\usepackage{array}
\usepackage{verbatim}
\usepackage{epsfig}
\usepackage{graphicx, graphics, color, epsfig, float}
\usepackage[normalem]{ulem}
\usepackage[colorlinks=true,citecolor=blue,linkcolor=blue,urlcolor=blue]{hyperref}
\usepackage[capitalise]{cleveref}
\usepackage[dvipsnames]{xcolor}
\usepackage{multirow}
\usepackage{graphicx} 
\usepackage{dcolumn} 
\usepackage{slashed}
\usepackage{xpatch}
\usepackage[mathlines]{lineno} 
\usepackage{comment}
\usepackage{soul}
\usepackage{xcolor}
\usepackage{xfrac}
\usepackage{tabularx}
\usepackage{mathrsfs}
\usepackage{physics}
\usepackage[section]{placeins}
\usepackage{dsfont}
\usepackage{hhline}
\usepackage{amssymb}
\usepackage{booktabs}
\usepackage{tikz}
\usepackage{makecell}

\begin{document}

\preprint{JLAB-THY-26-4766}

\title{Impact of Future Dihadron Production Measurements on the \\ Transversity Distributions and Tensor Charges of the Nucleon}

\newcommand*{\TU}{Department of Physics, SERC, Temple University, Philadelphia, Pennsylvania 19122, USA}\affiliation{\TU}
\newcommand*{\WM}{Department of Physics, College of William \& Mary, Williamsburg, VA 23185, USA}\affiliation{\WM}
\newcommand*{\Duke}{Department of Physics, Duke University, Durham, North Carolina 27708, USA}\affiliation{\Duke}
\newcommand*{\LVC}{Department of Physics, Lebanon Valley College, Annville, Pennsylvania 17003, USA}\affiliation{\LVC}
\newcommand*{\PSU}{Division of Science, Penn State University Berks, Reading, Pennsylvania 19610, USA}\affiliation{\PSU}
\newcommand*{\JLAB}{Jefferson Lab, Newport News, VA 23606, USA \\
        \vspace*{0.2cm}
        {\bf JAM Collaboration \\ {\footnotesize \ (DiFF Analysis Group)}
        \vspace*{0.2cm} }}\affiliation{\JLAB}

\author{Y.~Sawaya}\email{yorgo.sawaya@temple.edu}\affiliation{\TU}
\author{C.~Cocuzza}\email{cjcocuzza@wm.edu}\affiliation{\WM}
\author{G.~Matousek}\email{gregory.matousek@duke.edu}\affiliation{\Duke}
\author{M.~McEneaney}\email{matthew.mceneaney@duke.edu}\affiliation{\Duke}
\author{A.~Metz}\email{metza@temple.edu}\affiliation{\TU}
\author{D.~Pitonyak}\email{pitonyak@lvc.edu}\affiliation{\LVC}
\author{A.~Prokudin}\email{prokudin@jlab.org}\affiliation{\PSU}\affiliation{\JLAB}
\author{N.~Sato}\email{nsato@jlab.org}\affiliation{\JLAB}
\author{A.~Vossen}\email{anselm.vossen@duke.edu}\affiliation{\Duke}

\begin{abstract}
We assess the impact of future measurements of dihadron production in semi-inclusive deep-inelastic scattering from the CLAS12 and proposed SoLID experiments at Jefferson Lab, as well as from the ePIC experiment at the future Electron-Ion Collider (EIC), on the transversity parton distribution functions (PDFs) and the corresponding tensor charges of the nucleon.
To this end, we generate pseudo-data for these experiments for a proton target (CLAS12 and ePIC) and a $^3$He target (SoLID and ePIC), and we include these pseudo-data in the JAMDiFF global analysis of existing experimental dihadron data.
We find that future data from Jefferson Lab will significantly reduce uncertainties in the transversity PDFs in the region of intermediate-to-large quark momentum fractions $x$, while the EIC will provide strong constraints across the entire range of $x$, allowing for the first experimental test of the predicted small-$x$ behavior of the transversity PDFs.
In discussing the reduction of uncertainties in the tensor charges, we also compare the results from the data analyses with those from lattice QCD, highlighting scenarios in which compatibility or tension between the two would arise.
\end{abstract}

\maketitle

\section{Introduction}
The transversity distribution $h_1^q$~\cite{Ralston:1979ys} is one of the three leading-twist quark parton distribution functions (PDFs) of the nucleon, alongside the unpolarized ($f_1^q$) and helicity ($g_1^q$) PDFs. 
It describes the distribution of transversely polarized quarks (of flavor $q$) in a transversely polarized nucleon, providing essential information on the spin structure of the nucleon in quantum chromodynamics (QCD)~\cite{Barone:2001sp, Lorce:2025aqp}. 
The first moment of the transversity PDFs defines the corresponding tensor charges,
\begin{equation}
\delta q (\mu)= \int_0^1 dx \, h_1^{q_v}(x;\mu) \,, 
\label{eq:deltaq}
\end{equation}
where $h_1^{q_v} = h_1^q - h_1^{\bar{q}}$ are the valence distributions, and $\mu$ is the renormalization scale.
The nucleon tensor charges are important for several areas of nuclear physics. 
For instance, extracting certain beyond-the-Standard-Model (BSM) couplings from the beta decay of free neutrons requires precise knowledge of the isovector combination $(\delta u - \delta d)$ of the up and down quark tensor charges~\cite{Herczeg:2001vk,Erler:2004cx,Severijns:2006dr,Cirigliano:2013xha,Courtoy:2015haa,Gonzalez-Alonso:2018omy}.
Moreover, the individual tensor charges $\delta u$ and $\delta d$ are key inputs in BSM studies that relate quark electric dipole moments to that of the nucleon~\cite{Erler:2004cx,Pospelov:2005pr,Yamanaka:2017mef,Liu:2017olr}.  
From the theoretical side, information on the nucleon tensor charges has been obtained from calculations in lattice QCD (LQCD)~\cite{Gupta:2018qil, Gupta:2018lvp, Yamanaka:2018uud, Hasan:2019noy, Alexandrou:2019brg, Harris:2019bih, Horkel:2020hpi, Alexandrou:2021oih, Park:2021ypf, Tsuji:2022ric, Bali:2023sdi, QCDSFUKQCDCSSM:2023qlx,Gao:2023ktu, Djukanovic:2024krw, Alexandrou:2024ozj, Wang:2025nsd} and in various models~\cite{He:1994gz, Barone:1996un, Schweitzer:2001sr, Gamberg:2001qc, Pasquini:2005dk, Wakamatsu:2007nc, Lorce:2007fa, Yamanaka:2013zoa, Pitschmann:2014jxa, Xu:2015kta, Wang:2018kto, Liu:2019wzj}. 

Since the transversity PDF is chiral-odd, it can only appear in observables in combination with another chiral-odd non-perturbative parton correlation function. 
Because of this complication, it is the least constrained of the three leading-twist quark PDFs of the nucleon. 
So far, efforts to extract the transversity PDFs from experimental data essentially rely on one of two methods. 
The first is measuring the Collins effect~\cite{Collins:1992kk} in single-hadron semi-inclusive deep-inelastic scattering (SIDIS), where the transversity couples to the transverse momentum dependent and chiral-odd Collins fragmentation function (FF)~\cite{Anselmino:2007fs, Anselmino:2008jk, Anselmino:2013vqa, Anselmino:2015sxa, Kang:2015msa, Lin:2017stx, DAlesio:2020vtw, Cammarota:2020qcw, Gamberg:2022kdb, Boglione:2021aha, Zeng:2024gun}. 
The second approach is using transverse-spin observables for dihadron production, which can be described in collinear factorization~\cite{Collins:1993kq,Jaffe:1997hf,Bianconi:1999cd,Radici:2001na,Bacchetta:2002ux,Bacchetta:2008wb,Bacchetta:2011ip,Bacchetta:2012ty,Radici:2015mwa,Radici:2016lam,Radici:2018iag,Cocuzza:2023oam,Cocuzza:2023vqs}. 
In this case, the hadronization process is parameterized in terms of dihadron FFs (DiFFs)~\cite{Collins:1993kq,Jaffe:1997hf,Bianconi:1999cd,Radici:2001na,Courtoy:2012ry,Pitonyak:2023gjx,Cocuzza:2023vqs,Rogers:2024nhb,Pitonyak:2025lin,Mahaut:2025hie}. 
For the observables of interest, $h_1^q$ couples to the chiral-odd DiFF $H_1^{\sphericalangle \, h_1 h_2 / q}$, which is sometimes referred to as the interference FF; see, for instance, Ref.~\cite{Metz:2016swz} for a review on FFs.
We also note that, recently, promising observables involving energy correlators have been proposed as a means to probe the transversity PDFs experimentally~\cite{Kang:2023big,Liu:2024kqt,Gao:2025evv,Cao:2025icu,Kang:2026pro}.
Beyond their role in transversity extractions, DiFFs are also of
interest in their own right as non-perturbative objects encoding spin and
momentum correlations in the hadronization process.
Furthermore, they have recently been employed for precision studies of the
Standard Model and searches for physics beyond it, including probes of
light-quark dipole moments and Yukawa couplings, quark entanglement, and
the color-octet mechanism in quarkonium
decays~\cite{Wen:2024cfu,Wen:2024nff,Cheng:2025cuv,Cao:2025qua,Cao:2025wfg,He:2026pxa}.

In the present work, we study the impact of future dihadron production measurements on the transversity PDFs and tensor charges of the nucleon.
For this purpose, we utilize the JAMDiFF framework~\cite{Cocuzza:2023oam,Cocuzza:2023vqs} that was previously developed and used for a comprehensive analysis of dihadron observables in electron-positron annihilation~\cite{Belle:2011cur, Belle:2017rwm}, SIDIS~\cite{HERMES:2008mcr, COMPASS:2023cgk}, and proton-proton collisions~\cite{STAR:2015jkc,STAR:2017wsi}. 
A simultaneous fit of these data provided the DiFFs for $\pi^+ \pi^-$ production and the nucleon transversity PDFs for light quarks and antiquarks. 
In particular, it was found that the extracted transversity PDFs and tensor charges are compatible with information obtained via single-hadron production and tensor charge results from LQCD~\cite{Cocuzza:2023oam,Cocuzza:2023vqs}.

Dihadron SIDIS measurements with transversely polarized targets are planned at Jefferson Lab, using the CEBAF Large Acceptance Spectrometer for operation at $12\,\textrm{GeV}$ beam energy (CLAS12)~\cite{Burkert:2020akg} and the Solenoidal Large Intensity Device (SoLID)~\cite{SoLID:2014proposal,Arrington:2022hgr}, and at the EIC with the ePIC detector~\cite{Accardi:2012qut, AbdulKhalek:2021gbh}.
We consider both a proton target (for CLAS12 and ePIC) and a $^3$He target (for SoLID and ePIC).
These two targets offer complementary sensitivity to the up and down quark transversity PDFs, making them essential for full flavor separation.
The baseline for our study is provided by the JAMDiFF global analysis~\cite{Cocuzza:2023vqs,Cocuzza:2023oam}, from which we generate pseudo-data for the mentioned experiments.  
To estimate their impact on the transversity PDFs and tensor charges, these pseudo-data are then included in a global fit together with all datasets used in Ref.~\cite{Cocuzza:2023vqs,Cocuzza:2023oam}.
An important aspect of our analysis is investigating whether a tension arises with tensor charge results obtained from LQCD.
We also mention that a related impact study, based on transversity-sensitive observables in single-hadron production, was presented in Ref.~\cite{Gamberg:2021lgx}.

Overall, our analysis shows that future dihadron SIDIS measurements will significantly improve the knowledge of the transversity PDFs and the tensor charges.
Specifically, CLAS12 data will probe a previously unexplored kinematic region and provide important constraints on the up quark transversity in the region of intermediate-to-large~$x$.
Depending on the measurement outcome, a tension with tensor charge results from LQCD may emerge.
SoLID with a $^3$He target provides access to a different combination of the transversity PDFs, enabling improved flavor separation and, in particular, significant constraints on the down quark transversity. 
This also substantially improves the determination of $\delta d$, allowing for a stringent test of its consistency with LQCD. 
Future EIC data taken with ePIC extend the kinematic coverage for both targets to low $x$, strongly reducing uncertainties on the transversity PDFs and tensor charges, and providing the first experimental test of the predicted small-$x$ behavior of the transversity PDFs~\cite{Kovchegov:2018zeq}. 

The paper is organized as follows: In Sec.~\ref{sec:methodology}, we describe our methodology, including the generation of the pseudo-data and the approach used for assessing the impact of a dataset.
In Sec.~\ref{sec:results}, we present the impact of the CLAS12, SoLID, and ePIC pseudo-data on the transversity PDFs and tensor charges, and we conclude in Sec.~\ref{sec:conclusion}.

\section{Methodology}
\label{sec:methodology}
Our study uses the JAMDiFF analysis framework presented in Refs.~\cite{Cocuzza:2023oam,Cocuzza:2023vqs}, which employs Bayesian inference and Monte-Carlo replica sampling techniques. 
We refer to these JAMDiFF papers for details about the parameterization of the transversity PDFs and the DiFFs, as well as the definition of the $\chi^2$ function.
The JAMDiFF analysis extracts three independent transversity PDFs: $h_1^{u_v}$,~$h_1^{d_v}$,~$h_1^{\bar{u}}$.
For $\bar{u}$ and $\bar{d}$ quark flavors we assume the relation between the antiquark transversity PDFs $h_1^{\bar{u}} = - h_1^{\bar{d}}$ motivated by large-$N_c$ arguments~\cite{Pobylitsa:2003ty}. All other sea quark transversities are assumed to be vanishing motivated by the lattice QCD studies~\cite{Gupta:2018lvp,Alexandrou:2019brg,Alexandrou:2021oih}  of tensor charges.

Furthermore, the DiFF $H_1^{\sphericalangle \, \pi^+ \pi^-/u}$ is extracted, together with the (unpolarized) DiFFs $D_1^{\pi^+ \pi^-/i}$ for four quark flavors and the gluon, that is, $i = u,\, s,\, c,\, b,\, g$.
As discussed in Ref.~\cite{Cocuzza:2023vqs}, all other DiFFs for $\pi^+ \pi^-$ production are either related to these or vanish.
The transversity PDFs and the DiFFs are defined at an initial scale and evolved to the relevant experimental scales using leading-order (LO) DGLAP evolution. 
The dihadron production data incorporated in the JAMDiFF analysis include the unpolarized cross section~\cite{Belle:2017rwm} and the Artru-Collins asymmetry~\cite{Belle:2011cur} in electron-positron annihilation from Belle, the transverse single-spin asymmetry (TSSA)
in SIDIS from HERMES~\cite{HERMES:2008mcr} and COMPASS~\cite{COMPASS:2023cgk}, and the TSSA in proton-proton collisions from STAR~\cite{STAR:2015jkc,STAR:2017wsi}.
In addition, PYTHIA simulations of electron-positron annihilation were used to achieve flavor separation of the unpolarized DiFFs.
In the present work, we adopt the exact same framework and datasets, supplemented by the pseudo-data for future experiments. 
We further note that COMPASS has recently reported a high-statistics measurement of the dihadron TSSA with a transversely polarized deuteron target~\cite{COMPASS:2025dwu}. These data were shown to provide non-negligible constraints on the transversity PDFs in the $x$-region probed by COMPASS, which partially overlaps with that of ePIC, albeit at lower $Q^2$ (see Fig.~\ref{fig:kin}). However, like the previous COMPASS dihadron measurements of Refs.~\cite{COMPASS:2012bfl,COMPASS:2014ysd}, these asymmetries are reported for pairs of oppositely charged unidentified hadrons, whereas the JAMDiFF framework and the present impact study apply to identified $\pi^+ \pi^-$ pairs. For this reason, of the COMPASS dihadron measurements, only the identified $\pi^+ \pi^-$ asymmetries of Ref.~\cite{COMPASS:2023cgk} enter our baseline.

As in the JAMDiFF work~\cite{Cocuzza:2023oam,Cocuzza:2023vqs}, we incorporate theoretical constraints to guide the fit.
The first is the Soffer positivity bound~\cite{Soffer:1994ww},
\begin{equation}
    |h_1^q(x;\mu)| \leq \frac{1}{2} \big[ f_1^q(x;\mu) + g_1^q(x;\mu) \big] \,,
\end{equation}
which mainly constrains the transversity PDFs in the large-$x$ region.
The second, which follows from a theoretical calculation~\cite{Kovchegov:2018zeq}, constrains the small-$x$ behavior of the transversity PDFs at the initial scale $\mu_0 = 1\,\mathrm{GeV}$,
\begin{equation}
    h_1^q(x;\mu_0)\big|_{x\to 0} \sim  x^\alpha  \;\;
    {\rm with}\;\;\alpha \simeq 0.17 \pm 0.085 \,.
    \label{eq:alpha}
\end{equation}
More details on the choice of the numerical range for the parameter $\alpha$ can be found in Ref.~\cite{Cocuzza:2023vqs}.

In our study, we use two distinct baselines.
The first, labeled JAMDiFF, was obtained from a fit to experimental dihadron data while enforcing the low-$x$ constraint in Eq.~\eqref{eq:alpha}, that is, it is the same as the analysis in Refs.~\cite{Cocuzza:2023oam,Cocuzza:2023vqs}.
The second, labeled JAMDiFF-x, is obtained from a fit to the same data but with the low-$x$ constraint relaxed, allowing $\alpha \in [-0.99, 5.00]$, where the lower limit for $\alpha$ is chosen such that the lowest moment of the transversity PDFs is finite. 
For the CLAS12 and SoLID analyses, we adopt the JAMDiFF baseline. 
Since these experiments do not probe the small-$x$ region, the choice of baseline has no effect on the assessed impact, as we have verified explicitly. 
The JAMDiFF-x baseline is instead employed for the ePIC analysis, ensuring that the impact of the measurements at small $x$ is assessed without the theoretical constraint determining the behavior in that region.
We emphasize that we consider the constraint in Eq.~\eqref{eq:alpha} to be very well motivated. 
However, in the case of ePIC, it is important to exploit the capability of the EIC to test the small-$x$ behavior of the transversity PDFs.

In what follows, we first describe the procedure for generating the pseudo-data used in our fits, and then outline the method employed to quantify their impact on the extracted transversity PDFs and tensor charges.

\subsection{Pseudo-data}
\label{sec:pseudo-data}
We consider pseudo-data for $\pi^+ \pi^-$ production in SIDIS for three experiments:~the CLAS12 experiment at Jefferson Lab, which plans to take data with a transversely polarized target in the near future, the proposed SoLID experiment at Jefferson Lab, and the future ePIC experiment at the EIC. 
For each experiment, pseudo-data are generated and included in the fit alongside the data sets used in the JAMDiFF analysis~\cite{Cocuzza:2023oam,Cocuzza:2023vqs}.

To generate the pseudo-data, we begin with projected kinematic points and their associated uncertainties for the observable, estimated from the expected detector acceptance and integrated luminosity. 
For each point, the TSSA for dihadron production in SIDIS~\cite{Bacchetta:2002ux,Bacchetta:2003vn,Bacchetta:2011ip,Cocuzza:2023oam,Cocuzza:2023vqs},
\begin{equation}
A_{UT} = \frac{\sum_q e_q^2 \, h_1^q(x; \mu) \, H_1^{\sphericalangle, q}(z, M_h; \mu)}{\sum_q e_q^2 \, f_1^q(x; \mu) \, D_1^q(z, M_h; \mu)} \,,
\label{eq:A_UT}
\end{equation}
is evaluated for all JAMDiFF replicas, and a value is then assigned to each kinematic point based on the resulting ensemble (which is either the mean of all replicas or the (mean $\pm\; 1\sigma$) -- see discussion of ``low'' and ``high'' scenarios below).
In Eq.~\eqref{eq:A_UT}, the sum runs over both quarks and antiquarks, $e_q$ is the quark charge in units of the elementary charge, $z$ is the fraction of the momentum of the fragmenting quark carried by the dihadron, and $M_h = \sqrt{P_h^2}$ is the dihadron invariant mass, with $P_h$ the four-momentum of the dihadron. The DiFFs are also understood to be for $\pi^+\pi^-$ production, so we drop the explicit superscript for brevity.
The expression in Eq.~\eqref{eq:A_UT} follows the COMPASS convention for $A_{UT}$, which does not include the quark depolarization factor, whereas this factor is included in the HERMES convention (see Eq.~(5) in Ref.~\cite{Cocuzza:2023vqs}).
We remark that, in the above expression and the rest of this work, we make no distinction between the momentum fraction $x$ and the Bjorken variable $x_B = Q^2 / (2P \cdot q)$, with $P$ $(q)$ the four-momentum of the nucleon (virtual photon) and $Q^2 = - q^2$.
In the LO approximation used here, the difference between these two variables is of ${\cal O}(1/Q^2)$.
The same applies to the difference between $z$ and $z_h = P \cdot P_h / P \cdot q$.

For a proton target, the numerator of Eq.~\eqref{eq:A_UT} can be evaluated according to
\begin{equation}
\label{eq:protonAUT}
    \sum_q e_q^2 \, h_1^{q/p} \, H_1^{\sphericalangle, q} = \frac{1}{9} \Big(4h_1^{u_v/p} - h_1^{d_v/p} \Big) \, H_1^{\sphericalangle,u} \,,
\end{equation}
meaning that the proton TSSA $A_{UT}^p$ is mostly sensitive to $h_1^{u_v/p}$.
For a $^3\mathrm{He}$ target, which is taken to be equivalent to a neutron target (modulo a polarization factor of 0.873)~\cite{SoLID:2014proposal}, we find 
\begin{equation}
\label{eq:neutronAUT}
    \sum_q e_q^2 \, h_1^{q/n} \, H_1^{\sphericalangle, q} = \frac{1}{9} \Big(4h_1^{d_v/p} - h_1^{u_v/p} \Big) \, H_1^{\sphericalangle,u} \,,
\end{equation}
meaning that the neutron TSSA $A_{UT}^n$ is mostly sensitive to $h_1^{d_v/p}$.
To obtain Eqs.~\eqref{eq:protonAUT} and~\eqref{eq:neutronAUT}, we used for $\pi^+ \pi^-$ production that $H_1^{\sphericalangle,d}=-H_1^{\sphericalangle,u}$, $H_1^{\sphericalangle,q}=0$ for $q=s, \,\bar{s}, \, c, \, \bar{c}, \, b, \, \bar{b}$, and the isospin relations $h_1^{u_v/n} = h_1^{d_v/p}$, $h_1^{d_v/n} = h_1^{u_v/p}$; see also Ref.~\cite{Cocuzza:2023vqs} for more details.
All results for the transversity PDFs presented in this paper correspond to a proton target.

The following kinematic cuts are applied to the pseudo-data: $W^2 > 4 \, \text{GeV}^2$ (where $W$ is the center-of-mass energy of the virtual photon-nucleon system) to avoid the resonance region, as well as $0.25 < z < 0.8$ and $M_h/Q < 0.75$ to exclude regions where the leading-order, leading-twist formalism may not apply. 
We note that the cut on $M_h/Q$, which is similar in spirit to the cut
applied to the Belle data in the JAMDiFF analysis (see Eq.~(35) in
Ref.~\cite{Cocuzza:2023vqs}), was not imposed on the SIDIS data in that
analysis, where it would remove only three COMPASS data points.
It is introduced here because the pseudo-data, in particular for SoLID,
contain points for which $M_h$ would otherwise be comparable to the hard
scale $Q$.
Together, the projected kinematic points, the corresponding calculated values for $A_{UT}$, and the associated estimated experimental errors constitute the pseudo-data.
To explore different possible outcomes of the future measurements, we consider more than one scenario for the pseudo-data. 
In the default scenario, each pseudo-data point is assigned the mean value of $A_{UT}$ from the replica ensemble. 
In the alternative scenarios, following an approach similar to that of Ref.~\cite{Zhou:2021llj}, pseudo-data are instead generated from the $\pm 1\sigma$ boundaries of the JAMDiFF uncertainty band for the observable, labeled ``high'' and ``low'' according to the magnitude of the asymmetry. 
We note that for observables that are predominantly negative, such as the proton TSSA, the ``low'' scenario corresponds to the upper boundary of the uncertainty band.
These alternative scenarios are applied selectively to the three experiments where the resulting shift in the pseudo-data could lead to qualitatively different physics conclusions, such as a change in compatibility with tensor charges from LQCD.

The pseudo-data are then added to the original JAMDiFF datasets to perform a new fit. 
Table~\ref{tab:exp_data} summarizes the pseudo-data that we consider, their binnings, numbers of points, luminosities, and systematic errors. 
The one-dimensional (1D) binning used for CLAS12 and ePIC reduces complexity and allows for a more direct assessment of the impact on the transversity PDFs. 
Extensive tests were conducted to verify that the choice of binning has no effect on the qualitative conclusions of this paper.
For SoLID, only 4D binning was available to us~\cite{SoLID:2014proposal}.
\begin{table}[t]
\centering
\setlength{\tabcolsep}{10pt}
\begin{tabular}{cccccccc}
\hhline{========}
\vspace{-0.2cm}
& & & & & & & \\
\makecell{Experiment \\ ~} 
& \makecell{Energy \\ {[GeV]}} 
& \makecell{Target \\ ~}
& \makecell{Binning \\ ~}
& \makecell{$N_{\text{dat}}$ \\ ~}
& \makecell{Luminosity \\ {[fb$^{-1}$]}}
& \makecell{Normalization \\ error [\%]}
& \makecell{Uncorrelated \\ error [\%]} \\
\vspace{-0.2cm}
& & & & & & & \\
\hline
\vspace{-0.2cm}
& & & & & & & \\
CLAS12~\cite{Burkert:2020akg} & 10.6 & \rm p & 1D & 9 & 43.2 & 5 & 5 \\
\vspace{-0.2cm}
& & & & & & & \\
\hline
\vspace{-0.2cm}
& & & & & & & \\
SoLID~\cite{SoLID:2014proposal,Arrington:2022hgr} & 8.8, 11 & $^3$He & 4D & 715 & 6220.8 & 3 & 6.8 \\
\vspace{-0.2cm}
& & & & & & & \\
\hline
\vspace{-0.2cm}
& & & & & & & \\
\multirow{2}{*}{ePIC~\cite{AbdulKhalek:2021gbh}} 
& 10 $\times$ 100 & \rm p & 1D & 18 & 10 & 2.3 & 5 \\
& 10 $\times$ 166 & $^3$He & 1D & 19 & 8.65 & 2.3 & 5 \\
\hhline{========}
\end{tabular}%
\caption{Summary of the pseudo-data sets considered in this study. For the fixed-target experiments CLAS12 and SoLID, the energies of the electron beam are shown. In the case of SoLID, projections at the beam energies $8.8\,\textrm{GeV}$ and $11\,\textrm{GeV}$ are combined. 
For the ePIC collider experiment, we consider an electron beam energy of $10 \, \textrm{GeV}$ for both the proton and $^3$He targets. The CLAS12 and ePIC pseudo-data are binned in $x$, while the SoLID pseudo-data are binned in $x, z, M_h, Q^2$. 
We list separately the normalization uncertainties on $A_{UT}$ and other (uncorrelated) systematic errors.}
\label{tab:exp_data}
\end{table}

The projected uncertainties for CLAS12 were estimated from an already recorded dataset with a longitudinally polarized target, assuming the run time approved by the Jefferson Lab Program Advisory Committee and conservative extrapolations of running conditions from existing data. 
Systematic uncertainties are derived from existing SIDIS analyses with the CLAS12 detector and are taken to be 5\% point-to-point and 5\% scale uncertainties. 
The former covers backgrounds from resonances as well as bin migration effects, whereas the latter is due to uncertainties determining target polarization and dilution factors.
For SoLID and ePIC, the uncertainties were derived from best estimates available at the time of writing of this paper. 

For the ePIC data a 5\% point-to-point and 2.3\% scale uncertainty systematics are used. 
These are derived from conservative estimates based on previous experiments and expected machine performance. 
The point-to-point systematics cover kinematic smearing and radiative effects, whereas the scale uncertainty covers expected uncertainties on beam polarization, luminosity, and electron purity. 
The assumed systematics are somewhat more conservative than the values assumed in the EIC Yellow Report~\cite{AbdulKhalek:2021gbh}. For the $^3$He sample, we assume 100\% tagging efficiency of the spectator protons. 
The dataset size and single energy configurations, $10\times 100 \; \textrm{GeV}$ for $e + p$ and $10 \times 166 \; \textrm{GeV}$ for $e \, +$ $^3$He, are based on what can reasonably be achieved in the first ten years of running at the EIC. 
The precision of the EIC dataset profits from the increased multiplicity at the higher center-of-mass energy and the hermetic acceptance of the project detector. Systematic effects, for example due to target polarization and dilution as well as background from baryonic resonances, are also expected to be smaller compared to fixed-target experiments.
The systematic uncertainties for SoLID, 3\% scale uncertainties and 6.8\% point-to-point, were provided by the collaboration~\cite{SoLID:2014proposal}.

\begin{figure}[t]
    \centering
    \includegraphics[width=0.625\linewidth]{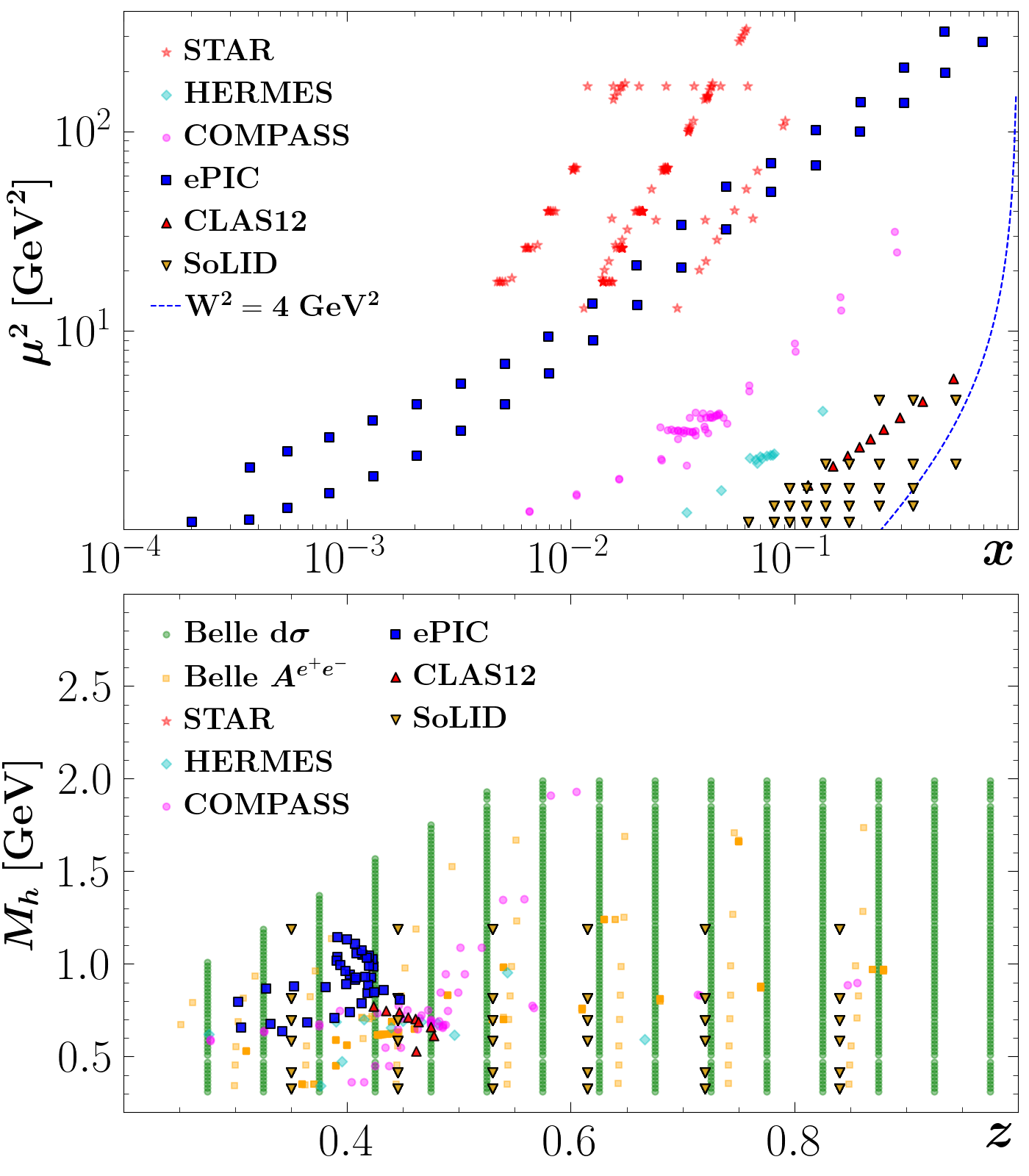}
    \caption{Coverage in $x$, $\mu^2$, $z$, and $M_h$ of projected pseudo-data, as well as of the dihadron data from Belle~\cite{Belle:2017rwm,Belle:2011cur}, HERMES~\cite{HERMES:2008mcr}, COMPASS~\cite{COMPASS:2023cgk}, and STAR~\cite{STAR:2015jkc,STAR:2017wsi} that were used in the JAMDiFF analysis~\cite{Cocuzza:2023oam,Cocuzza:2023vqs}. The scale $\mu^2$ represents $Q^2$ for SIDIS, and $P_{hT}^2$ for proton-proton scattering, with $P_{hT}$ denoting the dihadron transverse momentum.} 
    \label{fig:kin}
\end{figure}

The kinematic coverages of the different datasets used in the JAMDiFF analysis~\cite{Cocuzza:2023oam,Cocuzza:2023vqs} and for the pseudo-data are shown in Fig.~\ref{fig:kin}. 
As illustrated in the figure, the SoLID coverage in $x$ is similar to that of CLAS12.
However, given the different targets, these two experiments are complementary, as they are primarily sensitive to different quark flavors; see Eqs.~\eqref{eq:protonAUT} and~\eqref{eq:neutronAUT}.
By contrast, the ePIC collider experiment will probe a much wider kinematic range than any existing or planned fixed-target experiment, with both proton and $^3$He targets enabling simultaneous constraints on the up and down quark transversity PDFs.
Furthermore, compared to CLAS12 and SoLID, the ePIC experiment will provide data at significantly higher $Q^2$ for overlapping regions in $x$, thereby complementing the Jefferson Lab measurements. 

\subsection{Quantifying the Impact}
\label{sec:quant_imp}
To estimate the constraining power of future dihadron data, we compare results from the baseline fit with those from a fit including the respective additional pseudo-data. 
In particular, we determine how the pseudo-data modify the uncertainties of the extracted transversity PDFs and tensor charges.
In the case of the transversity PDFs, this effect is visualized through uncertainty-band plots of $h_1^{u_v}$ and $h_1^{d_v}$ before and after the pseudo-data are added, and quantified by plotting the ratio of uncertainties (before/after) as in Fig.~\ref{fig:PDFs}.

The JAMDiFF analysis found that the analyzed transversity-related dihadron data are compatible with LQCD results for the tensor charges~\cite{Cocuzza:2023vqs,Cocuzza:2023oam}.
Here we revisit this question of compatibility in detail after the inclusion of the pseudo-data.
If the pseudo-data make the fit fail when LQCD tensor charges are used as input --- in the sense that the $\chi^2$ values (and corresponding $Z$ scores, where $Z = \Phi^{-1}(p) = \sqrt{2} \, {\rm erf}^{-1}(2p-1)$, with the $p$ value of the resulting $\chi^2$ computed relative to the expected $\chi^2$ distribution) for the LQCD tensor charges and/or the pseudo-data are unacceptably large --- this suggests a possible tension.
However, the small number of LQCD points might lead the fit to ignore them in favor of a marginally better description of the experimental data. 
This could falsely suggest a tension, even though a simultaneous fit with slightly worse but still acceptable $\chi^2$ may in fact be possible.
To avoid this situation and determine whether a genuine tension with the LQCD results exists, we multiply the LQCD contribution to the $\chi^2$ function by a weight that ensures the lattice results carry an importance that is comparable to the experimental data in the fit.
The weight is calculated as
\begin{equation}
   w = \frac{N_{\mathrm{tot}}}{N_{\rm LQCD}} \,,\label{e:LQCDweight}
\end{equation}
where $N_{\rm tot}$ is the total number of data points in the fit and $N_{\rm LQCD}=4$ is the number of LQCD data points. 
(Here we utilize the same LQCD tensor charge results as those employed in the JAMDiFF analysis~\cite{Cocuzza:2023vqs,Cocuzza:2023oam}.)
A similar weighting procedure was used in Refs.~\cite{NNPDF:2021njg,Echevarria:2020hpy,Bhattacharya:2021twu}.
The weighting guarantees that the fit does not disregard the few LQCD data points.

We note that our study is designed to evaluate an experiment's impact on the precision (size of uncertainties) of the extracted transversity PDFs, rather than on their accuracy (central values), since the underlying assumption is that future measurements will be consistent with pseudo-data generated from a given baseline. 
To more thoroughly assess the potential of future experiments to find agreement or tension with LQCD results, we consider, in addition to the default prediction, the ``high'' and ``low'' scenarios described in Sec.~\ref{sec:pseudo-data}.

\section{Results}
\label{sec:results}
In what follows, we present the impact of the three future experiments separately. 
For each, we examine the resulting reduction in the uncertainty bands of the transversity PDFs and tensor charges.
These results are supplemented by a $\chi^2$ table showing the quality of the various fits.
We emphasize that we are fitting all nine parton correlation functions (three transversity PDFs and six DiFFs) listed at the beginning of Sec.~\ref{sec:methodology}, even though our main interest here is in $h_1^{u_v}$ and $h_1^{d_v}$.
The DiFFs extracted in the JAMDiFF analysis~\cite{Cocuzza:2023oam,Cocuzza:2023vqs} are already well constrained by the Belle and PYTHIA data, with the exception of the unpolarized gluon DiFF, which enters only through evolution. 
For this reason, including the pseudo-data leaves the DiFFs essentially unchanged, even with the 4D binning of SoLID.
For the antiquark transversity PDFs, we likewise find no significant change relative to JAMDiFF for any of the three pseudo-data sets, in any of the default, ``low'', and ``high'' scenarios. 
This is because the SIDIS TSSAs probe only the valence distributions; see Eqs.~\eqref{eq:protonAUT} and \eqref{eq:neutronAUT}. 
Because of the proton-proton data~\cite{STAR:2015jkc,STAR:2017wsi}, an improvement of $h_1^{u_v}$ and $h_1^{d_v}$ could, in principle, also result in improved antiquark transversity PDFs.
However, we observe only a negligible effect of this type because of the limited precision of the proton-proton data.

\subsection{CLAS12}
\label{sec:CLAS}
\begin{figure}[t]
    \centering
    {\Large  \bf{CLAS12} $\boldsymbol{A_{UT}^p[\%]}$}
    
    \vspace{0.3cm}
    \includegraphics[width=0.6\linewidth]{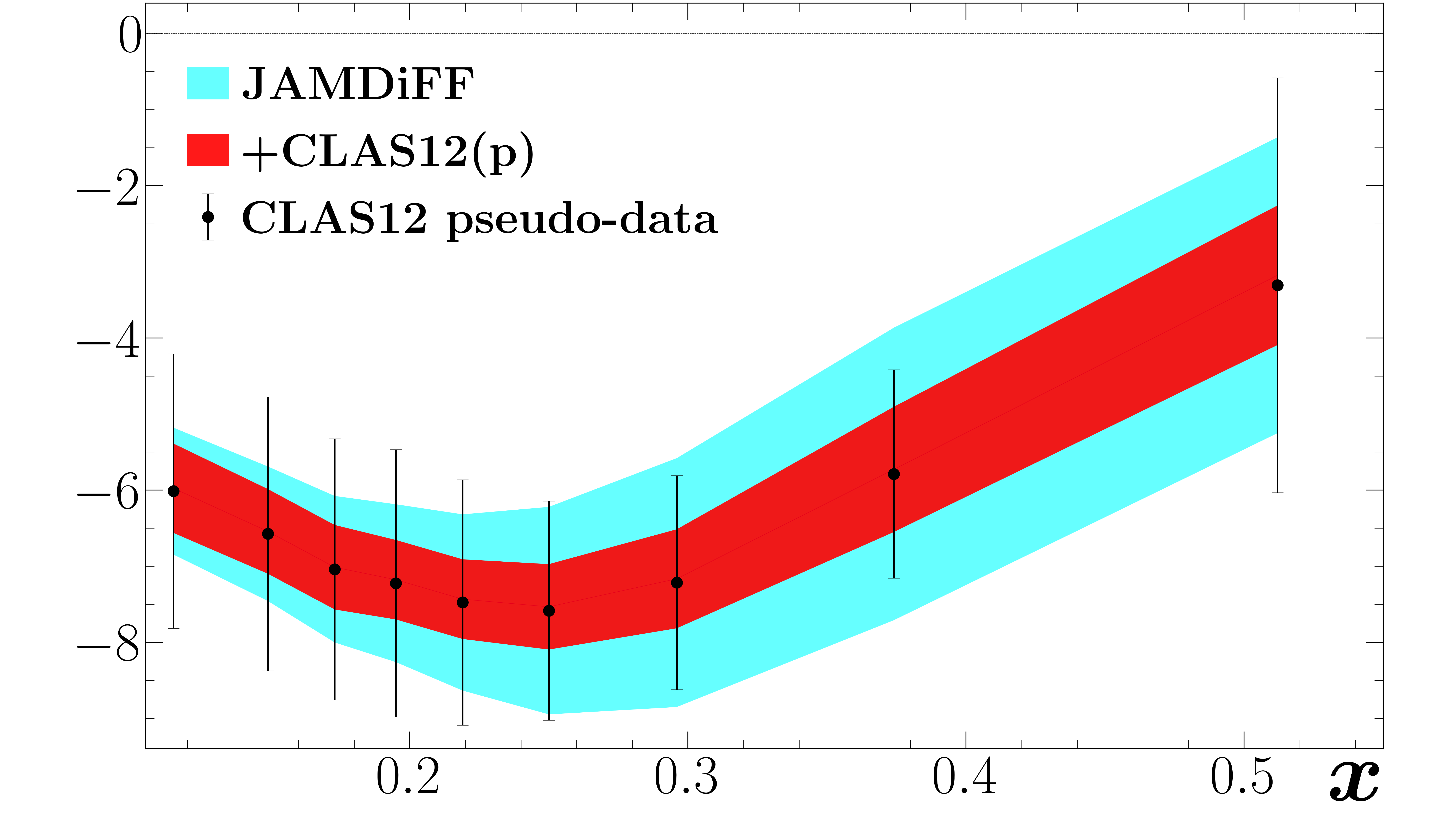}
    \caption{1$\sigma$ uncertainty bands of the TSSA $A_{UT}^p$ for CLAS12 kinematics: results based on the JAMDiFF analysis~\cite{Cocuzza:2023oam,Cocuzza:2023vqs} (cyan) and after including the CLAS12 pseudo-data in the fit (red). }
    \label{fig:CLAS}
\end{figure}
The dihadron SIDIS data in the JAMDiFF analysis~\cite{HERMES:2008mcr,COMPASS:2023cgk} roughly cover the range $0.005 \lesssim x \lesssim 0.3$ (see Fig.~\ref{fig:kin}) and have limited precision.
CLAS12 will provide data in the region of intermediate-to-high $x$ with improved precision and therefore has significant potential to improve our knowledge on the transversity PDFs and the tensor charges.

\begin{figure}[t]
    \centering
    \includegraphics[width=0.9\linewidth]{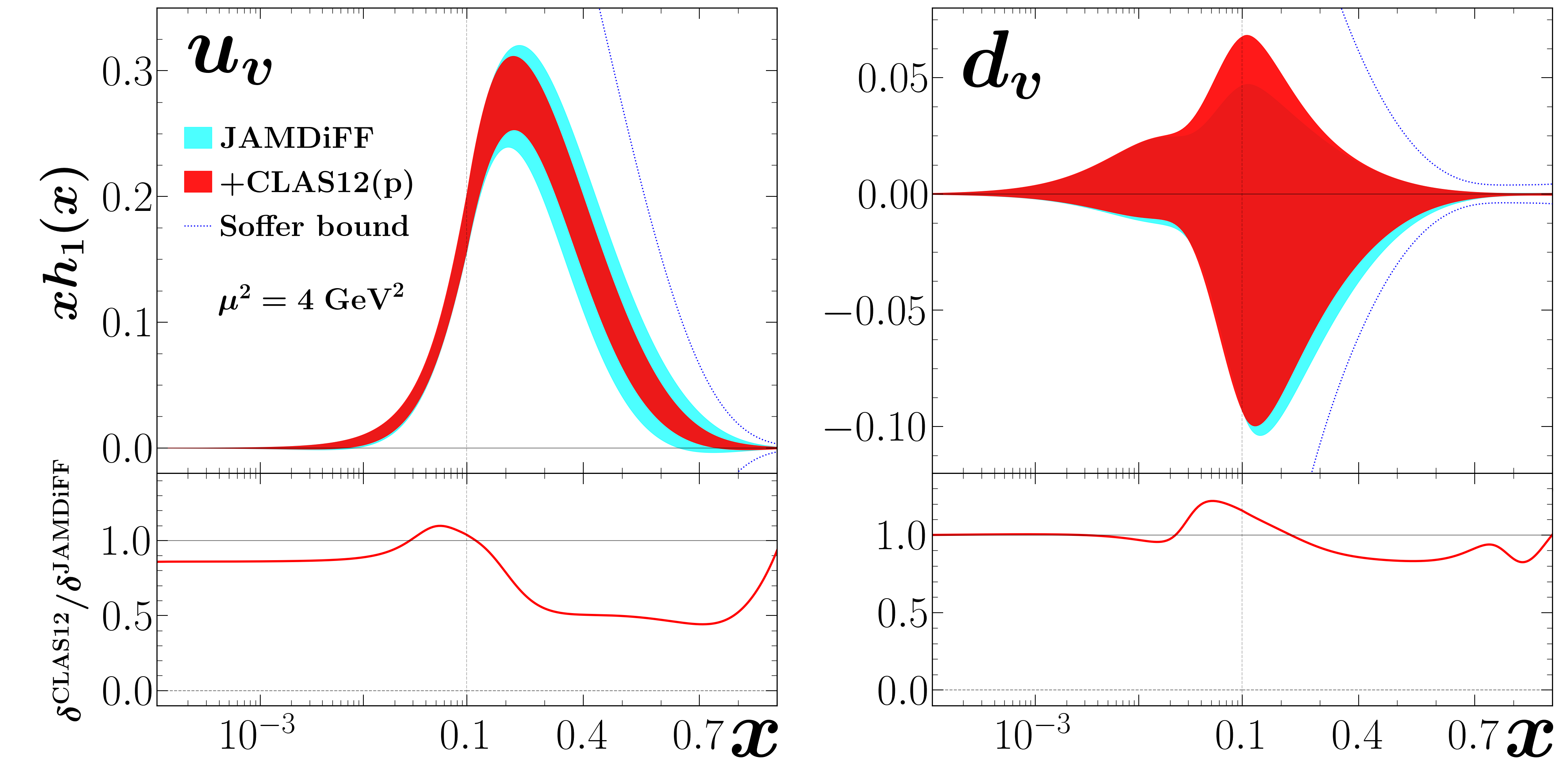}
\caption{1$\sigma$ uncertainty bands of $h_1^{u_v}$ and $h_1^{d_v}$: results based on the JAMDiFF analysis~\cite{Cocuzza:2023oam,Cocuzza:2023vqs} (cyan) and after including the CLAS12 pseudo-data in the fit (red). The Soffer bound is indicated by the blue dotted lines. The lower panels show the ratio of the uncertainties for the transversity PDFs (with/without CLAS12 pseudo-data).
Note that below $x = 0.1$ we switch from a linear scale to a logarithmic scale.}
    \label{fig:PDFs}
\end{figure}

As shown in Fig.~\ref{fig:CLAS}, the inclusion of the CLAS12 pseudo-data in the fit narrows the uncertainty bands of the theoretical prediction for $A_{UT}^p$. 
According to Eq.~\eqref{eq:protonAUT}, $A_{UT}^p$ is mostly sensitive to $h_1^{u_v}$.
Furthermore, the magnitude of $h_1^{u_v}$ is larger than that of $h_1^{d_v}$ in the kinematic range of CLAS12~\cite{Cocuzza:2023oam,Cocuzza:2023vqs}.
One might, therefore, expect that any improvement on $A_{UT}^p$ would primarily translate into an improved determination of $h_1^{u_v}$.
However, new precise data for $A_{UT}^p$ better constrain the linear combination $4 h_1^{u_v} - h_1^{d_v}$, leading to a strong correlation between $h_1^{u_v}$ and $h_1^{d_v}$, but not necessarily to a proportional reduction in their uncertainties when the errors of $4h_1^{u_v}$ and $h_1^{d_v}$ are comparable before the inclusion of the new data.
In the region $x \gtrsim 0.2$, however, the errors from the JAMDiFF analysis for $4h_1^{u_v}$ are much larger than those for $h_1^{d_v}$~\cite{Cocuzza:2023oam,Cocuzza:2023vqs}, causing a substantial reduction in the uncertainties of $h_1^{u_v}$, as shown in Fig.~\ref{fig:PDFs}.
This quantifies the potential of a future CLAS12 dihadron measurement.

Another observed effect is that, in some kinematic regions, the uncertainty ratio can exceed unity; see the lower panels in Fig.~\ref{fig:PDFs}.
We find this phenomenon even for the observable itself, although to a lesser extent and only outside of the kinematic region of the pseudo-data.
This reflects the fit adjusting to accommodate the pseudo-data, which can come at the cost of slightly increased uncertainty where the pseudo-data have less impact.
This behavior has been previously observed in QCD analyses; see, e.g., Ref.~\cite{Zhou:2021llj} and Fig.~7.4 in Ref.~\cite{AbdulKhalek:2021gbh}.

\begin{figure}[t]
    \hspace*{2.3cm}
    \includegraphics[width=0.9\linewidth]{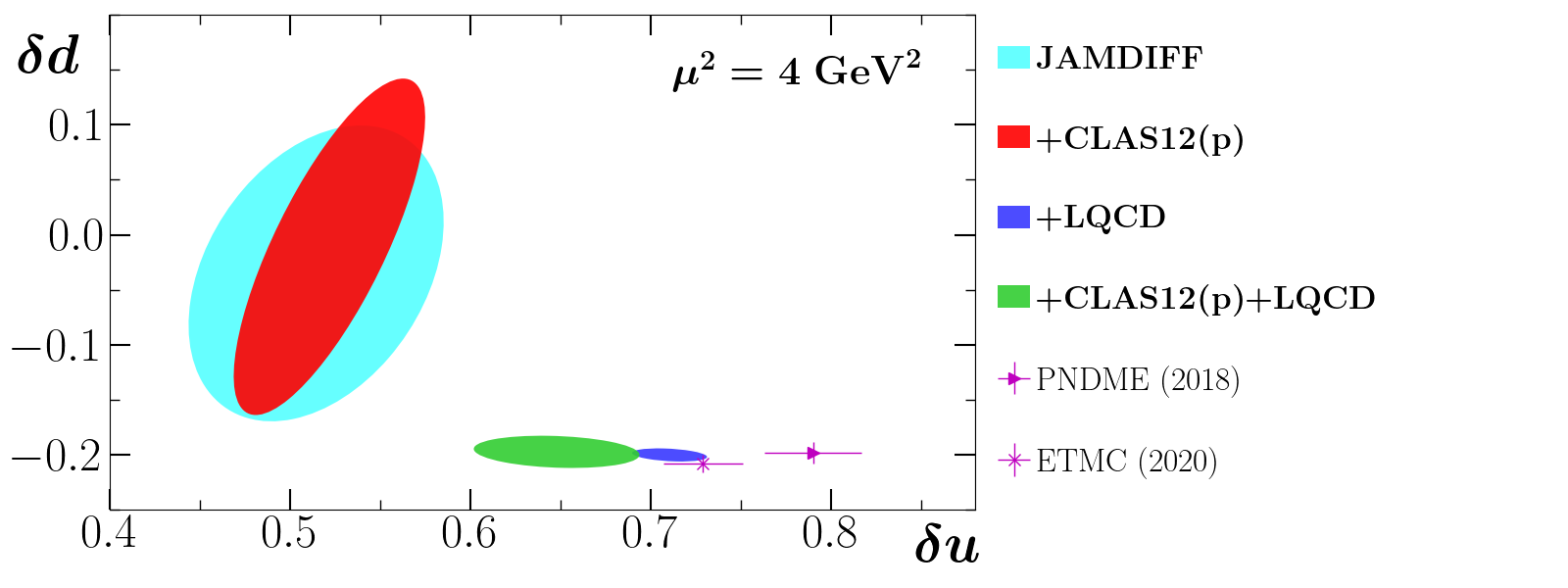}
    \caption{$1\sigma$ confidence ellipses for the tensor charges $\delta u$ and $\delta d$: results based on the JAMDiFF analysis~\cite{Cocuzza:2023oam,Cocuzza:2023vqs} without LQCD (cyan) and with LQCD (blue), and after including the CLAS12 pseudo-data in the fit without LQCD (red) and with LQCD (green). Also shown are the LQCD tensor charge results by PNDME~\cite{Gupta:2018lvp} and ETMC~\cite{Alexandrou:2021oih}.}
    \label{fig:gT}
\end{figure}

In Fig.~\ref{fig:gT} we show the impact of the pseudo-data on the tensor charges. 
We find only a modest improvement for $\delta u$ and no improvement for $\delta d$, but the linear combination $4\delta u - \delta d$ becomes much better constrained through the CLAS12 pseudo-data.
The quality of this fit is shown in the ``no LQCD'' column of Table~\ref{tab:chi2_clas}, which lists the values of the reduced $\chi^2$
($\chi_{\rm red}^2$) for the various fits. 
The value $\chi_{\rm red}^2 = 0.00$ (to two decimal places) for CLAS12 reflects the fit reproducing the pseudo-data nearly exactly. 
This outcome is expected, since the pseudo-data were generated in the default scenario (see Sec.~\ref{sec:pseudo-data}) and the projected CLAS12 errors are comparable to the JAMDiFF uncertainty bands for the asymmetry. 
However, when pseudo-data have very small uncertainties compared to the baseline bands and span a large kinematic range, the fit can lead to (considerable) deviations from $\chi_{\rm red}^2 = 0$ for the pseudo-data.

\begin{table}[b]
\centering
\setlength{\tabcolsep}{8pt}
\renewcommand{\arraystretch}{1.15}
\begin{tabular}{l c c | c c c c}
\toprule\toprule
\multicolumn{3}{c|}{} & \multicolumn{4}{c}{$\chi^2_{\text{red}}$} \\
\cmidrule(l){4-7}
Experiment & Target & $N_{\text{dat}}$ & no LQCD & w/ LQCD & w/ LQCD (weighted) & ``low'' w/ LQCD \\
\midrule
\textbf{CLAS12} \cite{Burkert:2020akg} & \bf{p} & \textbf{9} & \textbf{0.00} & \textbf{0.22} & \textbf{2.24} & \textbf{2.85} \\
\midrule
Belle ($d\sigma$) \cite{Belle:2017rwm}& -- & 1094 & 1.04 & 1.03 & 1.05 & 1.06 \\[0.09cm]
Belle (asym) \cite{Belle:2011cur}& -- & 183 & 0.64 & 0.79 & 0.68 & 0.65 \\
\midrule
HERMES \cite{HERMES:2008mcr} & p & 12 & 1.07 & 1.04 & 1.15 & 1.06 \\
\midrule
COMPASS \cite{COMPASS:2023cgk} & p, D & 47 & 0.83 & 1.17 & 1.20 & 1.09 \\
\midrule
STAR \cite{STAR:2015jkc,STAR:2017wsi} & pp & 130 & 1.13 & 1.23 & 1.36 & 1.17 \\
\midrule
$\delta u$ (ETMC) \cite{Alexandrou:2021oih} & -- & 1 & -- & 13.62 & 0.00 & 3.86 \\
$\delta u$ (PNDME) \cite{Gupta:2018lvp} & -- & 1 & -- & 27.73 & 5.07 & 14.90 \\
$\delta d$ (ETMC) \cite{Alexandrou:2021oih} & -- & 1 & -- & 1.93 & 1.48 & 1.28 \\
$\delta d$ (PNDME) \cite{Gupta:2018lvp} & -- & 1 & -- & 0.01 & 0.00 & 0.01 \\
\bottomrule
\end{tabular}
\caption{Summary of $\chi_{\rm red}^2$ values after including the CLAS12 pseudo-data in the fit:~without LQCD, with LQCD, with LQCD and re-weighting, with LQCD  and pseudo-data for the ``low'' scenario. For Belle, we list separately the results for the unpolarized cross-section ($d\sigma$) and the Artru-Collins asymmetry (asym). }
\label{tab:chi2_clas}
\end{table}

To examine the compatibility between experimental data and LQCD calculations of the tensor charges, we follow the JAMDiFF analysis~\cite{Cocuzza:2023oam,Cocuzza:2023vqs} and consider fits both without and with LQCD inputs. 
The JAMDiFF fit to experimental data yields tensor charges that suggest tension with LQCD results~\cite{Cocuzza:2023oam,Cocuzza:2023vqs}, especially for $\delta u$; see the cyan ellipse in comparison to the LQCD results in Fig.~\ref{fig:gT}.
However, after including the LQCD results in the fit, compatibility was achieved, and potential reasons were presented for why such a significant shift in $\delta u$ could occur~\cite{Cocuzza:2023oam,Cocuzza:2023vqs}.
Here we follow the same procedure and test whether both experimental data and LQCD results can be simultaneously described.
This, and further tests of agreement, are outlined in Sec.~\ref{sec:quant_imp}. 
The $\chi_{\rm red}^2$ values for these various scenarios are summarized in Tab.~\ref{tab:chi2_clas}.

We find that when LQCD tensor charges are included, the fit seemingly cannot accommodate the CLAS12 pseudo-data generated from the default JAMDiFF prediction together with the LQCD tensor charges; see the green ellipse in Fig.~\ref{fig:gT} and the $\chi_{\rm red}^2$ values in the ``w/ LQCD'' column in Tab.~\ref{tab:chi2_clas}. 
However, applying the weighting procedure leads to good agreement with both LQCD calculations, and a $\chi_{\rm red}^2$ value of $2.24$ for the CLAS12 pseudo-data ($Z= 2.12$). 
In the default scenario, the results remain therefore inconclusive.
Since the future CLAS12 data may not agree with the default scenario, we further consider one of the alternative scenarios described in Sec.~\ref{sec:pseudo-data}. 
For the proton TSSA measured by CLAS12, the ``low'' scenario corresponds to a smaller $h_1^{u_v}$, pushing the resulting tensor charge $\delta u$ further from LQCD predictions and making a simultaneous fit more demanding. 
Indeed, in this case the fit fails to describe the pseudo-data and the LQCD tensor charges; see the last column in Tab.~\ref{tab:chi2_clas}. 
While the weighting method enforces a good description of the LQCD results, it also causes the $\chi_{\rm red}^2$ for the CLAS12 pseudo-data to increase further (from 2.85 to 3.88), and their $Z$ score to increase to 3.84.  
Note also that the changes of the $\chi_{\rm red}^2$ values for the existing dihadron data are small across the various fits. 
Overall, we find that, regardless of the measurement outcome, CLAS12 can provide very valuable new information on whether or not a tension exists between phenomenological extractions and LQCD predictions of the tensor charges. 
We note that this observed tension relies on the use of theoretical constraints (the Soffer bound and the low-$x$ asymptotics; see Sec.~\ref{sec:methodology}). 
Without them, the fit retains enough flexibility in regions unconstrained by data to accommodate the LQCD tensor charge results.
Finally, we point out that, in the JAMDiFF analysis without (with) LQCD input for the tensor charges, 67\% (70\%) of the value for $\delta u$ comes from the $x$-range covered by CLAS12. 
This further underlines the importance of this experiment.  

\subsection{SoLID}
\begin{figure}[b]
    \centering
    {\hspace{0.8cm} \large  \bf{SoLID $\boldsymbol{A_{UT}^n[\%]}$ for $Q^2 = 1.325~\mathrm{GeV}^2$}}

    \includegraphics[width=0.7\linewidth]{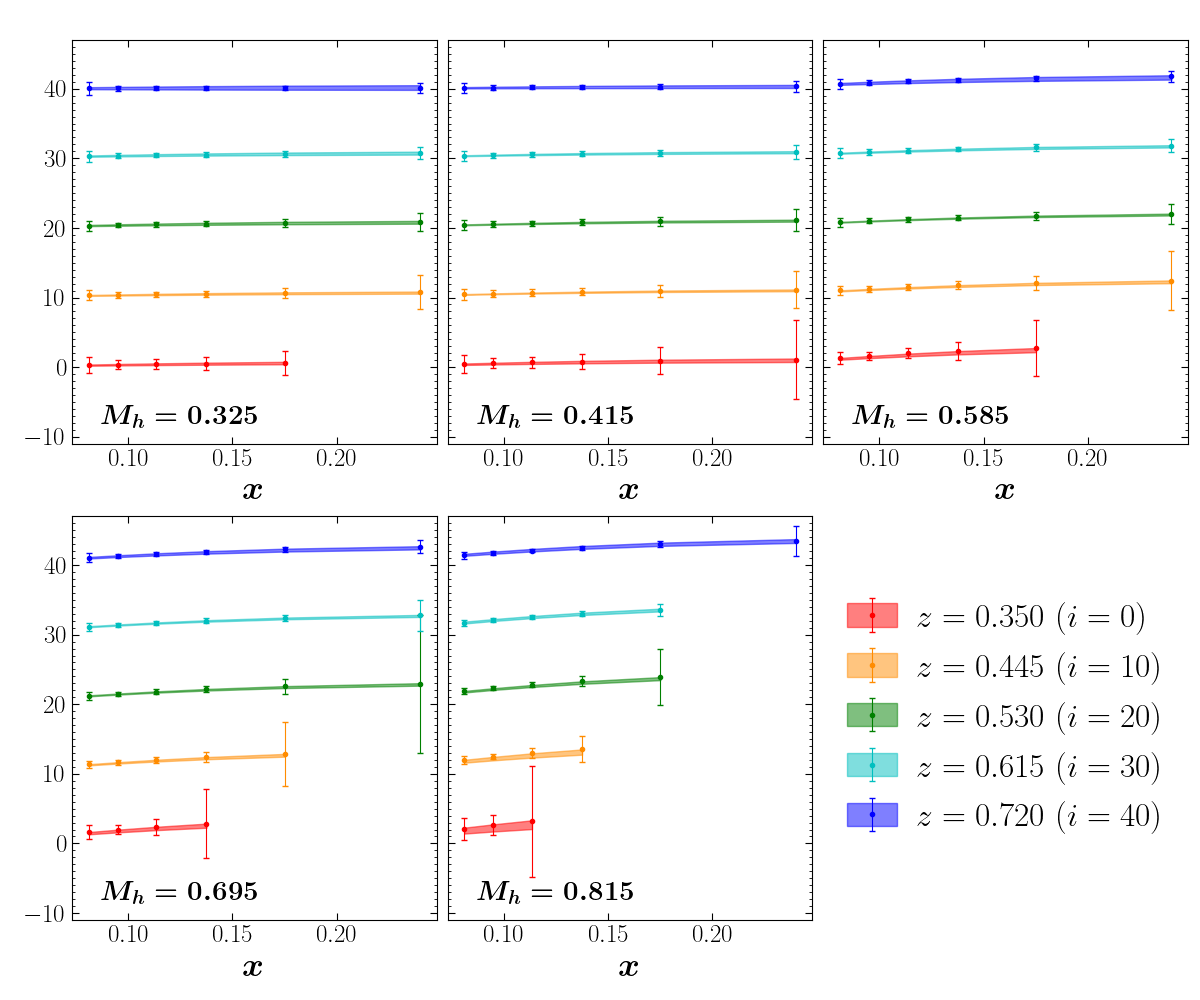}
    \caption{$1\sigma$ uncertainty bands of the TSSA $A_{UT}^n$ for SoLID kinematics at $Q^2 = 1.325~\text{GeV}^2$ after including the SoLID pseudo-data in the fit. Panels correspond to bins in $M_h$ and curves to bins in $z$, with each curve offset vertically by the value $i$ indicated in the legend.}
    \label{fig:SoLID}
\end{figure}

\begin{figure}[t]
    \includegraphics[width=0.9\linewidth]{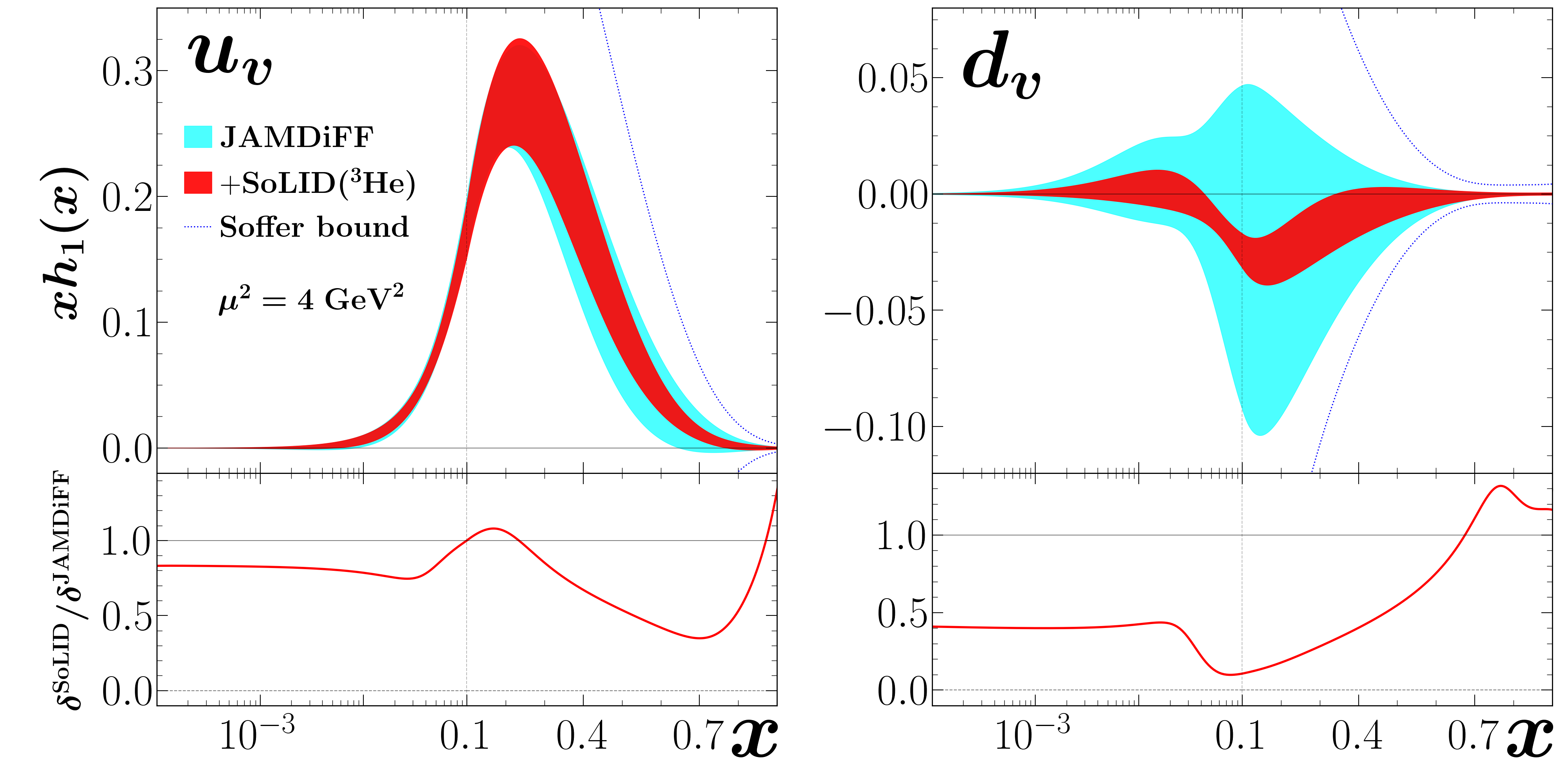}
\caption{1$\sigma$ uncertainty bands of $h_1^{u_v}$ and $h_1^{d_v}$:~results based on the JAMDiFF analysis~\cite{Cocuzza:2023oam,Cocuzza:2023vqs} (cyan) and after including the SoLID pseudo-data in the fit (red). The Soffer bound is indicated by the blue dotted lines. The lower panels show the ratio of the uncertainties for the transversity PDFs (with/without SoLID pseudo-data).}
    \label{fig:sPDFs}
\end{figure}

The SoLID pseudo-data are binned in four kinematic variables~\cite{SoLID:2014proposal}:~$x$, $Q^2$, $z$, and $M_h$. 
Figure~\ref{fig:SoLID} shows a subset of the $A_{UT}^n$ pseudo-data as a function of $x$, for different values of $z$ (corresponding to the different curves) and different values of $M_h$ (corresponding to the different panels). 
Only the $Q^2=1.325~\mathrm{GeV}^2$ bin is shown.
According to Eq.~\eqref{eq:neutronAUT}, $A_{UT}^n$ is sensitive to the linear combination $4 h_1^{d_v} - h_1^{u_v}$.
In the JAMDiFF analysis, for $x \lesssim 0.4$, the uncertainty of $4h_1^{d_v}$ is significantly larger than that of $h_1^{u_v}$, whereas the opposite is true for $x \gtrsim 0.5$. 
Therefore, $h_1^{d_v}$ benefits the most from the inclusion of SoLID pseudo-data at lower $x$, while $h_1^{u_v}$ benefits more at higher $x$; see Fig.~\ref{fig:sPDFs}.
We note that the uncertainties of $h_1^{d_v}$ are reduced even for
$x \lesssim 0.01$, which is below the kinematic reach of SoLID (see
Fig.~\ref{fig:kin}).
Since no pseudo-data constrain this region, the behavior of the transversity PDFs there is determined by the parameterization, which correlates their values at small $x$ with their values at the lowest $x$ covered by SoLID.
The improved precision around $x \simeq 0.06$ therefore propagates to smaller $x$.

\begin{figure}[t]
    \hspace*{2.3cm}
    \includegraphics[width=0.9\linewidth]{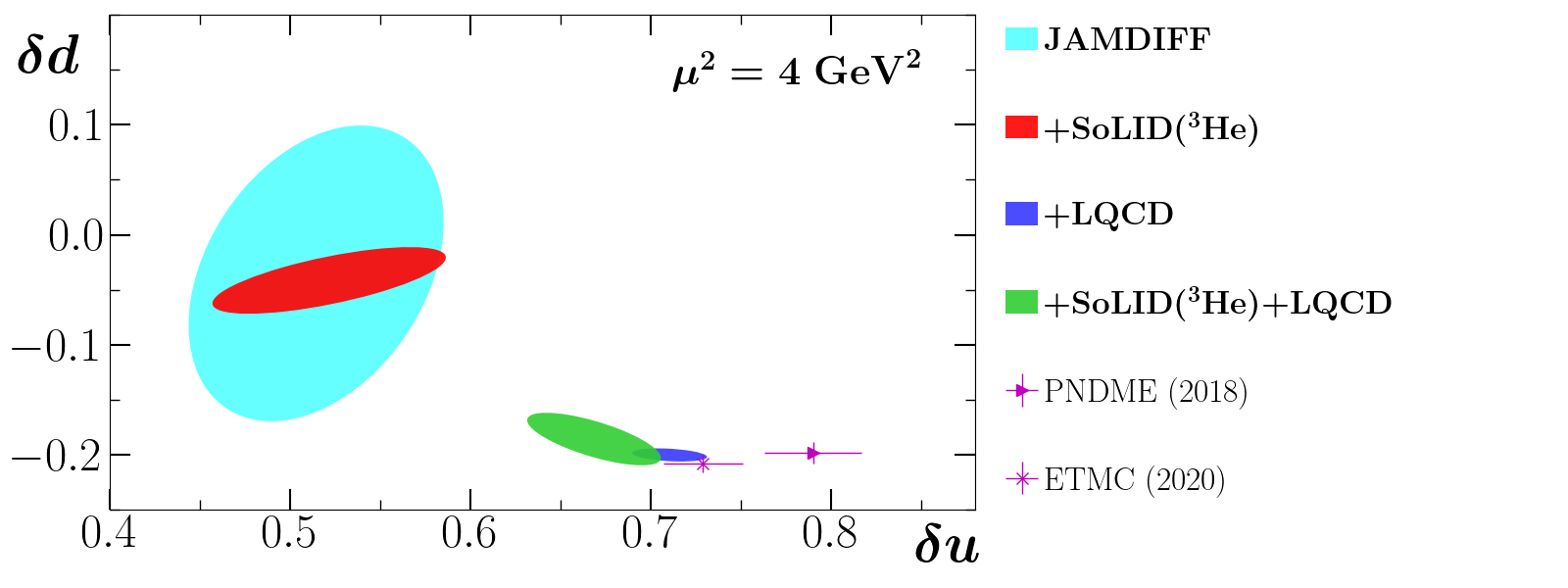}
    \caption{$1\sigma$ confidence ellipses for the tensor charges $\delta u$ and $\delta d$:~results based on the JAMDiFF analysis~\cite{Cocuzza:2023oam,Cocuzza:2023vqs} without LQCD (cyan) and with LQCD (blue), and after including the SoLID pseudo-data in the fit without LQCD (red) and with LQCD (green). Also shown are the two LQCD tensor charge calculations by ETMC~\cite{Alexandrou:2021oih} and PNDME~\cite{Gupta:2018lvp}.}
    \label{fig:gT-s}
\end{figure}

The resulting impact on the tensor charges is shown in Fig.~\ref{fig:gT-s}, where $\delta d$ is significantly better constrained, while $\delta u$ remains largely unchanged. 
The enhanced precision for $\delta d$ enables a more stringent comparison with LQCD predictions.
In the simultaneous fit to SoLID pseudo-data and LQCD, the tensor charge ellipse (Fig.~\ref{fig:gT-s}) overlaps with the ETMC~\cite{Alexandrou:2021oih} result within errors but not with the PNDME~\cite{Gupta:2018lvp} result.
While the agreement with ETMC is not reflected in the $\chi_{\rm red}^2$ value~(Tab.~\ref{tab:chi2_solid}) (which does not account for theoretical uncertainty bands), it is evidenced by the $Z$ score of $2.51$ for the ETMC tensor charge $\delta u$. 
As a further test, we apply the weighting procedure of Sec.~\ref{sec:quant_imp} and find that, in this case, the fit describes both inputs well ($Z = -0.47 \; (1.60)$ for the ETMC (PNDME) $\delta u$ calculation). 
We therefore find that it is possible to fit SoLID pseudo-data and LQCD tensor charges simultaneously. 
This illustrates the possibility described in Sec.~\ref{sec:quant_imp} that weighting LQCD can reveal an apparent tension to be spurious.
While this compatibility may be unexpected from the red ellipse in Fig.~\ref{fig:gT-s} alone, it mirrors the behavior observed for $\delta u$ in the JAMDiFF analysis~\cite{Cocuzza:2023oam,Cocuzza:2023vqs}, showing that only a fit including the LQCD results can determine whether a genuine tension exists.
For $\delta d$, this compatibility is largely due to the lack of precise measurements constraining $h_1^{d_v}$ at lower values of $x$.
The lowest SoLID point sits at $x = 0.06$, and with minimal other data constraining $h_1^{d_v}$ below that point, the PDF is able to increase in magnitude at $x \lesssim 0.1$ without degrading the $\chi_{\rm red}^2$ for the pseudo-data or experimental data. 
When LQCD results and SoLID pseudo-data are both included in the fit, the region $x<0.1$ contributes an integral of $-0.16$, accounting for $\sim 84\%$ of the total $\delta d \approx -0.19$.

Lastly, we consider the case where the pseudo-data are generated from the low-magnitude boundary of $A^n_{UT}$. 
Roughly speaking, this corresponds to the upper and lower boundaries of the JAMDiFF band for $h_1^{d_v}$ and $ h_1^{u_v}$ respectively, pulling both of them away from LQCD predictions. 
We find that the fit fails to simultaneously describe the pseudo-data and the LQCD tensor charges, as shown by the large $\chi_{\rm red}^2$ values, and infinite $Z$ scores for all LQCD tensor charges. 
This qualitative outcome is further confirmed by the weighting method, which leads to reasonable agreement with the LQCD results but yields a $\chi_{\rm red}^2$ of 5.65 for the SoLID pseudo-data, indicating that the fit fails. 
In summary, future SoLID data will be able to provide crucial new insights into the nucleon tensor charges, especially $\delta d$.

\begin{table}[t]
\centering
\setlength{\tabcolsep}{8pt}
\renewcommand{\arraystretch}{1.15}
\begin{tabular}{l c c | c c c c}
\toprule\toprule
\multicolumn{3}{c|}{} & \multicolumn{4}{c}{$\chi^2_{\text{red}}$} \\
\cmidrule(l){4-7}
Experiment & Target & $N_{\text{dat}}$ & no LQCD & w/ LQCD & w/ LQCD (weighted) & ``low'' w/ LQCD \\
\midrule
\textbf{SoLID} \cite{SoLID:2014proposal,Arrington:2022hgr} & $\boldsymbol{^3\mathrm{He}}$ & \textbf{570} & \textbf{0.03} & \textbf{0.04} & \textbf{0.09} & \textbf{3.20} \\
\midrule
Belle ($d\sigma$) \cite{Belle:2017rwm}& -- & 1094 & 1.11 & 1.00 & 1.00 & 1.00 \\[0.09cm]
Belle (asym) \cite{Belle:2011cur}& -- & 183 & 0.77 & 0.79 & 0.82 & 0.88 \\
\midrule
HERMES \cite{HERMES:2008mcr} & p & 12 & 1.04 & 1.06 & 1.11 & 0.92 \\
\midrule
COMPASS \cite{COMPASS:2023cgk} & p, D & 47 & 0.83 & 1.12 & 1.31 & 1.13 \\
\midrule
STAR \cite{STAR:2015jkc,STAR:2017wsi} & pp & 130 & 1.18 & 1.22 & 1.28 & 1.43 \\
\midrule
$\delta u$ (ETMC) \cite{Alexandrou:2021oih} & -- & 1 & -- & 7.55 & 0.17 & 152.41 \\
$\delta u$ (PNDME) \cite{Gupta:2018lvp} & -- & 1 & -- & 20.23 & 3.70 & 151.75 \\
$\delta d$ (ETMC) \cite{Alexandrou:2021oih} & -- & 1 & -- & 8.63 & 0.22 & 506.70 \\
$\delta d$ (PNDME) \cite{Gupta:2018lvp} & -- & 1 & -- & 1.57 & 0.36 & 253.84 \\
\bottomrule
\end{tabular}
\caption{Summary of $\chi_{\rm red}^2$ values after including the SoLID pseudo-data in the fit: without LQCD, with LQCD, with LQCD and re-weighting, with LQCD and pseudo-data for the ``low'' scenario. For Belle, we list separately the results for the unpolarized cross section and the Artru-Collins asymmetry.
}
\label{tab:chi2_solid}
\end{table}
\FloatBarrier
\subsection{ePIC}
We now turn to the impact of pseudo-data from ePIC at the future EIC, which will provide access to the internal structure of the nucleon and nuclei across a wide kinematic range, with particularly strong coverage at smaller values of $x$~\cite{AbdulKhalek:2021gbh,Accardi:2012qut}. 
For this reason, we use the JAMDiFF-x baseline discussed in Sec.~\ref{sec:methodology}, for which the theoretical constraint on the small-$x$ behavior in Eq.~\eqref{eq:alpha} is relaxed. 
The JAMDiFF-x transversity PDFs are nearly identical to JAMDiFF across most of the $x$-range, but their uncertainties increase strongly for $x \lesssim 10^{-2}$. 
As mentioned in Sec.~\ref{sec:methodology}, using JAMDiFF-x avoids a baseline that is theoretically well constrained at small $x$, allowing a proper assessment of the impact of the ePIC experiment in that region. 
\begin{figure}[t]
\centering
\begin{tikzpicture}
    \node[inner sep=0] (img) {\includegraphics[width=0.9\linewidth]{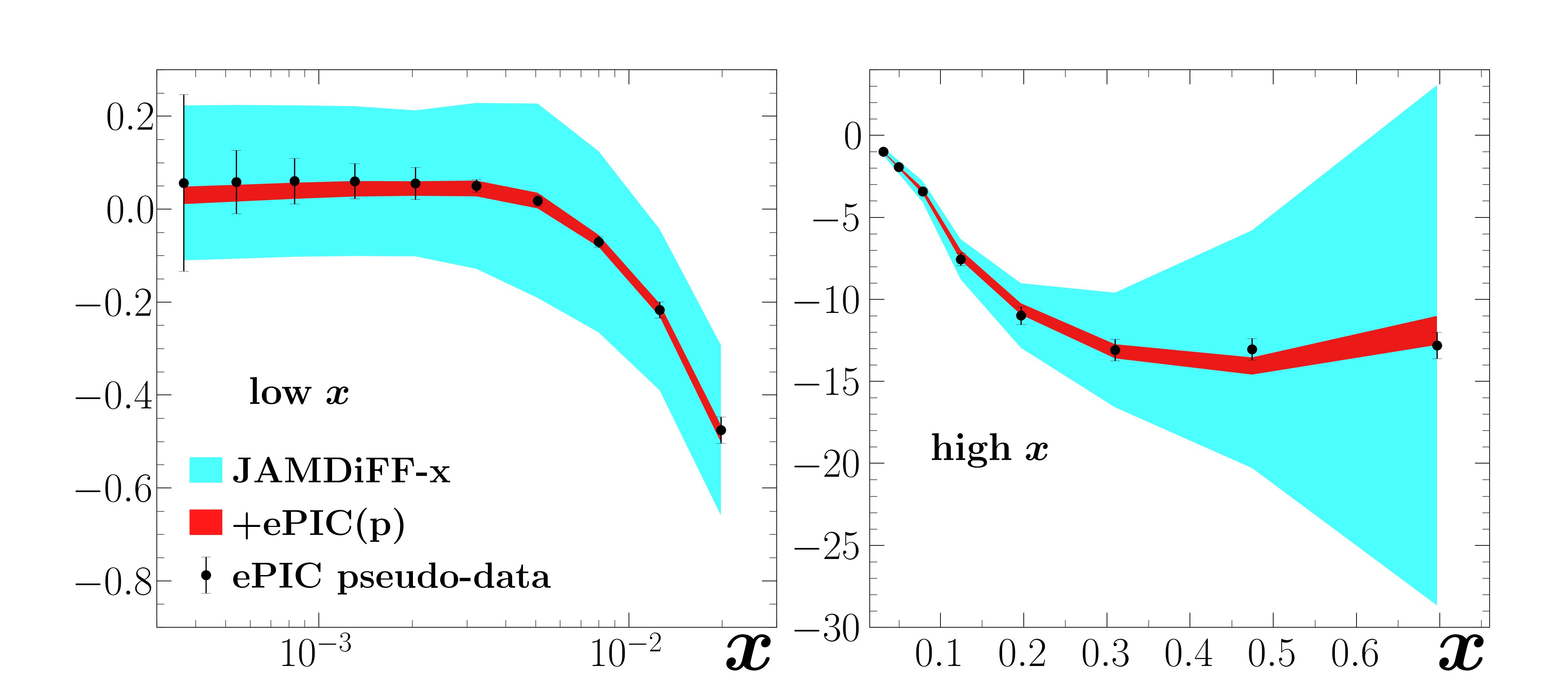}};
    \node[anchor=north,xshift =0.6cm, yshift=0.3cm] at (img.north)
        {\Large \bfseries ePIC $\boldsymbol{A_{UT}^p[\%]}$};
\end{tikzpicture}
\caption{1$\sigma$ uncertainty bands of the TSSA $A_{UT}^p$ for ePIC kinematics at low $x$ (left) and high $x$ (right):~results based on JAMDiFF-x (cyan) and after including the ePIC pseudo-data in the fit (red).}
    \label{fig:eEIC}
\end{figure}
Fig.~\ref{fig:eEIC} shows the TSSA $A_{UT}^p$ for ePIC kinematics, illustrating the significant extension of the accessible region to low $x$.
The resulting impact on the PDFs is shown in Fig.~\ref{fig:eePDFs}, along with the ratio of uncertainties before and after including the ePIC proton pseudo-data. 
As in the CLAS12 case, $A_{UT}^p$ mostly constrains $h_1^{u_v}$, since in the JAMDiFF-x baseline the errors on $4h_1^{u_v}$ are substantially larger than those on $h_1^{d_v}$ for most of the $x$-range.
\begin{figure}[t]
    \centering
    \includegraphics[width=0.9\linewidth]{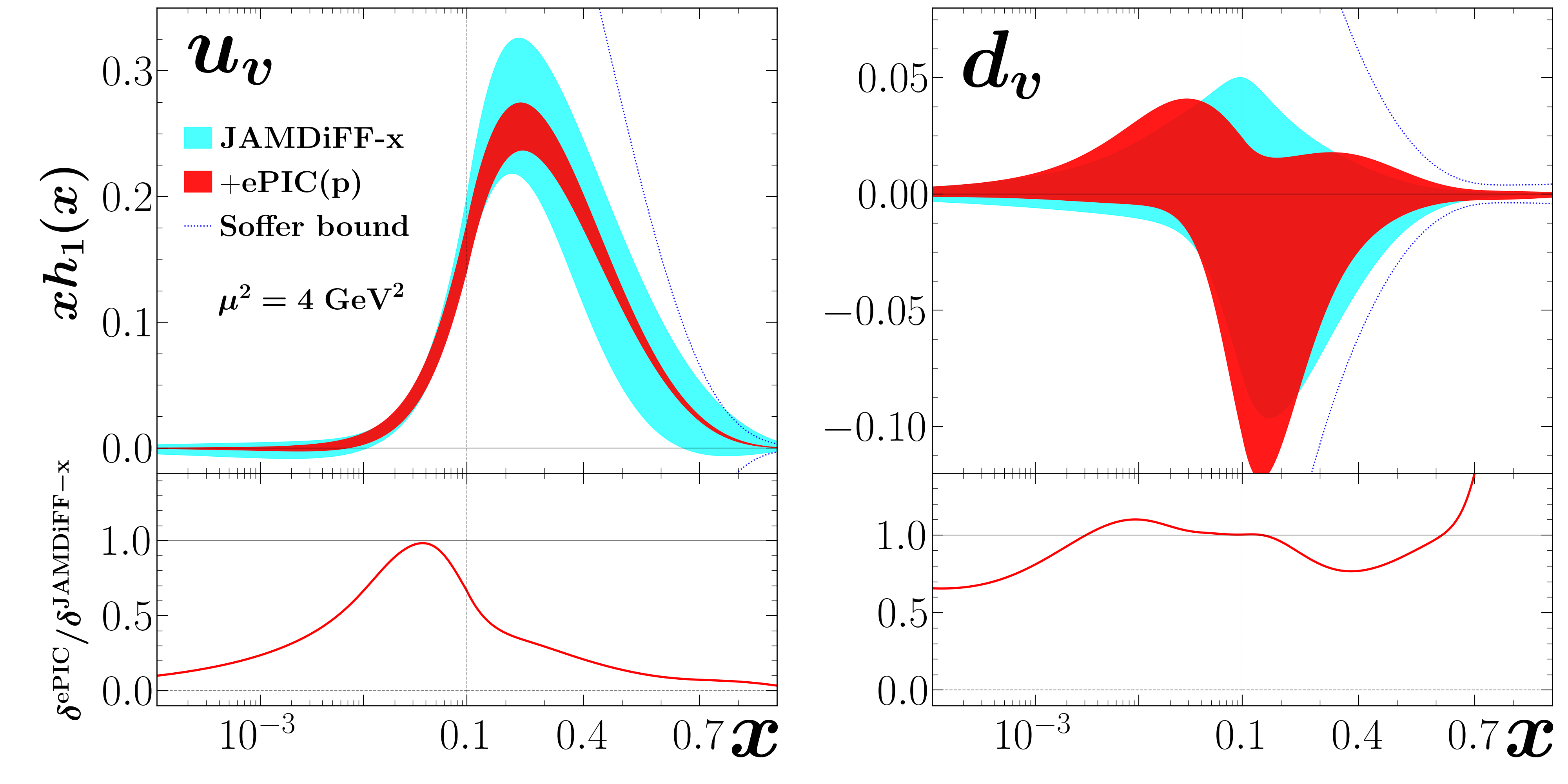}
    \caption{1$\sigma$ uncertainty bands of $h_1^{u_v}$ and $h_1^{d_v}$:~results based on JAMDiFF-x (cyan) and after including the ePIC proton pseudo-data in the fit (red). The Soffer bound is indicated by the blue dotted lines. The lower panels show the ratio of the uncertainties for the transversity PDFs (with/without ePIC pseudo-data). }
    \label{fig:eePDFs}
\end{figure}

\begin{figure}[t]
    \centering
    \includegraphics[width=0.9\linewidth]{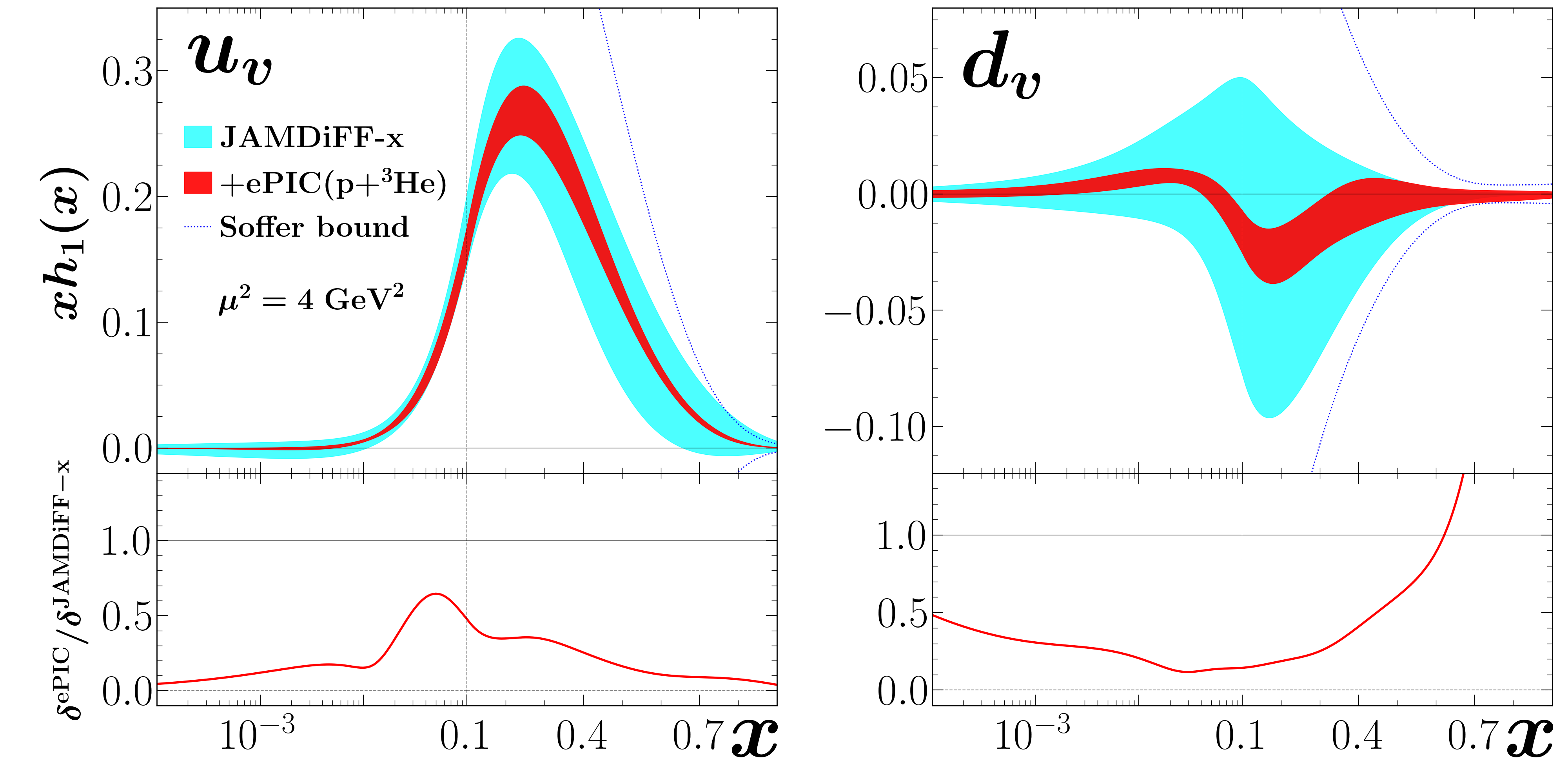}
\caption{1$\sigma$ uncertainty bands of $h_1^{u_v}$ and $h_1^{d_v}$:~results based on JAMDiFF-x (cyan) and after including the ePIC proton and $^3$He pseudo-data in the fit (red). The Soffer bound is indicated by the blue dotted lines. The lower panels show the ratio of the uncertainties for the transversity PDFs (with/without ePIC pseudo-data).}
    \label{fig:eePDFsph}
\end{figure}

In addition to a proton target, ePIC is expected to measure dihadron SIDIS on a $^3\mathrm{He}$ target. 
These measurements are essential for constraining the down quark transversity and de-correlating it from the up quark distribution. 
Similar to the discussion in Sec.~\ref{sec:CLAS}, the addition of a high-precision proton dataset constraining the linear combination $4 h_1^{u_v} - h_1^{d_v}$ (see Eq.~\eqref{eq:protonAUT}) leads to a strong correlation between the two functions. 
The further addition of data taken with a $^3\mathrm{He}$ target constrains the linear combination $4 h_1^{d_v} - h_1^{u_v}$ (see Eq.~\eqref{eq:neutronAUT}), enabling a simultaneous precise extraction of $h_1^{u_v}$ and $h_1^{d_v}$.
This can be quantified with the correlation coefficient $\rho$ of the tensor charges $\delta u$ and $\delta d$, which gives an indication of the correlation of the underlying transversity PDFs:
\begin{equation}
    \rho = \frac{\langle\delta u \cdot \delta d\rangle-\langle\delta u\rangle\langle\delta d\rangle}{\sigma(\delta u)\sigma(\delta d)} =
    \left\{
    \begin{array}{l}
    0.31 \text{ for JAMDiFF-x}\\
    0.65 \text{ for JAMDiFF-x$\,+\,$ePIC(}\mathrm{p}) \\
    0.59 \text{ for JAMDiFF-x$\,+\,$ePIC(}^3\mathrm{He}) \\
    \vspace{5pt}
    -0.17 \text{ for JAMDiFF-x$\,+\,$ePIC(\rm p+$^3$\rm He)}
    \end{array}
    \right. 
\end{equation}
Fitting datasets with diverse dihadron observables and kinematic coverages, JAMDiFF-x leads to $h_1^{u_v}$ and $h_1^{d_v}$ that are reasonably well decorrelated. 
Either ePIC proton or $^3\rm He$ dataset individually causes a significant rise of $\rho$, while their combination restores a low correlation, as was also observed in Ref.~\cite{Gamberg:2021lgx}.
This motivates the general need for SIDIS measurements on both proton and $^3$He (or deuteron) targets, which also emphasizes the complementary role of CLAS12 and SoLID data.
The combined impact of using both targets is shown in Fig.~\ref{fig:eePDFsph}, where the two transversity PDFs exhibit significantly reduced uncertainties across the full kinematic range, with the exception of $h_1^{d_v}$ for $x \gtrsim 0.5$.
The quality of this fit is shown in the ``no LQCD'' column of Tab.~\ref{tab:chi2_epic}. 
As alluded to in Sec.~\ref{sec:CLAS}, the very small size of the projected errors of the ePIC proton data,  combined with the large kinematic coverage, causes a significant deviation from $\chi^2_{\rm red} =0$ for the pseudo-data. 
This is not the case for the $^3$He pseudo-data as they have larger errors, especially in the high-$x$ region.

\begin{figure}[t]
    \hspace*{2.3cm}
    \includegraphics[width=0.9\linewidth]{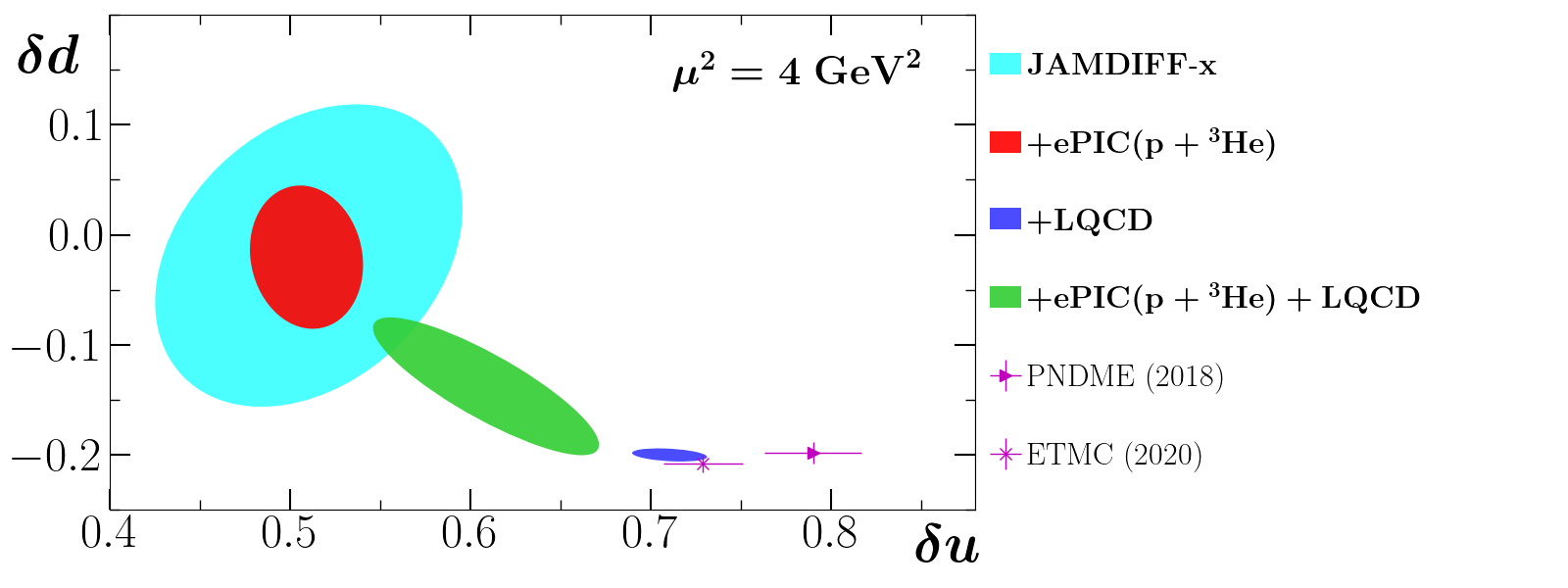}
    \caption{$1\sigma$ confidence ellipses for the tensor charges $\delta u$ and $\delta d$:~results based on JAMDiFF-x without LQCD (cyan) and with LQCD (blue), and after including the ePIC pseudo-data in the fit without LQCD (red) and with LQCD (green). Also shown are the two LQCD tensor charge calculations by ETMC~\cite{Alexandrou:2021oih} and PNDME~\cite{Gupta:2018lvp}.}
    \label{fig:egT}
\end{figure}

Table~\ref{tab:chi2_epic} summarizes the quality of the fit for the various scenarios. 
Based on the ``w/ LQCD'' column and Fig.~\ref{fig:egT}, it is evident that the fit fails to accommodate the LQCD tensor charges and the ePIC pseudo-data simultaneously. 
Applying the weighting procedure forces agreement with LQCD, but at the cost of a large increase in the $\chi_{\rm red}^2$ for the ePIC pseudo-data, confirming that the two inputs are incompatible. This is further evidenced by the $Z$ scores of 4.82 and 6.49 for the ePIC proton and $^3$He pseudo-data, respectively.

We additionally consider the high-magnitude scenario (labeled ``high'') for both $A_{UT}^p$ and $A_{UT}^n$. These correspond to the upper boundary of $h_1^{u_v}$, and the lower boundary of $h_1^{d_v}$. 
Unlike the alternative scenarios for CLAS12 and SoLID, this shifts the tensor charges closer to the LQCD predictions, making it more difficult to observe a tension between the two. 
However, even in this case, the fit fails, as reflected by the numbers in the last column of Tab.~\ref{tab:chi2_epic}. 
This is further confirmed by the weighting method for the ``high'' scenario, which leads to agreement with the LQCD tensor charges but results in $\chi_{\rm red}^2 = 3.38$ for the ePIC proton dataset ($Z = 4.67$).
Overall, this shows that future ePIC data can definitively establish or rule out the compatibility of phenomenological extractions with LQCD predictions of the tensor charges.
This conclusion holds without the use of the low-$x$ theoretical constraint, relying entirely on the constraining power of the data.

\begin{table}[t]
\centering
\setlength{\tabcolsep}{8pt}
\renewcommand{\arraystretch}{1.15}
\begin{tabular}{l c c | c c c c}
\toprule\toprule
\multicolumn{3}{c|}{} & \multicolumn{4}{c}{$\chi^2_{\text{red}}$} \\
\cmidrule(l){4-7}
Experiment & Target & $N_{\text{dat}}$ & no LQCD & w/ LQCD & w/ LQCD (weighted) & ``high'' w/ LQCD \\
\midrule
 & \bf{p} & \textbf{18} & \textbf{0.79} & \textbf{1.28} & \textbf{3.49} & \textbf{3.72} \\
\multirow{-2}{*}{\textbf{ePIC} \cite{AbdulKhalek:2021gbh}} & $\boldsymbol{^3\mathrm{He}}$ & \textbf{19} & \textbf{0.05} & \textbf{1.19} & \textbf{4.70} & \textbf{0.68} \\
\midrule
Belle ($d\sigma$) \cite{Belle:2017rwm}& -- & 1094 & 1.04 & 1.03 & 1.04 & 1.03 \\[0.09cm]
Belle (asym) \cite{Belle:2011cur}& -- & 183 & 0.63 & 0.79 & 1.12 & 0.67 \\
\midrule
HERMES \cite{HERMES:2008mcr} & p & 12 & 1.07 & 1.00 & 1.03 & 1.04 \\
\midrule
COMPASS \cite{COMPASS:2023cgk} & p, D & 47 & 0.85 & 0.88 & 1.22 & 0.94 \\
\midrule
STAR \cite{STAR:2015jkc,STAR:2017wsi} & pp & 130 & 1.14 & 1.14 & 1.23 & 1.15 \\
\midrule
$\delta u$ (ETMC) \cite{Alexandrou:2021oih} & -- & 1 & -- & 29.93 & 0.67 & 14.62 \\
$\delta u$ (PNDME) \cite{Gupta:2018lvp} & -- & 1 & -- & 45.12 & 2.54 & 28.89 \\
$\delta d$ (ETMC) \cite{Alexandrou:2021oih} & -- & 1 & -- & 86.93 & 0.48 & 28.94 \\
$\delta d$ (PNDME) \cite{Gupta:2018lvp} & -- & 1 & -- & 36.51 & 0.18 & 9.52 \\
\bottomrule
\end{tabular}
\caption{Summary of $\chi_{\rm red}^2$ values after including the ePIC pseudo-data in the fit:~without LQCD, with LQCD, with LQCD and re-weighting, with LQCD and pseudo-data for the ``high'' scenario. For Belle, we list separately the results for the unpolarized cross section and the Artru-Collins asymmetry. }
\label{tab:chi2_epic}
\end{table}

\section{Conclusion}
\label{sec:conclusion}
We have evaluated the impact of future dihadron SIDIS measurements at the CLAS12 (proton target) and SoLID ($^3$He target) experiments at Jefferson Lab, and the ePIC (proton and $^3$He targets) experiment at the EIC, on the transversity PDFs and tensor charges of the nucleon. 
For our study, we have generated pseudo-data for these experiments and analyzed them together with existing dihadron data within the JAMDiFF framework.

We have found that the three experiments could play complementary roles. 
CLAS12 will provide near-term data in a previously unexplored region, strongly constraining the up quark transversity in the region of intermediate-to-high~$x$.
This also leads to valuable new information on the tensor charge $\delta u$, which can be compared with LQCD results.
In the JAMDiFF analysis~\cite{Cocuzza:2023oam, Cocuzza:2023vqs}, approximately two thirds of the value for $\delta u$ originates from the $x$-range covered by CLAS12, which by itself highlights the significance of the experiment. 
The information gained from CLAS12 can also act as valuable guidance for future ePIC measurements. 
SoLID with a $^3$He target will complement proton data from CLAS12, providing unprecedented constraints on the down quark transversity and tensor charge.
This also enables a stringent comparison with LQCD results for $\delta d$. 
The ePIC experiment extends coverage to low $x$ for both targets, resulting in the first direct experimental constraints on the small-$x$ behavior of the transversity PDFs and permitting a simultaneous precise extraction of the distributions for both up and down quarks. 
In addition, ePIC dihadron data can definitively answer the question whether or not phenomenological extractions and LQCD predictions of the tensor charges are compatible.

Our final comment concerns the DiFFs.
As the experimental precision of dihadron SIDIS measurements of the TSSA improves, the current uncertainties in the DiFFs will eventually become the primary limiting factor in achieving higher precision for the transversity PDFs and tensor charges. We also recall that the flavor separation of the unpolarized DiFFs relies on PYTHIA~\cite{Cocuzza:2023oam, Cocuzza:2023vqs}. 
We therefore encourage new measurements that would enable this flavor separation and allow even more precise extractions of the DiFFs, such as measurements of the unpolarized cross section and the Artru-Collins asymmetry in electron-positron annihilation and, in particular, measurements of the unpolarized cross section in SIDIS and proton-proton collisions.

\begin{acknowledgments}
We thank Marco Contalbrigo for a helpful discussion.  
This work was supported by the National Science Foundation and the US Department of Energy, Office of Science under Grants No.~PHY-2412792 (A.M.~and Y.S.), No.~PHY-2308567, PHY-2608477 (D.P.), No.~PHY-2310031, No.~PHY-2335114 (A.P.) and DE-SC0024505 (G.M, M.M, A.V.).
The work of N.S. was supported by the DOE, Office of Science, Office of Nuclear Physics in the Early Career Program.
The work of Y.S.~was also supported by a JSA Graduate Fellowship. 
This work was supported in part by the US Department of Energy Contract No.~DE-AC05-06OR23177, under which Jefferson Science Associates LLC operates Jefferson Lab, and Award DE-SC0023646 of the Quark-Gluon Tomography (QGT) Topical Collaboration.
\end{acknowledgments}

\bibliography{dihadron_impact}

\begin{thebibliography}{105}%
\makeatletter
\providecommand \@ifxundefined [1]{%
 \@ifx{#1\undefined}
}%
\providecommand \@ifnum [1]{%
 \ifnum #1\expandafter \@firstoftwo
 \else \expandafter \@secondoftwo
 \fi
}%
\providecommand \@ifx [1]{%
 \ifx #1\expandafter \@firstoftwo
 \else \expandafter \@secondoftwo
 \fi
}%
\providecommand \natexlab [1]{#1}%
\providecommand \enquote  [1]{``#1''}%
\providecommand \bibnamefont  [1]{#1}%
\providecommand \bibfnamefont [1]{#1}%
\providecommand \citenamefont [1]{#1}%
\providecommand \href@noop [0]{\@secondoftwo}%
\providecommand \href [0]{\begingroup \@sanitize@url \@href}%
\providecommand \@href[1]{\@@startlink{#1}\@@href}%
\providecommand \@@href[1]{\endgroup#1\@@endlink}%
\providecommand \@sanitize@url [0]{\catcode `\\12\catcode `\$12\catcode
  `\&12\catcode `\#12\catcode `\^12\catcode `\_12\catcode `\%12\relax}%
\providecommand \@@startlink[1]{}%
\providecommand \@@endlink[0]{}%
\providecommand \url  [0]{\begingroup\@sanitize@url \@url }%
\providecommand \@url [1]{\endgroup\@href {#1}{\urlprefix }}%
\providecommand \urlprefix  [0]{URL }%
\providecommand \Eprint [0]{\href }%
\providecommand \doibase [0]{http://dx.doi.org/}%
\providecommand \selectlanguage [0]{\@gobble}%
\providecommand \bibinfo  [0]{\@secondoftwo}%
\providecommand \bibfield  [0]{\@secondoftwo}%
\providecommand \translation [1]{[#1]}%
\providecommand \BibitemOpen [0]{}%
\providecommand \bibitemStop [0]{}%
\providecommand \bibitemNoStop [0]{.\EOS\space}%
\providecommand \EOS [0]{\spacefactor3000\relax}%
\providecommand \BibitemShut  [1]{\csname bibitem#1\endcsname}%
\let\auto@bib@innerbib\@empty
\bibitem [{\citenamefont {Ralston}\ and\ \citenamefont
  {Soper}(1979)}]{Ralston:1979ys}%
  \BibitemOpen
  \bibfield  {author} {\bibinfo {author} {\bibfnamefont {J.~P.}\ \bibnamefont
  {Ralston}}\ and\ \bibinfo {author} {\bibfnamefont {D.~E.}\ \bibnamefont
  {Soper}},\ }\href {\doibase 10.1016/0550-3213(79)90082-8} {\bibfield
  {journal} {\bibinfo  {journal} {Nucl. Phys. B}\ }\textbf {\bibinfo {volume}
  {152}},\ \bibinfo {pages} {109} (\bibinfo {year} {1979})}\BibitemShut
  {NoStop}%
\bibitem [{\citenamefont {Barone}\ \emph {et~al.}(2002)\citenamefont {Barone},
  \citenamefont {Drago},\ and\ \citenamefont {Ratcliffe}}]{Barone:2001sp}%
  \BibitemOpen
  \bibfield  {author} {\bibinfo {author} {\bibfnamefont {V.}~\bibnamefont
  {Barone}}, \bibinfo {author} {\bibfnamefont {A.}~\bibnamefont {Drago}}, \
  and\ \bibinfo {author} {\bibfnamefont {P.~G.}\ \bibnamefont {Ratcliffe}},\
  }\href {\doibase 10.1016/S0370-1573(01)00051-5} {\bibfield  {journal}
  {\bibinfo  {journal} {Phys. Rept.}\ }\textbf {\bibinfo {volume} {359}},\
  \bibinfo {pages} {1} (\bibinfo {year} {2002})},\ \Eprint
  {http://arxiv.org/abs/hep-ph/0104283} {arXiv:hep-ph/0104283 [hep-ph]}
  \BibitemShut {NoStop}%
\bibitem [{\citenamefont {Lorc{\'e}}\ \emph {et~al.}(2026)\citenamefont
  {Lorc{\'e}}, \citenamefont {Metz}, \citenamefont {Pasquini},\ and\
  \citenamefont {Schweitzer}}]{Lorce:2025aqp}%
  \BibitemOpen
  \bibfield  {author} {\bibinfo {author} {\bibfnamefont {C.}~\bibnamefont
  {Lorc{\'e}}}, \bibinfo {author} {\bibfnamefont {A.}~\bibnamefont {Metz}},
  \bibinfo {author} {\bibfnamefont {B.}~\bibnamefont {Pasquini}}, \ and\
  \bibinfo {author} {\bibfnamefont {P.}~\bibnamefont {Schweitzer}},\ }in\
  \href@noop {} {\emph {\bibinfo {booktitle} {Encyclopedia of Particle
  Physics}}},\ Vol.~\bibinfo {volume} {2}\ (\bibinfo  {publisher} {Elsevier},\
  \bibinfo {address} {Amsterdam},\ \bibinfo {year} {2026})\ pp.\ \bibinfo
  {pages} {167--218},\ \Eprint {http://arxiv.org/abs/2507.12664}
  {arXiv:2507.12664 [hep-ph]} \BibitemShut {NoStop}%
\bibitem [{\citenamefont {Herczeg}(2001)}]{Herczeg:2001vk}%
  \BibitemOpen
  \bibfield  {author} {\bibinfo {author} {\bibfnamefont {P.}~\bibnamefont
  {Herczeg}},\ }\href {\doibase 10.1016/S0146-6410(01)00149-1} {\bibfield
  {journal} {\bibinfo  {journal} {Prog. Part. Nucl. Phys.}\ }\textbf {\bibinfo
  {volume} {46}},\ \bibinfo {pages} {413} (\bibinfo {year} {2001})}\BibitemShut
  {NoStop}%
\bibitem [{\citenamefont {Erler}\ and\ \citenamefont
  {Ramsey-Musolf}(2005)}]{Erler:2004cx}%
  \BibitemOpen
  \bibfield  {author} {\bibinfo {author} {\bibfnamefont {J.}~\bibnamefont
  {Erler}}\ and\ \bibinfo {author} {\bibfnamefont {M.~J.}\ \bibnamefont
  {Ramsey-Musolf}},\ }\href {\doibase 10.1016/j.ppnp.2004.08.001} {\bibfield
  {journal} {\bibinfo  {journal} {Prog. Part. Nucl. Phys.}\ }\textbf {\bibinfo
  {volume} {54}},\ \bibinfo {pages} {351} (\bibinfo {year} {2005})},\ \Eprint
  {http://arxiv.org/abs/hep-ph/0404291} {arXiv:hep-ph/0404291} \BibitemShut
  {NoStop}%
\bibitem [{\citenamefont {Severijns}\ \emph {et~al.}(2006)\citenamefont
  {Severijns}, \citenamefont {Beck},\ and\ \citenamefont
  {Naviliat-Cuncic}}]{Severijns:2006dr}%
  \BibitemOpen
  \bibfield  {author} {\bibinfo {author} {\bibfnamefont {N.}~\bibnamefont
  {Severijns}}, \bibinfo {author} {\bibfnamefont {M.}~\bibnamefont {Beck}}, \
  and\ \bibinfo {author} {\bibfnamefont {O.}~\bibnamefont {Naviliat-Cuncic}},\
  }\href {\doibase 10.1103/RevModPhys.78.991} {\bibfield  {journal} {\bibinfo
  {journal} {Rev. Mod. Phys.}\ }\textbf {\bibinfo {volume} {78}},\ \bibinfo
  {pages} {991} (\bibinfo {year} {2006})},\ \Eprint
  {http://arxiv.org/abs/nucl-ex/0605029} {arXiv:nucl-ex/0605029} \BibitemShut
  {NoStop}%
\bibitem [{\citenamefont {Cirigliano}\ \emph {et~al.}(2013)\citenamefont
  {Cirigliano}, \citenamefont {Gardner},\ and\ \citenamefont
  {Holstein}}]{Cirigliano:2013xha}%
  \BibitemOpen
  \bibfield  {author} {\bibinfo {author} {\bibfnamefont {V.}~\bibnamefont
  {Cirigliano}}, \bibinfo {author} {\bibfnamefont {S.}~\bibnamefont {Gardner}},
  \ and\ \bibinfo {author} {\bibfnamefont {B.}~\bibnamefont {Holstein}},\
  }\href {\doibase 10.1016/j.ppnp.2013.03.005} {\bibfield  {journal} {\bibinfo
  {journal} {Prog. Part. Nucl. Phys.}\ }\textbf {\bibinfo {volume} {71}},\
  \bibinfo {pages} {93} (\bibinfo {year} {2013})},\ \Eprint
  {http://arxiv.org/abs/1303.6953} {arXiv:1303.6953 [hep-ph]} \BibitemShut
  {NoStop}%
\bibitem [{\citenamefont {Courtoy}\ \emph {et~al.}(2015)\citenamefont
  {Courtoy}, \citenamefont {Bae\ss{}ler}, \citenamefont {Gonz\'alez-Alonso},\
  and\ \citenamefont {Liuti}}]{Courtoy:2015haa}%
  \BibitemOpen
  \bibfield  {author} {\bibinfo {author} {\bibfnamefont {A.}~\bibnamefont
  {Courtoy}}, \bibinfo {author} {\bibfnamefont {S.}~\bibnamefont
  {Bae\ss{}ler}}, \bibinfo {author} {\bibfnamefont {M.}~\bibnamefont
  {Gonz\'alez-Alonso}}, \ and\ \bibinfo {author} {\bibfnamefont
  {S.}~\bibnamefont {Liuti}},\ }\href {\doibase 10.1103/PhysRevLett.115.162001}
  {\bibfield  {journal} {\bibinfo  {journal} {Phys. Rev. Lett.}\ }\textbf
  {\bibinfo {volume} {115}},\ \bibinfo {pages} {162001} (\bibinfo {year}
  {2015})},\ \Eprint {http://arxiv.org/abs/1503.06814} {arXiv:1503.06814
  [hep-ph]} \BibitemShut {NoStop}%
\bibitem [{\citenamefont {Gonz\'alez-Alonso}\ \emph {et~al.}(2019)\citenamefont
  {Gonz\'alez-Alonso}, \citenamefont {Naviliat-Cuncic},\ and\ \citenamefont
  {Severijns}}]{Gonzalez-Alonso:2018omy}%
  \BibitemOpen
  \bibfield  {author} {\bibinfo {author} {\bibfnamefont {M.}~\bibnamefont
  {Gonz\'alez-Alonso}}, \bibinfo {author} {\bibfnamefont {O.}~\bibnamefont
  {Naviliat-Cuncic}}, \ and\ \bibinfo {author} {\bibfnamefont {N.}~\bibnamefont
  {Severijns}},\ }\href {\doibase 10.1016/j.ppnp.2018.08.002} {\bibfield
  {journal} {\bibinfo  {journal} {Prog. Part. Nucl. Phys.}\ }\textbf {\bibinfo
  {volume} {104}},\ \bibinfo {pages} {165} (\bibinfo {year} {2019})},\ \Eprint
  {http://arxiv.org/abs/1803.08732} {arXiv:1803.08732 [hep-ph]} \BibitemShut
  {NoStop}%
\bibitem [{\citenamefont {Pospelov}\ and\ \citenamefont
  {Ritz}(2005)}]{Pospelov:2005pr}%
  \BibitemOpen
  \bibfield  {author} {\bibinfo {author} {\bibfnamefont {M.}~\bibnamefont
  {Pospelov}}\ and\ \bibinfo {author} {\bibfnamefont {A.}~\bibnamefont
  {Ritz}},\ }\href {\doibase 10.1016/j.aop.2005.04.002} {\bibfield  {journal}
  {\bibinfo  {journal} {Annals Phys.}\ }\textbf {\bibinfo {volume} {318}},\
  \bibinfo {pages} {119} (\bibinfo {year} {2005})},\ \Eprint
  {http://arxiv.org/abs/hep-ph/0504231} {arXiv:hep-ph/0504231} \BibitemShut
  {NoStop}%
\bibitem [{\citenamefont {Yamanaka}\ \emph {et~al.}(2017)\citenamefont
  {Yamanaka}, \citenamefont {Sahoo}, \citenamefont {Yoshinaga}, \citenamefont
  {Sato}, \citenamefont {Asahi},\ and\ \citenamefont {Das}}]{Yamanaka:2017mef}%
  \BibitemOpen
  \bibfield  {author} {\bibinfo {author} {\bibfnamefont {N.}~\bibnamefont
  {Yamanaka}}, \bibinfo {author} {\bibfnamefont {B.~K.}\ \bibnamefont {Sahoo}},
  \bibinfo {author} {\bibfnamefont {N.}~\bibnamefont {Yoshinaga}}, \bibinfo
  {author} {\bibfnamefont {T.}~\bibnamefont {Sato}}, \bibinfo {author}
  {\bibfnamefont {K.}~\bibnamefont {Asahi}}, \ and\ \bibinfo {author}
  {\bibfnamefont {B.~P.}\ \bibnamefont {Das}},\ }\href {\doibase
  10.1140/epja/i2017-12237-2} {\bibfield  {journal} {\bibinfo  {journal} {Eur.
  Phys. J. A}\ }\textbf {\bibinfo {volume} {53}},\ \bibinfo {pages} {54}
  (\bibinfo {year} {2017})},\ \Eprint {http://arxiv.org/abs/1703.01570}
  {arXiv:1703.01570 [hep-ph]} \BibitemShut {NoStop}%
\bibitem [{\citenamefont {Liu}\ \emph {et~al.}(2018)\citenamefont {Liu},
  \citenamefont {Zhao},\ and\ \citenamefont {Gao}}]{Liu:2017olr}%
  \BibitemOpen
  \bibfield  {author} {\bibinfo {author} {\bibfnamefont {T.}~\bibnamefont
  {Liu}}, \bibinfo {author} {\bibfnamefont {Z.}~\bibnamefont {Zhao}}, \ and\
  \bibinfo {author} {\bibfnamefont {H.}~\bibnamefont {Gao}},\ }\href {\doibase
  10.1103/PhysRevD.97.074018} {\bibfield  {journal} {\bibinfo  {journal} {Phys.
  Rev. D}\ }\textbf {\bibinfo {volume} {97}},\ \bibinfo {pages} {074018}
  (\bibinfo {year} {2018})},\ \Eprint {http://arxiv.org/abs/1704.00113}
  {arXiv:1704.00113 [hep-ph]} \BibitemShut {NoStop}%
\bibitem [{\citenamefont {Gupta}\ \emph
  {et~al.}(2018{\natexlab{a}})\citenamefont {Gupta}, \citenamefont {Jang},
  \citenamefont {Yoon}, \citenamefont {Lin}, \citenamefont {Cirigliano},\ and\
  \citenamefont {Bhattacharya}}]{Gupta:2018qil}%
  \BibitemOpen
  \bibfield  {author} {\bibinfo {author} {\bibfnamefont {R.}~\bibnamefont
  {Gupta}}, \bibinfo {author} {\bibfnamefont {Y.-C.}\ \bibnamefont {Jang}},
  \bibinfo {author} {\bibfnamefont {B.}~\bibnamefont {Yoon}}, \bibinfo {author}
  {\bibfnamefont {H.-W.}\ \bibnamefont {Lin}}, \bibinfo {author} {\bibfnamefont
  {V.}~\bibnamefont {Cirigliano}}, \ and\ \bibinfo {author} {\bibfnamefont
  {T.}~\bibnamefont {Bhattacharya}},\ }\href {\doibase
  10.1103/PhysRevD.98.034503} {\bibfield  {journal} {\bibinfo  {journal} {Phys.
  Rev. D}\ }\textbf {\bibinfo {volume} {98}},\ \bibinfo {pages} {034503}
  (\bibinfo {year} {2018}{\natexlab{a}})},\ \Eprint
  {http://arxiv.org/abs/1806.09006} {arXiv:1806.09006 [hep-lat]} \BibitemShut
  {NoStop}%
\bibitem [{\citenamefont {Gupta}\ \emph
  {et~al.}(2018{\natexlab{b}})\citenamefont {Gupta}, \citenamefont {Yoon},
  \citenamefont {Bhattacharya}, \citenamefont {Cirigliano}, \citenamefont
  {Jang},\ and\ \citenamefont {Lin}}]{Gupta:2018lvp}%
  \BibitemOpen
  \bibfield  {author} {\bibinfo {author} {\bibfnamefont {R.}~\bibnamefont
  {Gupta}}, \bibinfo {author} {\bibfnamefont {B.}~\bibnamefont {Yoon}},
  \bibinfo {author} {\bibfnamefont {T.}~\bibnamefont {Bhattacharya}}, \bibinfo
  {author} {\bibfnamefont {V.}~\bibnamefont {Cirigliano}}, \bibinfo {author}
  {\bibfnamefont {Y.-C.}\ \bibnamefont {Jang}}, \ and\ \bibinfo {author}
  {\bibfnamefont {H.-W.}\ \bibnamefont {Lin}},\ }\href {\doibase
  10.1103/PhysRevD.98.091501} {\bibfield  {journal} {\bibinfo  {journal} {Phys.
  Rev. D}\ }\textbf {\bibinfo {volume} {98}},\ \bibinfo {pages} {091501}
  (\bibinfo {year} {2018}{\natexlab{b}})},\ \Eprint
  {http://arxiv.org/abs/1808.07597} {arXiv:1808.07597 [hep-lat]} \BibitemShut
  {NoStop}%
\bibitem [{\citenamefont {Yamanaka}\ \emph {et~al.}(2018)\citenamefont
  {Yamanaka}, \citenamefont {Hashimoto}, \citenamefont {Kaneko},\ and\
  \citenamefont {Ohki}}]{Yamanaka:2018uud}%
  \BibitemOpen
  \bibfield  {author} {\bibinfo {author} {\bibfnamefont {N.}~\bibnamefont
  {Yamanaka}}, \bibinfo {author} {\bibfnamefont {S.}~\bibnamefont {Hashimoto}},
  \bibinfo {author} {\bibfnamefont {T.}~\bibnamefont {Kaneko}}, \ and\ \bibinfo
  {author} {\bibfnamefont {H.}~\bibnamefont {Ohki}} (\bibinfo {collaboration}
  {JLQCD}),\ }\href {\doibase 10.1103/PhysRevD.98.054516} {\bibfield  {journal}
  {\bibinfo  {journal} {Phys. Rev. D}\ }\textbf {\bibinfo {volume} {98}},\
  \bibinfo {pages} {054516} (\bibinfo {year} {2018})},\ \Eprint
  {http://arxiv.org/abs/1805.10507} {arXiv:1805.10507 [hep-lat]} \BibitemShut
  {NoStop}%
\bibitem [{\citenamefont {Hasan}\ \emph {et~al.}(2019)\citenamefont {Hasan},
  \citenamefont {Green}, \citenamefont {Meinel}, \citenamefont {Engelhardt},
  \citenamefont {Krieg}, \citenamefont {Negele}, \citenamefont {Pochinsky},\
  and\ \citenamefont {Syritsyn}}]{Hasan:2019noy}%
  \BibitemOpen
  \bibfield  {author} {\bibinfo {author} {\bibfnamefont {N.}~\bibnamefont
  {Hasan}}, \bibinfo {author} {\bibfnamefont {J.}~\bibnamefont {Green}},
  \bibinfo {author} {\bibfnamefont {S.}~\bibnamefont {Meinel}}, \bibinfo
  {author} {\bibfnamefont {M.}~\bibnamefont {Engelhardt}}, \bibinfo {author}
  {\bibfnamefont {S.}~\bibnamefont {Krieg}}, \bibinfo {author} {\bibfnamefont
  {J.}~\bibnamefont {Negele}}, \bibinfo {author} {\bibfnamefont
  {A.}~\bibnamefont {Pochinsky}}, \ and\ \bibinfo {author} {\bibfnamefont
  {S.}~\bibnamefont {Syritsyn}},\ }\href {\doibase 10.1103/PhysRevD.99.114505}
  {\bibfield  {journal} {\bibinfo  {journal} {Phys. Rev. D}\ }\textbf {\bibinfo
  {volume} {99}},\ \bibinfo {pages} {114505} (\bibinfo {year} {2019})},\
  \Eprint {http://arxiv.org/abs/1903.06487} {arXiv:1903.06487 [hep-lat]}
  \BibitemShut {NoStop}%
\bibitem [{\citenamefont {Alexandrou}\ \emph {et~al.}(2020)\citenamefont
  {Alexandrou}, \citenamefont {Bacchio}, \citenamefont {Constantinou},
  \citenamefont {Finkenrath}, \citenamefont {Hadjiyiannakou}, \citenamefont
  {Jansen}, \citenamefont {Koutsou},\ and\ \citenamefont {Vaquero
  Aviles-Casco}}]{Alexandrou:2019brg}%
  \BibitemOpen
  \bibfield  {author} {\bibinfo {author} {\bibfnamefont {C.}~\bibnamefont
  {Alexandrou}}, \bibinfo {author} {\bibfnamefont {S.}~\bibnamefont {Bacchio}},
  \bibinfo {author} {\bibfnamefont {M.}~\bibnamefont {Constantinou}}, \bibinfo
  {author} {\bibfnamefont {J.}~\bibnamefont {Finkenrath}}, \bibinfo {author}
  {\bibfnamefont {K.}~\bibnamefont {Hadjiyiannakou}}, \bibinfo {author}
  {\bibfnamefont {K.}~\bibnamefont {Jansen}}, \bibinfo {author} {\bibfnamefont
  {G.}~\bibnamefont {Koutsou}}, \ and\ \bibinfo {author} {\bibfnamefont
  {A.}~\bibnamefont {Vaquero Aviles-Casco}},\ }\href {\doibase
  10.1103/PhysRevD.102.054517} {\bibfield  {journal} {\bibinfo  {journal}
  {Phys. Rev. D}\ }\textbf {\bibinfo {volume} {102}},\ \bibinfo {pages}
  {054517} (\bibinfo {year} {2020})},\ \Eprint
  {http://arxiv.org/abs/1909.00485} {arXiv:1909.00485 [hep-lat]} \BibitemShut
  {NoStop}%
\bibitem [{\citenamefont {Harris}\ \emph {et~al.}(2019)\citenamefont {Harris},
  \citenamefont {von Hippel}, \citenamefont {Junnarkar}, \citenamefont {Meyer},
  \citenamefont {Ottnad}, \citenamefont {Wilhelm}, \citenamefont {Wittig},\
  and\ \citenamefont {Wrang}}]{Harris:2019bih}%
  \BibitemOpen
  \bibfield  {author} {\bibinfo {author} {\bibfnamefont {T.}~\bibnamefont
  {Harris}}, \bibinfo {author} {\bibfnamefont {G.}~\bibnamefont {von Hippel}},
  \bibinfo {author} {\bibfnamefont {P.}~\bibnamefont {Junnarkar}}, \bibinfo
  {author} {\bibfnamefont {H.~B.}\ \bibnamefont {Meyer}}, \bibinfo {author}
  {\bibfnamefont {K.}~\bibnamefont {Ottnad}}, \bibinfo {author} {\bibfnamefont
  {J.}~\bibnamefont {Wilhelm}}, \bibinfo {author} {\bibfnamefont
  {H.}~\bibnamefont {Wittig}}, \ and\ \bibinfo {author} {\bibfnamefont
  {L.}~\bibnamefont {Wrang}},\ }\href {\doibase 10.1103/PhysRevD.100.034513}
  {\bibfield  {journal} {\bibinfo  {journal} {Phys. Rev. D}\ }\textbf {\bibinfo
  {volume} {100}},\ \bibinfo {pages} {034513} (\bibinfo {year} {2019})},\
  \Eprint {http://arxiv.org/abs/1905.01291} {arXiv:1905.01291 [hep-lat]}
  \BibitemShut {NoStop}%
\bibitem [{\citenamefont {Horkel}\ \emph {et~al.}(2020)\citenamefont {Horkel},
  \citenamefont {Bi}, \citenamefont {Constantinou}, \citenamefont {Draper},
  \citenamefont {Liang}, \citenamefont {Liu}, \citenamefont {Liu},\ and\
  \citenamefont {Yang}}]{Horkel:2020hpi}%
  \BibitemOpen
  \bibfield  {author} {\bibinfo {author} {\bibfnamefont {D.}~\bibnamefont
  {Horkel}}, \bibinfo {author} {\bibfnamefont {Y.}~\bibnamefont {Bi}}, \bibinfo
  {author} {\bibfnamefont {M.}~\bibnamefont {Constantinou}}, \bibinfo {author}
  {\bibfnamefont {T.}~\bibnamefont {Draper}}, \bibinfo {author} {\bibfnamefont
  {J.}~\bibnamefont {Liang}}, \bibinfo {author} {\bibfnamefont {K.-F.}\
  \bibnamefont {Liu}}, \bibinfo {author} {\bibfnamefont {Z.}~\bibnamefont
  {Liu}}, \ and\ \bibinfo {author} {\bibfnamefont {Y.-B.}\ \bibnamefont {Yang}}
  (\bibinfo {collaboration} {\ensuremath{\chi}QCD}),\ }\href {\doibase
  10.1103/PhysRevD.101.094501} {\bibfield  {journal} {\bibinfo  {journal}
  {Phys. Rev. D}\ }\textbf {\bibinfo {volume} {101}},\ \bibinfo {pages}
  {094501} (\bibinfo {year} {2020})},\ \Eprint
  {http://arxiv.org/abs/2002.06699} {arXiv:2002.06699 [hep-lat]} \BibitemShut
  {NoStop}%
\bibitem [{\citenamefont {Alexandrou}\ \emph {et~al.}(2021)\citenamefont
  {Alexandrou}, \citenamefont {Constantinou}, \citenamefont {Hadjiyiannakou},
  \citenamefont {Jansen},\ and\ \citenamefont
  {Manigrasso}}]{Alexandrou:2021oih}%
  \BibitemOpen
  \bibfield  {author} {\bibinfo {author} {\bibfnamefont {C.}~\bibnamefont
  {Alexandrou}}, \bibinfo {author} {\bibfnamefont {M.}~\bibnamefont
  {Constantinou}}, \bibinfo {author} {\bibfnamefont {K.}~\bibnamefont
  {Hadjiyiannakou}}, \bibinfo {author} {\bibfnamefont {K.}~\bibnamefont
  {Jansen}}, \ and\ \bibinfo {author} {\bibfnamefont {F.}~\bibnamefont
  {Manigrasso}},\ }\href {\doibase 10.1103/PhysRevD.104.054503} {\bibfield
  {journal} {\bibinfo  {journal} {Phys. Rev. D}\ }\textbf {\bibinfo {volume}
  {104}},\ \bibinfo {pages} {054503} (\bibinfo {year} {2021})},\ \Eprint
  {http://arxiv.org/abs/2106.16065} {arXiv:2106.16065 [hep-lat]} \BibitemShut
  {NoStop}%
\bibitem [{\citenamefont {Park}\ \emph {et~al.}(2022)\citenamefont {Park},
  \citenamefont {Gupta}, \citenamefont {Yoon}, \citenamefont {Mondal},
  \citenamefont {Bhattacharya}, \citenamefont {Jang}, \citenamefont {Jo\'o},\
  and\ \citenamefont {Winter}}]{Park:2021ypf}%
  \BibitemOpen
  \bibfield  {author} {\bibinfo {author} {\bibfnamefont {S.}~\bibnamefont
  {Park}}, \bibinfo {author} {\bibfnamefont {R.}~\bibnamefont {Gupta}},
  \bibinfo {author} {\bibfnamefont {B.}~\bibnamefont {Yoon}}, \bibinfo {author}
  {\bibfnamefont {S.}~\bibnamefont {Mondal}}, \bibinfo {author} {\bibfnamefont
  {T.}~\bibnamefont {Bhattacharya}}, \bibinfo {author} {\bibfnamefont {Y.-C.}\
  \bibnamefont {Jang}}, \bibinfo {author} {\bibfnamefont {B.}~\bibnamefont
  {Jo\'o}}, \ and\ \bibinfo {author} {\bibfnamefont {F.}~\bibnamefont {Winter}}
  (\bibinfo {collaboration} {Nucleon Matrix Elements (NME)}),\ }\href {\doibase
  10.1103/PhysRevD.105.054505} {\bibfield  {journal} {\bibinfo  {journal}
  {Phys. Rev. D}\ }\textbf {\bibinfo {volume} {105}},\ \bibinfo {pages}
  {054505} (\bibinfo {year} {2022})},\ \Eprint
  {http://arxiv.org/abs/2103.05599} {arXiv:2103.05599 [hep-lat]} \BibitemShut
  {NoStop}%
\bibitem [{\citenamefont {Tsuji}\ \emph {et~al.}(2022)\citenamefont {Tsuji},
  \citenamefont {Tsukamoto}, \citenamefont {Aoki}, \citenamefont {Ishikawa},
  \citenamefont {Kuramashi}, \citenamefont {Sasaki}, \citenamefont {Shintani},\
  and\ \citenamefont {Yamazaki}}]{Tsuji:2022ric}%
  \BibitemOpen
  \bibfield  {author} {\bibinfo {author} {\bibfnamefont {R.}~\bibnamefont
  {Tsuji}}, \bibinfo {author} {\bibfnamefont {N.}~\bibnamefont {Tsukamoto}},
  \bibinfo {author} {\bibfnamefont {Y.}~\bibnamefont {Aoki}}, \bibinfo {author}
  {\bibfnamefont {K.-I.}\ \bibnamefont {Ishikawa}}, \bibinfo {author}
  {\bibfnamefont {Y.}~\bibnamefont {Kuramashi}}, \bibinfo {author}
  {\bibfnamefont {S.}~\bibnamefont {Sasaki}}, \bibinfo {author} {\bibfnamefont
  {E.}~\bibnamefont {Shintani}}, \ and\ \bibinfo {author} {\bibfnamefont
  {T.}~\bibnamefont {Yamazaki}} (\bibinfo {collaboration} {PACS}),\ }\href
  {\doibase 10.1103/PhysRevD.106.094505} {\bibfield  {journal} {\bibinfo
  {journal} {Phys. Rev. D}\ }\textbf {\bibinfo {volume} {106}},\ \bibinfo
  {pages} {094505} (\bibinfo {year} {2022})},\ \Eprint
  {http://arxiv.org/abs/2207.11914} {arXiv:2207.11914 [hep-lat]} \BibitemShut
  {NoStop}%
\bibitem [{\citenamefont {Bali}\ \emph {et~al.}(2023)\citenamefont {Bali},
  \citenamefont {Collins}, \citenamefont {Heybrock}, \citenamefont
  {L{\"o}ffler}, \citenamefont {R{\"o}dl}, \citenamefont {S{\"o}ldner},\ and\
  \citenamefont {Weish{\"a}upl}}]{Bali:2023sdi}%
  \BibitemOpen
  \bibfield  {author} {\bibinfo {author} {\bibfnamefont {G.~S.}\ \bibnamefont
  {Bali}}, \bibinfo {author} {\bibfnamefont {S.}~\bibnamefont {Collins}},
  \bibinfo {author} {\bibfnamefont {S.}~\bibnamefont {Heybrock}}, \bibinfo
  {author} {\bibfnamefont {M.}~\bibnamefont {L{\"o}ffler}}, \bibinfo {author}
  {\bibfnamefont {R.}~\bibnamefont {R{\"o}dl}}, \bibinfo {author}
  {\bibfnamefont {W.}~\bibnamefont {S{\"o}ldner}}, \ and\ \bibinfo {author}
  {\bibfnamefont {S.}~\bibnamefont {Weish{\"a}upl}} (\bibinfo {collaboration}
  {RQCD}),\ }\href {\doibase 10.1103/PhysRevD.108.034512} {\bibfield  {journal}
  {\bibinfo  {journal} {Phys. Rev. D}\ }\textbf {\bibinfo {volume} {108}},\
  \bibinfo {pages} {034512} (\bibinfo {year} {2023})},\ \Eprint
  {http://arxiv.org/abs/2305.04717} {arXiv:2305.04717 [hep-lat]} \BibitemShut
  {NoStop}%
\bibitem [{\citenamefont {Smail}\ \emph {et~al.}(2023)\citenamefont {Smail}
  \emph {et~al.}}]{QCDSFUKQCDCSSM:2023qlx}%
  \BibitemOpen
  \bibfield  {author} {\bibinfo {author} {\bibfnamefont {R.~E.}\ \bibnamefont
  {Smail}} \emph {et~al.} (\bibinfo {collaboration} {QCDSF/UKQCD/CSSM}),\
  }\href {\doibase 10.1103/PhysRevD.108.094511} {\bibfield  {journal} {\bibinfo
   {journal} {Phys. Rev. D}\ }\textbf {\bibinfo {volume} {108}},\ \bibinfo
  {pages} {094511} (\bibinfo {year} {2023})},\ \Eprint
  {http://arxiv.org/abs/2304.02866} {arXiv:2304.02866 [hep-lat]} \BibitemShut
  {NoStop}%
\bibitem [{\citenamefont {Gao}\ \emph {et~al.}(2024)\citenamefont {Gao},
  \citenamefont {Hanlon}, \citenamefont {Mukherjee}, \citenamefont {Petreczky},
  \citenamefont {Shi}, \citenamefont {Syritsyn},\ and\ \citenamefont
  {Zhao}}]{Gao:2023ktu}%
  \BibitemOpen
  \bibfield  {author} {\bibinfo {author} {\bibfnamefont {X.}~\bibnamefont
  {Gao}}, \bibinfo {author} {\bibfnamefont {A.~D.}\ \bibnamefont {Hanlon}},
  \bibinfo {author} {\bibfnamefont {S.}~\bibnamefont {Mukherjee}}, \bibinfo
  {author} {\bibfnamefont {P.}~\bibnamefont {Petreczky}}, \bibinfo {author}
  {\bibfnamefont {Q.}~\bibnamefont {Shi}}, \bibinfo {author} {\bibfnamefont
  {S.}~\bibnamefont {Syritsyn}}, \ and\ \bibinfo {author} {\bibfnamefont
  {Y.}~\bibnamefont {Zhao}},\ }\href {\doibase 10.1103/PhysRevD.109.054506}
  {\bibfield  {journal} {\bibinfo  {journal} {Phys. Rev. D}\ }\textbf {\bibinfo
  {volume} {109}},\ \bibinfo {pages} {054506} (\bibinfo {year} {2024})},\
  \Eprint {http://arxiv.org/abs/2310.19047} {arXiv:2310.19047 [hep-lat]}
  \BibitemShut {NoStop}%
\bibitem [{\citenamefont {Djukanovic}\ \emph {et~al.}(2024)\citenamefont
  {Djukanovic}, \citenamefont {von Hippel}, \citenamefont {Meyer},
  \citenamefont {Ottnad},\ and\ \citenamefont {Wittig}}]{Djukanovic:2024krw}%
  \BibitemOpen
  \bibfield  {author} {\bibinfo {author} {\bibfnamefont {D.}~\bibnamefont
  {Djukanovic}}, \bibinfo {author} {\bibfnamefont {G.}~\bibnamefont {von
  Hippel}}, \bibinfo {author} {\bibfnamefont {H.~B.}\ \bibnamefont {Meyer}},
  \bibinfo {author} {\bibfnamefont {K.}~\bibnamefont {Ottnad}}, \ and\ \bibinfo
  {author} {\bibfnamefont {H.}~\bibnamefont {Wittig}},\ }\href {\doibase
  10.1103/PhysRevD.109.074507} {\bibfield  {journal} {\bibinfo  {journal}
  {Phys. Rev. D}\ }\textbf {\bibinfo {volume} {109}},\ \bibinfo {pages}
  {074507} (\bibinfo {year} {2024})},\ \Eprint
  {http://arxiv.org/abs/2402.03024} {arXiv:2402.03024 [hep-lat]} \BibitemShut
  {NoStop}%
\bibitem [{\citenamefont {Alexandrou}\ \emph {et~al.}(2025)\citenamefont
  {Alexandrou}, \citenamefont {Bacchio}, \citenamefont {Finkenrath},
  \citenamefont {Iona}, \citenamefont {Koutsou}, \citenamefont {Li},\ and\
  \citenamefont {Spanoudes}}]{Alexandrou:2024ozj}%
  \BibitemOpen
  \bibfield  {author} {\bibinfo {author} {\bibfnamefont {C.}~\bibnamefont
  {Alexandrou}}, \bibinfo {author} {\bibfnamefont {S.}~\bibnamefont {Bacchio}},
  \bibinfo {author} {\bibfnamefont {J.}~\bibnamefont {Finkenrath}}, \bibinfo
  {author} {\bibfnamefont {C.}~\bibnamefont {Iona}}, \bibinfo {author}
  {\bibfnamefont {G.}~\bibnamefont {Koutsou}}, \bibinfo {author} {\bibfnamefont
  {Y.}~\bibnamefont {Li}}, \ and\ \bibinfo {author} {\bibfnamefont
  {G.}~\bibnamefont {Spanoudes}},\ }\href {\doibase
  10.1103/PhysRevD.111.054505} {\bibfield  {journal} {\bibinfo  {journal}
  {Phys. Rev. D}\ }\textbf {\bibinfo {volume} {111}},\ \bibinfo {pages}
  {054505} (\bibinfo {year} {2025})},\ \Eprint
  {http://arxiv.org/abs/2412.01535} {arXiv:2412.01535 [hep-lat]} \BibitemShut
  {NoStop}%
\bibitem [{\citenamefont {Wang}\ \emph {et~al.}(2025)\citenamefont {Wang},
  \citenamefont {Hu}, \citenamefont {Ji}, \citenamefont {Jiang}, \citenamefont
  {Su}, \citenamefont {Sun},\ and\ \citenamefont {Yang}}]{Wang:2025nsd}%
  \BibitemOpen
  \bibfield  {author} {\bibinfo {author} {\bibfnamefont {J.-H.}\ \bibnamefont
  {Wang}}, \bibinfo {author} {\bibfnamefont {Z.-C.}\ \bibnamefont {Hu}},
  \bibinfo {author} {\bibfnamefont {X.}~\bibnamefont {Ji}}, \bibinfo {author}
  {\bibfnamefont {X.}~\bibnamefont {Jiang}}, \bibinfo {author} {\bibfnamefont
  {Y.}~\bibnamefont {Su}}, \bibinfo {author} {\bibfnamefont {P.}~\bibnamefont
  {Sun}}, \ and\ \bibinfo {author} {\bibfnamefont {Y.-B.}\ \bibnamefont {Yang}}
  (\bibinfo {collaboration} {CLQCD}),\ }\href@noop {} {\  (\bibinfo {year}
  {2025})},\ \Eprint {http://arxiv.org/abs/2511.02326} {arXiv:2511.02326
  [hep-lat]} \BibitemShut {NoStop}%
\bibitem [{\citenamefont {He}\ and\ \citenamefont {Ji}(1995)}]{He:1994gz}%
  \BibitemOpen
  \bibfield  {author} {\bibinfo {author} {\bibfnamefont {H.-x.}\ \bibnamefont
  {He}}\ and\ \bibinfo {author} {\bibfnamefont {X.-D.}\ \bibnamefont {Ji}},\
  }\href {\doibase 10.1103/PhysRevD.52.2960} {\bibfield  {journal} {\bibinfo
  {journal} {Phys. Rev. D}\ }\textbf {\bibinfo {volume} {52}},\ \bibinfo
  {pages} {2960} (\bibinfo {year} {1995})},\ \Eprint
  {http://arxiv.org/abs/hep-ph/9412235} {arXiv:hep-ph/9412235} \BibitemShut
  {NoStop}%
\bibitem [{\citenamefont {Barone}\ \emph {et~al.}(1997)\citenamefont {Barone},
  \citenamefont {Calarco},\ and\ \citenamefont {Drago}}]{Barone:1996un}%
  \BibitemOpen
  \bibfield  {author} {\bibinfo {author} {\bibfnamefont {V.}~\bibnamefont
  {Barone}}, \bibinfo {author} {\bibfnamefont {T.}~\bibnamefont {Calarco}}, \
  and\ \bibinfo {author} {\bibfnamefont {A.}~\bibnamefont {Drago}},\ }\href
  {\doibase 10.1016/S0370-2693(96)01397-4} {\bibfield  {journal} {\bibinfo
  {journal} {Phys. Lett. B}\ }\textbf {\bibinfo {volume} {390}},\ \bibinfo
  {pages} {287} (\bibinfo {year} {1997})},\ \Eprint
  {http://arxiv.org/abs/hep-ph/9605434} {arXiv:hep-ph/9605434} \BibitemShut
  {NoStop}%
\bibitem [{\citenamefont {Schweitzer}\ \emph {et~al.}(2001)\citenamefont
  {Schweitzer}, \citenamefont {Urbano}, \citenamefont {Polyakov}, \citenamefont
  {Weiss}, \citenamefont {Pobylitsa},\ and\ \citenamefont
  {Goeke}}]{Schweitzer:2001sr}%
  \BibitemOpen
  \bibfield  {author} {\bibinfo {author} {\bibfnamefont {P.}~\bibnamefont
  {Schweitzer}}, \bibinfo {author} {\bibfnamefont {D.}~\bibnamefont {Urbano}},
  \bibinfo {author} {\bibfnamefont {M.~V.}\ \bibnamefont {Polyakov}}, \bibinfo
  {author} {\bibfnamefont {C.}~\bibnamefont {Weiss}}, \bibinfo {author}
  {\bibfnamefont {P.~V.}\ \bibnamefont {Pobylitsa}}, \ and\ \bibinfo {author}
  {\bibfnamefont {K.}~\bibnamefont {Goeke}},\ }\href {\doibase
  10.1103/PhysRevD.64.034013} {\bibfield  {journal} {\bibinfo  {journal} {Phys.
  Rev. D}\ }\textbf {\bibinfo {volume} {64}},\ \bibinfo {pages} {034013}
  (\bibinfo {year} {2001})},\ \Eprint {http://arxiv.org/abs/hep-ph/0101300}
  {arXiv:hep-ph/0101300} \BibitemShut {NoStop}%
\bibitem [{\citenamefont {Gamberg}\ and\ \citenamefont
  {Goldstein}(2001)}]{Gamberg:2001qc}%
  \BibitemOpen
  \bibfield  {author} {\bibinfo {author} {\bibfnamefont {L.~P.}\ \bibnamefont
  {Gamberg}}\ and\ \bibinfo {author} {\bibfnamefont {G.~R.}\ \bibnamefont
  {Goldstein}},\ }\href {\doibase 10.1103/PhysRevLett.87.242001} {\bibfield
  {journal} {\bibinfo  {journal} {Phys. Rev. Lett.}\ }\textbf {\bibinfo
  {volume} {87}},\ \bibinfo {pages} {242001} (\bibinfo {year} {2001})},\
  \Eprint {http://arxiv.org/abs/hep-ph/0107176} {arXiv:hep-ph/0107176}
  \BibitemShut {NoStop}%
\bibitem [{\citenamefont {Pasquini}\ \emph {et~al.}(2005)\citenamefont
  {Pasquini}, \citenamefont {Pincetti},\ and\ \citenamefont
  {Boffi}}]{Pasquini:2005dk}%
  \BibitemOpen
  \bibfield  {author} {\bibinfo {author} {\bibfnamefont {B.}~\bibnamefont
  {Pasquini}}, \bibinfo {author} {\bibfnamefont {M.}~\bibnamefont {Pincetti}},
  \ and\ \bibinfo {author} {\bibfnamefont {S.}~\bibnamefont {Boffi}},\ }\href
  {\doibase 10.1103/PhysRevD.72.094029} {\bibfield  {journal} {\bibinfo
  {journal} {Phys. Rev. D}\ }\textbf {\bibinfo {volume} {72}},\ \bibinfo
  {pages} {094029} (\bibinfo {year} {2005})},\ \Eprint
  {http://arxiv.org/abs/hep-ph/0510376} {arXiv:hep-ph/0510376} \BibitemShut
  {NoStop}%
\bibitem [{\citenamefont {Wakamatsu}(2007)}]{Wakamatsu:2007nc}%
  \BibitemOpen
  \bibfield  {author} {\bibinfo {author} {\bibfnamefont {M.}~\bibnamefont
  {Wakamatsu}},\ }\href {\doibase 10.1016/j.physletb.2007.08.013} {\bibfield
  {journal} {\bibinfo  {journal} {Phys. Lett. B}\ }\textbf {\bibinfo {volume}
  {653}},\ \bibinfo {pages} {398} (\bibinfo {year} {2007})},\ \Eprint
  {http://arxiv.org/abs/0705.2917} {arXiv:0705.2917 [hep-ph]} \BibitemShut
  {NoStop}%
\bibitem [{\citenamefont {Lorce}(2009)}]{Lorce:2007fa}%
  \BibitemOpen
  \bibfield  {author} {\bibinfo {author} {\bibfnamefont {C.}~\bibnamefont
  {Lorce}},\ }\href {\doibase 10.1103/PhysRevD.79.074027} {\bibfield  {journal}
  {\bibinfo  {journal} {Phys. Rev. D}\ }\textbf {\bibinfo {volume} {79}},\
  \bibinfo {pages} {074027} (\bibinfo {year} {2009})},\ \Eprint
  {http://arxiv.org/abs/0708.4168} {arXiv:0708.4168 [hep-ph]} \BibitemShut
  {NoStop}%
\bibitem [{\citenamefont {Yamanaka}\ \emph {et~al.}(2013)\citenamefont
  {Yamanaka}, \citenamefont {Doi}, \citenamefont {Imai},\ and\ \citenamefont
  {Suganuma}}]{Yamanaka:2013zoa}%
  \BibitemOpen
  \bibfield  {author} {\bibinfo {author} {\bibfnamefont {N.}~\bibnamefont
  {Yamanaka}}, \bibinfo {author} {\bibfnamefont {T.~M.}\ \bibnamefont {Doi}},
  \bibinfo {author} {\bibfnamefont {S.}~\bibnamefont {Imai}}, \ and\ \bibinfo
  {author} {\bibfnamefont {H.}~\bibnamefont {Suganuma}},\ }\href {\doibase
  10.1103/PhysRevD.88.074036} {\bibfield  {journal} {\bibinfo  {journal} {Phys.
  Rev. D}\ }\textbf {\bibinfo {volume} {88}},\ \bibinfo {pages} {074036}
  (\bibinfo {year} {2013})},\ \Eprint {http://arxiv.org/abs/1307.4208}
  {arXiv:1307.4208 [hep-ph]} \BibitemShut {NoStop}%
\bibitem [{\citenamefont {Pitschmann}\ \emph {et~al.}(2015)\citenamefont
  {Pitschmann}, \citenamefont {Seng}, \citenamefont {Roberts},\ and\
  \citenamefont {Schmidt}}]{Pitschmann:2014jxa}%
  \BibitemOpen
  \bibfield  {author} {\bibinfo {author} {\bibfnamefont {M.}~\bibnamefont
  {Pitschmann}}, \bibinfo {author} {\bibfnamefont {C.-Y.}\ \bibnamefont
  {Seng}}, \bibinfo {author} {\bibfnamefont {C.~D.}\ \bibnamefont {Roberts}}, \
  and\ \bibinfo {author} {\bibfnamefont {S.~M.}\ \bibnamefont {Schmidt}},\
  }\href {\doibase 10.1103/PhysRevD.91.074004} {\bibfield  {journal} {\bibinfo
  {journal} {Phys. Rev. D}\ }\textbf {\bibinfo {volume} {91}},\ \bibinfo
  {pages} {074004} (\bibinfo {year} {2015})},\ \Eprint
  {http://arxiv.org/abs/1411.2052} {arXiv:1411.2052 [nucl-th]} \BibitemShut
  {NoStop}%
\bibitem [{\citenamefont {Xu}\ \emph {et~al.}(2015)\citenamefont {Xu},
  \citenamefont {Chen}, \citenamefont {Cloet}, \citenamefont {Roberts},
  \citenamefont {Segovia},\ and\ \citenamefont {Zong}}]{Xu:2015kta}%
  \BibitemOpen
  \bibfield  {author} {\bibinfo {author} {\bibfnamefont {S.-S.}\ \bibnamefont
  {Xu}}, \bibinfo {author} {\bibfnamefont {C.}~\bibnamefont {Chen}}, \bibinfo
  {author} {\bibfnamefont {I.~C.}\ \bibnamefont {Cloet}}, \bibinfo {author}
  {\bibfnamefont {C.~D.}\ \bibnamefont {Roberts}}, \bibinfo {author}
  {\bibfnamefont {J.}~\bibnamefont {Segovia}}, \ and\ \bibinfo {author}
  {\bibfnamefont {H.-S.}\ \bibnamefont {Zong}},\ }\href {\doibase
  10.1103/PhysRevD.92.114034} {\bibfield  {journal} {\bibinfo  {journal} {Phys.
  Rev. D}\ }\textbf {\bibinfo {volume} {92}},\ \bibinfo {pages} {114034}
  (\bibinfo {year} {2015})},\ \Eprint {http://arxiv.org/abs/1509.03311}
  {arXiv:1509.03311 [nucl-th]} \BibitemShut {NoStop}%
\bibitem [{\citenamefont {Wang}\ \emph {et~al.}(2018)\citenamefont {Wang},
  \citenamefont {Qin}, \citenamefont {Roberts},\ and\ \citenamefont
  {Schmidt}}]{Wang:2018kto}%
  \BibitemOpen
  \bibfield  {author} {\bibinfo {author} {\bibfnamefont {Q.-W.}\ \bibnamefont
  {Wang}}, \bibinfo {author} {\bibfnamefont {S.-X.}\ \bibnamefont {Qin}},
  \bibinfo {author} {\bibfnamefont {C.~D.}\ \bibnamefont {Roberts}}, \ and\
  \bibinfo {author} {\bibfnamefont {S.~M.}\ \bibnamefont {Schmidt}},\ }\href
  {\doibase 10.1103/PhysRevD.98.054019} {\bibfield  {journal} {\bibinfo
  {journal} {Phys. Rev. D}\ }\textbf {\bibinfo {volume} {98}},\ \bibinfo
  {pages} {054019} (\bibinfo {year} {2018})},\ \Eprint
  {http://arxiv.org/abs/1806.01287} {arXiv:1806.01287 [nucl-th]} \BibitemShut
  {NoStop}%
\bibitem [{\citenamefont {Liu}\ \emph {et~al.}(2019)\citenamefont {Liu},
  \citenamefont {Chang},\ and\ \citenamefont {Liu}}]{Liu:2019wzj}%
  \BibitemOpen
  \bibfield  {author} {\bibinfo {author} {\bibfnamefont {L.}~\bibnamefont
  {Liu}}, \bibinfo {author} {\bibfnamefont {L.}~\bibnamefont {Chang}}, \ and\
  \bibinfo {author} {\bibfnamefont {Y.-X.}\ \bibnamefont {Liu}},\ }\href
  {\doibase 10.1103/PhysRevD.99.074013} {\bibfield  {journal} {\bibinfo
  {journal} {Phys. Rev. D}\ }\textbf {\bibinfo {volume} {99}},\ \bibinfo
  {pages} {074013} (\bibinfo {year} {2019})},\ \Eprint
  {http://arxiv.org/abs/1902.10917} {arXiv:1902.10917 [nucl-th]} \BibitemShut
  {NoStop}%
\bibitem [{\citenamefont {Collins}(1993)}]{Collins:1992kk}%
  \BibitemOpen
  \bibfield  {author} {\bibinfo {author} {\bibfnamefont {J.~C.}\ \bibnamefont
  {Collins}},\ }\href {\doibase 10.1016/0550-3213(93)90262-N} {\bibfield
  {journal} {\bibinfo  {journal} {Nucl. Phys. B}\ }\textbf {\bibinfo {volume}
  {396}},\ \bibinfo {pages} {161} (\bibinfo {year} {1993})},\ \Eprint
  {http://arxiv.org/abs/hep-ph/9208213} {arXiv:hep-ph/9208213} \BibitemShut
  {NoStop}%
\bibitem [{\citenamefont {Anselmino}\ \emph {et~al.}(2007)\citenamefont
  {Anselmino}, \citenamefont {Boglione}, \citenamefont {D'Alesio},
  \citenamefont {Kotzinian}, \citenamefont {Murgia}, \citenamefont {Prokudin},\
  and\ \citenamefont {Turk}}]{Anselmino:2007fs}%
  \BibitemOpen
  \bibfield  {author} {\bibinfo {author} {\bibfnamefont {M.}~\bibnamefont
  {Anselmino}}, \bibinfo {author} {\bibfnamefont {M.}~\bibnamefont {Boglione}},
  \bibinfo {author} {\bibfnamefont {U.}~\bibnamefont {D'Alesio}}, \bibinfo
  {author} {\bibfnamefont {A.}~\bibnamefont {Kotzinian}}, \bibinfo {author}
  {\bibfnamefont {F.}~\bibnamefont {Murgia}}, \bibinfo {author} {\bibfnamefont
  {A.}~\bibnamefont {Prokudin}}, \ and\ \bibinfo {author} {\bibfnamefont
  {C.}~\bibnamefont {Turk}},\ }\href {\doibase 10.1103/PhysRevD.75.054032}
  {\bibfield  {journal} {\bibinfo  {journal} {Phys. Rev. D}\ }\textbf {\bibinfo
  {volume} {75}},\ \bibinfo {pages} {054032} (\bibinfo {year} {2007})},\
  \Eprint {http://arxiv.org/abs/hep-ph/0701006} {arXiv:hep-ph/0701006}
  \BibitemShut {NoStop}%
\bibitem [{\citenamefont {Anselmino}\ \emph {et~al.}(2009)\citenamefont
  {Anselmino}, \citenamefont {Boglione}, \citenamefont {D'Alesio},
  \citenamefont {Kotzinian}, \citenamefont {Murgia}, \citenamefont {Prokudin},\
  and\ \citenamefont {Melis}}]{Anselmino:2008jk}%
  \BibitemOpen
  \bibfield  {author} {\bibinfo {author} {\bibfnamefont {M.}~\bibnamefont
  {Anselmino}}, \bibinfo {author} {\bibfnamefont {M.}~\bibnamefont {Boglione}},
  \bibinfo {author} {\bibfnamefont {U.}~\bibnamefont {D'Alesio}}, \bibinfo
  {author} {\bibfnamefont {A.}~\bibnamefont {Kotzinian}}, \bibinfo {author}
  {\bibfnamefont {F.}~\bibnamefont {Murgia}}, \bibinfo {author} {\bibfnamefont
  {A.}~\bibnamefont {Prokudin}}, \ and\ \bibinfo {author} {\bibfnamefont
  {S.}~\bibnamefont {Melis}},\ }\href {\doibase
  10.1016/j.nuclphysbps.2009.03.117} {\bibfield  {journal} {\bibinfo  {journal}
  {Nucl. Phys. B Proc. Suppl.}\ }\textbf {\bibinfo {volume} {191}},\ \bibinfo
  {pages} {98} (\bibinfo {year} {2009})},\ \Eprint
  {http://arxiv.org/abs/0812.4366} {arXiv:0812.4366 [hep-ph]} \BibitemShut
  {NoStop}%
\bibitem [{\citenamefont {Anselmino}\ \emph {et~al.}(2013)\citenamefont
  {Anselmino}, \citenamefont {Boglione}, \citenamefont {D'Alesio},
  \citenamefont {Melis}, \citenamefont {Murgia},\ and\ \citenamefont
  {Prokudin}}]{Anselmino:2013vqa}%
  \BibitemOpen
  \bibfield  {author} {\bibinfo {author} {\bibfnamefont {M.}~\bibnamefont
  {Anselmino}}, \bibinfo {author} {\bibfnamefont {M.}~\bibnamefont {Boglione}},
  \bibinfo {author} {\bibfnamefont {U.}~\bibnamefont {D'Alesio}}, \bibinfo
  {author} {\bibfnamefont {S.}~\bibnamefont {Melis}}, \bibinfo {author}
  {\bibfnamefont {F.}~\bibnamefont {Murgia}}, \ and\ \bibinfo {author}
  {\bibfnamefont {A.}~\bibnamefont {Prokudin}},\ }\href {\doibase
  10.1103/PhysRevD.87.094019} {\bibfield  {journal} {\bibinfo  {journal} {Phys.
  Rev. D}\ }\textbf {\bibinfo {volume} {87}},\ \bibinfo {pages} {094019}
  (\bibinfo {year} {2013})},\ \Eprint {http://arxiv.org/abs/1303.3822}
  {arXiv:1303.3822 [hep-ph]} \BibitemShut {NoStop}%
\bibitem [{\citenamefont {Anselmino}\ \emph {et~al.}(2015)\citenamefont
  {Anselmino}, \citenamefont {Boglione}, \citenamefont {D'Alesio},
  \citenamefont {Gonzalez~Hernandez}, \citenamefont {Melis}, \citenamefont
  {Murgia},\ and\ \citenamefont {Prokudin}}]{Anselmino:2015sxa}%
  \BibitemOpen
  \bibfield  {author} {\bibinfo {author} {\bibfnamefont {M.}~\bibnamefont
  {Anselmino}}, \bibinfo {author} {\bibfnamefont {M.}~\bibnamefont {Boglione}},
  \bibinfo {author} {\bibfnamefont {U.}~\bibnamefont {D'Alesio}}, \bibinfo
  {author} {\bibfnamefont {J.~O.}\ \bibnamefont {Gonzalez~Hernandez}}, \bibinfo
  {author} {\bibfnamefont {S.}~\bibnamefont {Melis}}, \bibinfo {author}
  {\bibfnamefont {F.}~\bibnamefont {Murgia}}, \ and\ \bibinfo {author}
  {\bibfnamefont {A.}~\bibnamefont {Prokudin}},\ }\href {\doibase
  10.1103/PhysRevD.92.114023} {\bibfield  {journal} {\bibinfo  {journal} {Phys.
  Rev. D}\ }\textbf {\bibinfo {volume} {92}},\ \bibinfo {pages} {114023}
  (\bibinfo {year} {2015})},\ \Eprint {http://arxiv.org/abs/1510.05389}
  {arXiv:1510.05389 [hep-ph]} \BibitemShut {NoStop}%
\bibitem [{\citenamefont {Kang}\ \emph {et~al.}(2016)\citenamefont {Kang},
  \citenamefont {Prokudin}, \citenamefont {Sun},\ and\ \citenamefont
  {Yuan}}]{Kang:2015msa}%
  \BibitemOpen
  \bibfield  {author} {\bibinfo {author} {\bibfnamefont {Z.-B.}\ \bibnamefont
  {Kang}}, \bibinfo {author} {\bibfnamefont {A.}~\bibnamefont {Prokudin}},
  \bibinfo {author} {\bibfnamefont {P.}~\bibnamefont {Sun}}, \ and\ \bibinfo
  {author} {\bibfnamefont {F.}~\bibnamefont {Yuan}},\ }\href {\doibase
  10.1103/PhysRevD.93.014009} {\bibfield  {journal} {\bibinfo  {journal} {Phys.
  Rev. D}\ }\textbf {\bibinfo {volume} {93}},\ \bibinfo {pages} {014009}
  (\bibinfo {year} {2016})},\ \Eprint {http://arxiv.org/abs/1505.05589}
  {arXiv:1505.05589 [hep-ph]} \BibitemShut {NoStop}%
\bibitem [{\citenamefont {Lin}\ \emph {et~al.}(2018)\citenamefont {Lin},
  \citenamefont {Melnitchouk}, \citenamefont {Prokudin}, \citenamefont {Sato},\
  and\ \citenamefont {Shows}}]{Lin:2017stx}%
  \BibitemOpen
  \bibfield  {author} {\bibinfo {author} {\bibfnamefont {H.-W.}\ \bibnamefont
  {Lin}}, \bibinfo {author} {\bibfnamefont {W.}~\bibnamefont {Melnitchouk}},
  \bibinfo {author} {\bibfnamefont {A.}~\bibnamefont {Prokudin}}, \bibinfo
  {author} {\bibfnamefont {N.}~\bibnamefont {Sato}}, \ and\ \bibinfo {author}
  {\bibfnamefont {H.}~\bibnamefont {Shows}},\ }\href {\doibase
  10.1103/PhysRevLett.120.152502} {\bibfield  {journal} {\bibinfo  {journal}
  {Phys. Rev. Lett.}\ }\textbf {\bibinfo {volume} {120}},\ \bibinfo {pages}
  {152502} (\bibinfo {year} {2018})},\ \Eprint
  {http://arxiv.org/abs/1710.09858} {arXiv:1710.09858 [hep-ph]} \BibitemShut
  {NoStop}%
\bibitem [{\citenamefont {D'Alesio}\ \emph {et~al.}(2020)\citenamefont
  {D'Alesio}, \citenamefont {Flore},\ and\ \citenamefont
  {Prokudin}}]{DAlesio:2020vtw}%
  \BibitemOpen
  \bibfield  {author} {\bibinfo {author} {\bibfnamefont {U.}~\bibnamefont
  {D'Alesio}}, \bibinfo {author} {\bibfnamefont {C.}~\bibnamefont {Flore}}, \
  and\ \bibinfo {author} {\bibfnamefont {A.}~\bibnamefont {Prokudin}},\ }\href
  {\doibase 10.1016/j.physletb.2020.135347} {\bibfield  {journal} {\bibinfo
  {journal} {Phys. Lett. B}\ }\textbf {\bibinfo {volume} {803}},\ \bibinfo
  {pages} {135347} (\bibinfo {year} {2020})},\ \Eprint
  {http://arxiv.org/abs/2001.01573} {arXiv:2001.01573 [hep-ph]} \BibitemShut
  {NoStop}%
\bibitem [{\citenamefont {Cammarota}\ \emph {et~al.}(2020)\citenamefont
  {Cammarota}, \citenamefont {Gamberg}, \citenamefont {Kang}, \citenamefont
  {Miller}, \citenamefont {Pitonyak}, \citenamefont {Prokudin}, \citenamefont
  {Rogers},\ and\ \citenamefont {Sato}}]{Cammarota:2020qcw}%
  \BibitemOpen
  \bibfield  {author} {\bibinfo {author} {\bibfnamefont {J.}~\bibnamefont
  {Cammarota}}, \bibinfo {author} {\bibfnamefont {L.}~\bibnamefont {Gamberg}},
  \bibinfo {author} {\bibfnamefont {Z.-B.}\ \bibnamefont {Kang}}, \bibinfo
  {author} {\bibfnamefont {J.~A.}\ \bibnamefont {Miller}}, \bibinfo {author}
  {\bibfnamefont {D.}~\bibnamefont {Pitonyak}}, \bibinfo {author}
  {\bibfnamefont {A.}~\bibnamefont {Prokudin}}, \bibinfo {author}
  {\bibfnamefont {T.~C.}\ \bibnamefont {Rogers}}, \ and\ \bibinfo {author}
  {\bibfnamefont {N.}~\bibnamefont {Sato}} (\bibinfo {collaboration} {Jefferson
  Lab Angular Momentum}),\ }\href {\doibase 10.1103/PhysRevD.102.054002}
  {\bibfield  {journal} {\bibinfo  {journal} {Phys. Rev. D}\ }\textbf {\bibinfo
  {volume} {102}},\ \bibinfo {pages} {054002} (\bibinfo {year} {2020})},\
  \Eprint {http://arxiv.org/abs/2002.08384} {arXiv:2002.08384 [hep-ph]}
  \BibitemShut {NoStop}%
\bibitem [{\citenamefont {Gamberg}\ \emph {et~al.}(2022)\citenamefont
  {Gamberg}, \citenamefont {Malda}, \citenamefont {Miller}, \citenamefont
  {Pitonyak}, \citenamefont {Prokudin},\ and\ \citenamefont
  {Sato}}]{Gamberg:2022kdb}%
  \BibitemOpen
  \bibfield  {author} {\bibinfo {author} {\bibfnamefont {L.}~\bibnamefont
  {Gamberg}}, \bibinfo {author} {\bibfnamefont {M.}~\bibnamefont {Malda}},
  \bibinfo {author} {\bibfnamefont {J.~A.}\ \bibnamefont {Miller}}, \bibinfo
  {author} {\bibfnamefont {D.}~\bibnamefont {Pitonyak}}, \bibinfo {author}
  {\bibfnamefont {A.}~\bibnamefont {Prokudin}}, \ and\ \bibinfo {author}
  {\bibfnamefont {N.}~\bibnamefont {Sato}} (\bibinfo {collaboration} {Jefferson
  Lab Angular Momentum}),\ }\href {\doibase 10.1103/PhysRevD.106.034014}
  {\bibfield  {journal} {\bibinfo  {journal} {Phys. Rev. D}\ }\textbf {\bibinfo
  {volume} {106}},\ \bibinfo {pages} {034014} (\bibinfo {year} {2022})},\
  \Eprint {http://arxiv.org/abs/2205.00999} {arXiv:2205.00999 [hep-ph]}
  \BibitemShut {NoStop}%
\bibitem [{\citenamefont {Boglione}\ \emph {et~al.}(2021)\citenamefont
  {Boglione}, \citenamefont {D'Alesio}, \citenamefont {Flore}, \citenamefont
  {Gonzalez-Hernandez}, \citenamefont {Murgia},\ and\ \citenamefont
  {Prokudin}}]{Boglione:2021aha}%
  \BibitemOpen
  \bibfield  {author} {\bibinfo {author} {\bibfnamefont {M.}~\bibnamefont
  {Boglione}}, \bibinfo {author} {\bibfnamefont {U.}~\bibnamefont {D'Alesio}},
  \bibinfo {author} {\bibfnamefont {C.}~\bibnamefont {Flore}}, \bibinfo
  {author} {\bibfnamefont {J.~O.}\ \bibnamefont {Gonzalez-Hernandez}}, \bibinfo
  {author} {\bibfnamefont {F.}~\bibnamefont {Murgia}}, \ and\ \bibinfo {author}
  {\bibfnamefont {A.}~\bibnamefont {Prokudin}},\ }\href {\doibase
  10.1016/j.physletb.2021.136135} {\bibfield  {journal} {\bibinfo  {journal}
  {Phys. Lett. B}\ }\textbf {\bibinfo {volume} {815}},\ \bibinfo {pages}
  {136135} (\bibinfo {year} {2021})},\ \Eprint
  {http://arxiv.org/abs/2101.03955} {arXiv:2101.03955 [hep-ph]} \BibitemShut
  {NoStop}%
\bibitem [{\citenamefont {Zeng}\ \emph {et~al.}(2024)\citenamefont {Zeng},
  \citenamefont {Dong}, \citenamefont {Liue}, \citenamefont {Sun},\ and\
  \citenamefont {Zhao}}]{Zeng:2024gun}%
  \BibitemOpen
  \bibfield  {author} {\bibinfo {author} {\bibfnamefont {C.}~\bibnamefont
  {Zeng}}, \bibinfo {author} {\bibfnamefont {H.}~\bibnamefont {Dong}}, \bibinfo
  {author} {\bibfnamefont {T.}~\bibnamefont {Liue}}, \bibinfo {author}
  {\bibfnamefont {P.}~\bibnamefont {Sun}}, \ and\ \bibinfo {author}
  {\bibfnamefont {Y.}~\bibnamefont {Zhao}},\ }\href@noop {} {\  (\bibinfo
  {year} {2024})},\ \Eprint {http://arxiv.org/abs/2412.18324} {arXiv:2412.18324
  [hep-ph]} \BibitemShut {NoStop}%
\bibitem [{\citenamefont {Collins}\ \emph {et~al.}(1994)\citenamefont
  {Collins}, \citenamefont {Heppelmann},\ and\ \citenamefont
  {Ladinsky}}]{Collins:1993kq}%
  \BibitemOpen
  \bibfield  {author} {\bibinfo {author} {\bibfnamefont {J.~C.}\ \bibnamefont
  {Collins}}, \bibinfo {author} {\bibfnamefont {S.~F.}\ \bibnamefont
  {Heppelmann}}, \ and\ \bibinfo {author} {\bibfnamefont {G.~A.}\ \bibnamefont
  {Ladinsky}},\ }\href {\doibase 10.1016/0550-3213(94)90078-7} {\bibfield
  {journal} {\bibinfo  {journal} {Nucl. Phys. B}\ }\textbf {\bibinfo {volume}
  {420}},\ \bibinfo {pages} {565} (\bibinfo {year} {1994})},\ \Eprint
  {http://arxiv.org/abs/hep-ph/9305309} {arXiv:hep-ph/9305309} \BibitemShut
  {NoStop}%
\bibitem [{\citenamefont {Jaffe}\ \emph {et~al.}(1998)\citenamefont {Jaffe},
  \citenamefont {Jin},\ and\ \citenamefont {Tang}}]{Jaffe:1997hf}%
  \BibitemOpen
  \bibfield  {author} {\bibinfo {author} {\bibfnamefont {R.~L.}\ \bibnamefont
  {Jaffe}}, \bibinfo {author} {\bibfnamefont {X.-m.}\ \bibnamefont {Jin}}, \
  and\ \bibinfo {author} {\bibfnamefont {J.}~\bibnamefont {Tang}},\ }\href
  {\doibase 10.1103/PhysRevLett.80.1166} {\bibfield  {journal} {\bibinfo
  {journal} {Phys. Rev. Lett.}\ }\textbf {\bibinfo {volume} {80}},\ \bibinfo
  {pages} {1166} (\bibinfo {year} {1998})},\ \Eprint
  {http://arxiv.org/abs/hep-ph/9709322} {arXiv:hep-ph/9709322} \BibitemShut
  {NoStop}%
\bibitem [{\citenamefont {Bianconi}\ \emph {et~al.}(2000)\citenamefont
  {Bianconi}, \citenamefont {Boffi}, \citenamefont {Jakob},\ and\ \citenamefont
  {Radici}}]{Bianconi:1999cd}%
  \BibitemOpen
  \bibfield  {author} {\bibinfo {author} {\bibfnamefont {A.}~\bibnamefont
  {Bianconi}}, \bibinfo {author} {\bibfnamefont {S.}~\bibnamefont {Boffi}},
  \bibinfo {author} {\bibfnamefont {R.}~\bibnamefont {Jakob}}, \ and\ \bibinfo
  {author} {\bibfnamefont {M.}~\bibnamefont {Radici}},\ }\href {\doibase
  10.1103/PhysRevD.62.034008} {\bibfield  {journal} {\bibinfo  {journal} {Phys.
  Rev. D}\ }\textbf {\bibinfo {volume} {62}},\ \bibinfo {pages} {034008}
  (\bibinfo {year} {2000})},\ \Eprint {http://arxiv.org/abs/hep-ph/9907475}
  {arXiv:hep-ph/9907475} \BibitemShut {NoStop}%
\bibitem [{\citenamefont {Radici}\ \emph {et~al.}(2002)\citenamefont {Radici},
  \citenamefont {Jakob},\ and\ \citenamefont {Bianconi}}]{Radici:2001na}%
  \BibitemOpen
  \bibfield  {author} {\bibinfo {author} {\bibfnamefont {M.}~\bibnamefont
  {Radici}}, \bibinfo {author} {\bibfnamefont {R.}~\bibnamefont {Jakob}}, \
  and\ \bibinfo {author} {\bibfnamefont {A.}~\bibnamefont {Bianconi}},\ }\href
  {\doibase 10.1103/PhysRevD.65.074031} {\bibfield  {journal} {\bibinfo
  {journal} {Phys. Rev. D}\ }\textbf {\bibinfo {volume} {65}},\ \bibinfo
  {pages} {074031} (\bibinfo {year} {2002})},\ \Eprint
  {http://arxiv.org/abs/hep-ph/0110252} {arXiv:hep-ph/0110252} \BibitemShut
  {NoStop}%
\bibitem [{\citenamefont {Bacchetta}\ and\ \citenamefont
  {Radici}(2003)}]{Bacchetta:2002ux}%
  \BibitemOpen
  \bibfield  {author} {\bibinfo {author} {\bibfnamefont {A.}~\bibnamefont
  {Bacchetta}}\ and\ \bibinfo {author} {\bibfnamefont {M.}~\bibnamefont
  {Radici}},\ }\href {\doibase 10.1103/PhysRevD.67.094002} {\bibfield
  {journal} {\bibinfo  {journal} {Phys. Rev. D}\ }\textbf {\bibinfo {volume}
  {67}},\ \bibinfo {pages} {094002} (\bibinfo {year} {2003})},\ \Eprint
  {http://arxiv.org/abs/hep-ph/0212300} {arXiv:hep-ph/0212300} \BibitemShut
  {NoStop}%
\bibitem [{\citenamefont {Bacchetta}\ \emph {et~al.}(2009)\citenamefont
  {Bacchetta}, \citenamefont {Ceccopieri}, \citenamefont {Mukherjee},\ and\
  \citenamefont {Radici}}]{Bacchetta:2008wb}%
  \BibitemOpen
  \bibfield  {author} {\bibinfo {author} {\bibfnamefont {A.}~\bibnamefont
  {Bacchetta}}, \bibinfo {author} {\bibfnamefont {F.~A.}\ \bibnamefont
  {Ceccopieri}}, \bibinfo {author} {\bibfnamefont {A.}~\bibnamefont
  {Mukherjee}}, \ and\ \bibinfo {author} {\bibfnamefont {M.}~\bibnamefont
  {Radici}},\ }\href {\doibase 10.1103/PhysRevD.79.034029} {\bibfield
  {journal} {\bibinfo  {journal} {Phys. Rev. D}\ }\textbf {\bibinfo {volume}
  {79}},\ \bibinfo {pages} {034029} (\bibinfo {year} {2009})},\ \Eprint
  {http://arxiv.org/abs/0812.0611} {arXiv:0812.0611 [hep-ph]} \BibitemShut
  {NoStop}%
\bibitem [{\citenamefont {Bacchetta}\ \emph {et~al.}(2011)\citenamefont
  {Bacchetta}, \citenamefont {Courtoy},\ and\ \citenamefont
  {Radici}}]{Bacchetta:2011ip}%
  \BibitemOpen
  \bibfield  {author} {\bibinfo {author} {\bibfnamefont {A.}~\bibnamefont
  {Bacchetta}}, \bibinfo {author} {\bibfnamefont {A.}~\bibnamefont {Courtoy}},
  \ and\ \bibinfo {author} {\bibfnamefont {M.}~\bibnamefont {Radici}},\ }\href
  {\doibase 10.1103/PhysRevLett.107.012001} {\bibfield  {journal} {\bibinfo
  {journal} {Phys. Rev. Lett.}\ }\textbf {\bibinfo {volume} {107}},\ \bibinfo
  {pages} {012001} (\bibinfo {year} {2011})},\ \Eprint
  {http://arxiv.org/abs/1104.3855} {arXiv:1104.3855 [hep-ph]} \BibitemShut
  {NoStop}%
\bibitem [{\citenamefont {Bacchetta}\ \emph {et~al.}(2013)\citenamefont
  {Bacchetta}, \citenamefont {Courtoy},\ and\ \citenamefont
  {Radici}}]{Bacchetta:2012ty}%
  \BibitemOpen
  \bibfield  {author} {\bibinfo {author} {\bibfnamefont {A.}~\bibnamefont
  {Bacchetta}}, \bibinfo {author} {\bibfnamefont {A.}~\bibnamefont {Courtoy}},
  \ and\ \bibinfo {author} {\bibfnamefont {M.}~\bibnamefont {Radici}},\ }\href
  {\doibase 10.1007/JHEP03(2013)119} {\bibfield  {journal} {\bibinfo  {journal}
  {JHEP}\ }\textbf {\bibinfo {volume} {03}},\ \bibinfo {pages} {119} (\bibinfo
  {year} {2013})},\ \Eprint {http://arxiv.org/abs/1212.3568} {arXiv:1212.3568
  [hep-ph]} \BibitemShut {NoStop}%
\bibitem [{\citenamefont {Radici}\ \emph {et~al.}(2015)\citenamefont {Radici},
  \citenamefont {Courtoy}, \citenamefont {Bacchetta},\ and\ \citenamefont
  {Guagnelli}}]{Radici:2015mwa}%
  \BibitemOpen
  \bibfield  {author} {\bibinfo {author} {\bibfnamefont {M.}~\bibnamefont
  {Radici}}, \bibinfo {author} {\bibfnamefont {A.}~\bibnamefont {Courtoy}},
  \bibinfo {author} {\bibfnamefont {A.}~\bibnamefont {Bacchetta}}, \ and\
  \bibinfo {author} {\bibfnamefont {M.}~\bibnamefont {Guagnelli}},\ }\href
  {\doibase 10.1007/JHEP05(2015)123} {\bibfield  {journal} {\bibinfo  {journal}
  {JHEP}\ }\textbf {\bibinfo {volume} {05}},\ \bibinfo {pages} {123} (\bibinfo
  {year} {2015})},\ \Eprint {http://arxiv.org/abs/1503.03495} {arXiv:1503.03495
  [hep-ph]} \BibitemShut {NoStop}%
\bibitem [{\citenamefont {Radici}\ \emph {et~al.}(2016)\citenamefont {Radici},
  \citenamefont {Ricci}, \citenamefont {Bacchetta},\ and\ \citenamefont
  {Mukherjee}}]{Radici:2016lam}%
  \BibitemOpen
  \bibfield  {author} {\bibinfo {author} {\bibfnamefont {M.}~\bibnamefont
  {Radici}}, \bibinfo {author} {\bibfnamefont {A.~M.}\ \bibnamefont {Ricci}},
  \bibinfo {author} {\bibfnamefont {A.}~\bibnamefont {Bacchetta}}, \ and\
  \bibinfo {author} {\bibfnamefont {A.}~\bibnamefont {Mukherjee}},\ }\href
  {\doibase 10.1103/PhysRevD.94.034012} {\bibfield  {journal} {\bibinfo
  {journal} {Phys. Rev. D}\ }\textbf {\bibinfo {volume} {94}},\ \bibinfo
  {pages} {034012} (\bibinfo {year} {2016})},\ \Eprint
  {http://arxiv.org/abs/1604.06585} {arXiv:1604.06585 [hep-ph]} \BibitemShut
  {NoStop}%
\bibitem [{\citenamefont {Radici}\ and\ \citenamefont
  {Bacchetta}(2018)}]{Radici:2018iag}%
  \BibitemOpen
  \bibfield  {author} {\bibinfo {author} {\bibfnamefont {M.}~\bibnamefont
  {Radici}}\ and\ \bibinfo {author} {\bibfnamefont {A.}~\bibnamefont
  {Bacchetta}},\ }\href {\doibase 10.1103/PhysRevLett.120.192001} {\bibfield
  {journal} {\bibinfo  {journal} {Phys. Rev. Lett.}\ }\textbf {\bibinfo
  {volume} {120}},\ \bibinfo {pages} {192001} (\bibinfo {year} {2018})},\
  \Eprint {http://arxiv.org/abs/1802.05212} {arXiv:1802.05212 [hep-ph]}
  \BibitemShut {NoStop}%
\bibitem [{\citenamefont {Cocuzza}\ \emph
  {et~al.}(2024{\natexlab{a}})\citenamefont {Cocuzza}, \citenamefont {Metz},
  \citenamefont {Pitonyak}, \citenamefont {Prokudin}, \citenamefont {Sato},\
  and\ \citenamefont {Seidl}}]{Cocuzza:2023oam}%
  \BibitemOpen
  \bibfield  {author} {\bibinfo {author} {\bibfnamefont {C.}~\bibnamefont
  {Cocuzza}}, \bibinfo {author} {\bibfnamefont {A.}~\bibnamefont {Metz}},
  \bibinfo {author} {\bibfnamefont {D.}~\bibnamefont {Pitonyak}}, \bibinfo
  {author} {\bibfnamefont {A.}~\bibnamefont {Prokudin}}, \bibinfo {author}
  {\bibfnamefont {N.}~\bibnamefont {Sato}}, \ and\ \bibinfo {author}
  {\bibfnamefont {R.}~\bibnamefont {Seidl}} (\bibinfo {collaboration} {JAM}),\
  }\href {\doibase 10.1103/PhysRevLett.132.091901} {\bibfield  {journal}
  {\bibinfo  {journal} {Phys. Rev. Lett.}\ }\textbf {\bibinfo {volume} {132}},\
  \bibinfo {pages} {091901} (\bibinfo {year} {2024}{\natexlab{a}})},\ \Eprint
  {http://arxiv.org/abs/2306.12998} {arXiv:2306.12998 [hep-ph]} \BibitemShut
  {NoStop}%
\bibitem [{\citenamefont {Cocuzza}\ \emph
  {et~al.}(2024{\natexlab{b}})\citenamefont {Cocuzza}, \citenamefont {Metz},
  \citenamefont {Pitonyak}, \citenamefont {Prokudin}, \citenamefont {Sato},\
  and\ \citenamefont {Seidl}}]{Cocuzza:2023vqs}%
  \BibitemOpen
  \bibfield  {author} {\bibinfo {author} {\bibfnamefont {C.}~\bibnamefont
  {Cocuzza}}, \bibinfo {author} {\bibfnamefont {A.}~\bibnamefont {Metz}},
  \bibinfo {author} {\bibfnamefont {D.}~\bibnamefont {Pitonyak}}, \bibinfo
  {author} {\bibfnamefont {A.}~\bibnamefont {Prokudin}}, \bibinfo {author}
  {\bibfnamefont {N.}~\bibnamefont {Sato}}, \ and\ \bibinfo {author}
  {\bibfnamefont {R.}~\bibnamefont {Seidl}} (\bibinfo {collaboration}
  {Jefferson Lab Angular Momentum (JAM)}),\ }\href {\doibase
  10.1103/PhysRevD.109.034024} {\bibfield  {journal} {\bibinfo  {journal}
  {Phys. Rev. D}\ }\textbf {\bibinfo {volume} {109}},\ \bibinfo {pages}
  {034024} (\bibinfo {year} {2024}{\natexlab{b}})},\ \Eprint
  {http://arxiv.org/abs/2308.14857} {arXiv:2308.14857 [hep-ph]} \BibitemShut
  {NoStop}%
\bibitem [{\citenamefont {Courtoy}\ \emph {et~al.}(2012)\citenamefont
  {Courtoy}, \citenamefont {Bacchetta}, \citenamefont {Radici},\ and\
  \citenamefont {Bianconi}}]{Courtoy:2012ry}%
  \BibitemOpen
  \bibfield  {author} {\bibinfo {author} {\bibfnamefont {A.}~\bibnamefont
  {Courtoy}}, \bibinfo {author} {\bibfnamefont {A.}~\bibnamefont {Bacchetta}},
  \bibinfo {author} {\bibfnamefont {M.}~\bibnamefont {Radici}}, \ and\ \bibinfo
  {author} {\bibfnamefont {A.}~\bibnamefont {Bianconi}},\ }\href {\doibase
  10.1103/PhysRevD.85.114023} {\bibfield  {journal} {\bibinfo  {journal} {Phys.
  Rev. D}\ }\textbf {\bibinfo {volume} {85}},\ \bibinfo {pages} {114023}
  (\bibinfo {year} {2012})},\ \Eprint {http://arxiv.org/abs/1202.0323}
  {arXiv:1202.0323 [hep-ph]} \BibitemShut {NoStop}%
\bibitem [{\citenamefont {Pitonyak}\ \emph {et~al.}(2024)\citenamefont
  {Pitonyak}, \citenamefont {Cocuzza}, \citenamefont {Metz}, \citenamefont
  {Prokudin},\ and\ \citenamefont {Sato}}]{Pitonyak:2023gjx}%
  \BibitemOpen
  \bibfield  {author} {\bibinfo {author} {\bibfnamefont {D.}~\bibnamefont
  {Pitonyak}}, \bibinfo {author} {\bibfnamefont {C.}~\bibnamefont {Cocuzza}},
  \bibinfo {author} {\bibfnamefont {A.}~\bibnamefont {Metz}}, \bibinfo {author}
  {\bibfnamefont {A.}~\bibnamefont {Prokudin}}, \ and\ \bibinfo {author}
  {\bibfnamefont {N.}~\bibnamefont {Sato}},\ }\href {\doibase
  10.1103/PhysRevLett.132.011902} {\bibfield  {journal} {\bibinfo  {journal}
  {Phys. Rev. Lett.}\ }\textbf {\bibinfo {volume} {132}},\ \bibinfo {pages}
  {011902} (\bibinfo {year} {2024})},\ \Eprint
  {http://arxiv.org/abs/2305.11995} {arXiv:2305.11995 [hep-ph]} \BibitemShut
  {NoStop}%
\bibitem [{\citenamefont {Rogers}\ \emph {et~al.}(2025)\citenamefont {Rogers},
  \citenamefont {Radici}, \citenamefont {Courtoy},\ and\ \citenamefont
  {Rainaldi}}]{Rogers:2024nhb}%
  \BibitemOpen
  \bibfield  {author} {\bibinfo {author} {\bibfnamefont {T.~C.}\ \bibnamefont
  {Rogers}}, \bibinfo {author} {\bibfnamefont {M.}~\bibnamefont {Radici}},
  \bibinfo {author} {\bibfnamefont {A.}~\bibnamefont {Courtoy}}, \ and\
  \bibinfo {author} {\bibfnamefont {T.}~\bibnamefont {Rainaldi}},\ }\href
  {\doibase 10.1103/PhysRevD.111.056001} {\bibfield  {journal} {\bibinfo
  {journal} {Phys. Rev. D}\ }\textbf {\bibinfo {volume} {111}},\ \bibinfo
  {pages} {056001} (\bibinfo {year} {2025})},\ \Eprint
  {http://arxiv.org/abs/2412.12282} {arXiv:2412.12282 [hep-ph]} \BibitemShut
  {NoStop}%
\bibitem [{\citenamefont {Pitonyak}\ \emph {et~al.}(2026)\citenamefont
  {Pitonyak}, \citenamefont {Cocuzza}, \citenamefont {Metz}, \citenamefont
  {Prokudin},\ and\ \citenamefont {Sato}}]{Pitonyak:2025lin}%
  \BibitemOpen
  \bibfield  {author} {\bibinfo {author} {\bibfnamefont {D.}~\bibnamefont
  {Pitonyak}}, \bibinfo {author} {\bibfnamefont {C.}~\bibnamefont {Cocuzza}},
  \bibinfo {author} {\bibfnamefont {A.}~\bibnamefont {Metz}}, \bibinfo {author}
  {\bibfnamefont {A.}~\bibnamefont {Prokudin}}, \ and\ \bibinfo {author}
  {\bibfnamefont {N.}~\bibnamefont {Sato}},\ }\href {\doibase
  10.1103/lqgm-vdm1} {\bibfield  {journal} {\bibinfo  {journal} {Phys. Rev. D}\
  }\textbf {\bibinfo {volume} {113}},\ \bibinfo {pages} {038901} (\bibinfo
  {year} {2026})},\ \Eprint {http://arxiv.org/abs/2502.15817} {arXiv:2502.15817
  [hep-ph]} \BibitemShut {NoStop}%
\bibitem [{\citenamefont {Mahaut}\ \emph {et~al.}(2026)\citenamefont {Mahaut},
  \citenamefont {Polano}, \citenamefont {Bacchetta}, \citenamefont {Bertone},
  \citenamefont {Cerutti}, \citenamefont {Radici},\ and\ \citenamefont
  {Rossi}}]{Mahaut:2025hie}%
  \BibitemOpen
  \bibfield  {author} {\bibinfo {author} {\bibfnamefont {V.}~\bibnamefont
  {Mahaut}}, \bibinfo {author} {\bibfnamefont {L.}~\bibnamefont {Polano}},
  \bibinfo {author} {\bibfnamefont {A.}~\bibnamefont {Bacchetta}}, \bibinfo
  {author} {\bibfnamefont {V.}~\bibnamefont {Bertone}}, \bibinfo {author}
  {\bibfnamefont {M.}~\bibnamefont {Cerutti}}, \bibinfo {author} {\bibfnamefont
  {M.}~\bibnamefont {Radici}}, \ and\ \bibinfo {author} {\bibfnamefont
  {L.}~\bibnamefont {Rossi}} (\bibinfo {collaboration} {MAP (Multi-dimensional
  Analyses of Partonic distributions)}),\ }\href {\doibase
  10.1007/JHEP02(2026)051} {\bibfield  {journal} {\bibinfo  {journal} {JHEP}\
  }\textbf {\bibinfo {volume} {02}},\ \bibinfo {pages} {051} (\bibinfo {year}
  {2026})},\ \Eprint {http://arxiv.org/abs/2509.11855} {arXiv:2509.11855
  [hep-ph]} \BibitemShut {NoStop}%
\bibitem [{\citenamefont {Metz}\ and\ \citenamefont
  {Vossen}(2016)}]{Metz:2016swz}%
  \BibitemOpen
  \bibfield  {author} {\bibinfo {author} {\bibfnamefont {A.}~\bibnamefont
  {Metz}}\ and\ \bibinfo {author} {\bibfnamefont {A.}~\bibnamefont {Vossen}},\
  }\href {\doibase 10.1016/j.ppnp.2016.08.003} {\bibfield  {journal} {\bibinfo
  {journal} {Prog. Part. Nucl. Phys.}\ }\textbf {\bibinfo {volume} {91}},\
  \bibinfo {pages} {136} (\bibinfo {year} {2016})},\ \Eprint
  {http://arxiv.org/abs/1607.02521} {arXiv:1607.02521 [hep-ex]} \BibitemShut
  {NoStop}%
\bibitem [{\citenamefont {Kang}\ \emph {et~al.}(2024)\citenamefont {Kang},
  \citenamefont {Lee}, \citenamefont {Shao},\ and\ \citenamefont
  {Zhao}}]{Kang:2023big}%
  \BibitemOpen
  \bibfield  {author} {\bibinfo {author} {\bibfnamefont {Z.-B.}\ \bibnamefont
  {Kang}}, \bibinfo {author} {\bibfnamefont {K.}~\bibnamefont {Lee}}, \bibinfo
  {author} {\bibfnamefont {D.~Y.}\ \bibnamefont {Shao}}, \ and\ \bibinfo
  {author} {\bibfnamefont {F.}~\bibnamefont {Zhao}},\ }\href {\doibase
  10.1007/JHEP03(2024)153} {\bibfield  {journal} {\bibinfo  {journal} {JHEP}\
  }\textbf {\bibinfo {volume} {03}},\ \bibinfo {pages} {153} (\bibinfo {year}
  {2024})},\ \Eprint {http://arxiv.org/abs/2310.15159} {arXiv:2310.15159
  [hep-ph]} \BibitemShut {NoStop}%
\bibitem [{\citenamefont {Liu}\ and\ \citenamefont {Zhu}(2024)}]{Liu:2024kqt}%
  \BibitemOpen
  \bibfield  {author} {\bibinfo {author} {\bibfnamefont {X.}~\bibnamefont
  {Liu}}\ and\ \bibinfo {author} {\bibfnamefont {H.~X.}\ \bibnamefont {Zhu}},\
  }\href@noop {} {\  (\bibinfo {year} {2024})},\ \Eprint
  {http://arxiv.org/abs/2403.08874} {arXiv:2403.08874 [hep-ph]} \BibitemShut
  {NoStop}%
\bibitem [{\citenamefont {Gao}\ \emph {et~al.}(2026)\citenamefont {Gao},
  \citenamefont {Kang}, \citenamefont {Li},\ and\ \citenamefont
  {Shao}}]{Gao:2025evv}%
  \BibitemOpen
  \bibfield  {author} {\bibinfo {author} {\bibfnamefont {M.-S.}\ \bibnamefont
  {Gao}}, \bibinfo {author} {\bibfnamefont {Z.-B.}\ \bibnamefont {Kang}},
  \bibinfo {author} {\bibfnamefont {W.}~\bibnamefont {Li}}, \ and\ \bibinfo
  {author} {\bibfnamefont {D.~Y.}\ \bibnamefont {Shao}},\ }\href {\doibase
  10.1103/1lz2-3fm9} {\bibfield  {journal} {\bibinfo  {journal} {Phys. Rev.
  Lett.}\ }\textbf {\bibinfo {volume} {136}},\ \bibinfo {pages} {151902}
  (\bibinfo {year} {2026})},\ \Eprint {http://arxiv.org/abs/2509.15809}
  {arXiv:2509.15809 [hep-ph]} \BibitemShut {NoStop}%
\bibitem [{\citenamefont {Cao}\ \emph {et~al.}(2026{\natexlab{a}})\citenamefont
  {Cao}, \citenamefont {Yu}, \citenamefont {Yuan}, \citenamefont {Zhang},\ and\
  \citenamefont {Zhu}}]{Cao:2025icu}%
  \BibitemOpen
  \bibfield  {author} {\bibinfo {author} {\bibfnamefont {Q.-H.}\ \bibnamefont
  {Cao}}, \bibinfo {author} {\bibfnamefont {Z.}~\bibnamefont {Yu}}, \bibinfo
  {author} {\bibfnamefont {C.~P.}\ \bibnamefont {Yuan}}, \bibinfo {author}
  {\bibfnamefont {S.-T.}\ \bibnamefont {Zhang}}, \ and\ \bibinfo {author}
  {\bibfnamefont {H.~X.}\ \bibnamefont {Zhu}},\ }\href {\doibase
  10.1007/JHEP02(2026)244} {\bibfield  {journal} {\bibinfo  {journal} {JHEP}\
  }\textbf {\bibinfo {volume} {02}},\ \bibinfo {pages} {244} (\bibinfo {year}
  {2026}{\natexlab{a}})},\ \bibinfo {note} {[Erratum: JHEP 04, 206 (2026)]},\
  \Eprint {http://arxiv.org/abs/2509.18892} {arXiv:2509.18892 [hep-ph]}
  \BibitemShut {NoStop}%
\bibitem [{\citenamefont {Kang}\ \emph {et~al.}(2026)\citenamefont {Kang},
  \citenamefont {Metz}, \citenamefont {Pitonyak},\ and\ \citenamefont
  {Zhang}}]{Kang:2026pro}%
  \BibitemOpen
  \bibfield  {author} {\bibinfo {author} {\bibfnamefont {Z.-B.}\ \bibnamefont
  {Kang}}, \bibinfo {author} {\bibfnamefont {A.}~\bibnamefont {Metz}}, \bibinfo
  {author} {\bibfnamefont {D.}~\bibnamefont {Pitonyak}}, \ and\ \bibinfo
  {author} {\bibfnamefont {C.}~\bibnamefont {Zhang}},\ }\href@noop {} {\
  (\bibinfo {year} {2026})},\ \Eprint {http://arxiv.org/abs/2604.28131}
  {arXiv:2604.28131 [hep-ph]} \BibitemShut {NoStop}%
\bibitem [{\citenamefont {Wen}\ \emph {et~al.}(2026)\citenamefont {Wen},
  \citenamefont {Yan}, \citenamefont {Yu},\ and\ \citenamefont
  {Yuan}}]{Wen:2024cfu}%
  \BibitemOpen
  \bibfield  {author} {\bibinfo {author} {\bibfnamefont {X.-K.}\ \bibnamefont
  {Wen}}, \bibinfo {author} {\bibfnamefont {B.}~\bibnamefont {Yan}}, \bibinfo
  {author} {\bibfnamefont {Z.}~\bibnamefont {Yu}}, \ and\ \bibinfo {author}
  {\bibfnamefont {C.~P.}\ \bibnamefont {Yuan}},\ }\href {\doibase
  10.1007/s11433-026-2963-9} {\bibfield  {journal} {\bibinfo  {journal} {Sci.
  China Phys. Mech. Astron.}\ }\textbf {\bibinfo {volume} {69}},\ \bibinfo
  {pages} {271011} (\bibinfo {year} {2026})},\ \Eprint
  {http://arxiv.org/abs/2408.07255} {arXiv:2408.07255 [hep-ph]} \BibitemShut
  {NoStop}%
\bibitem [{\citenamefont {Wen}\ \emph {et~al.}(2025)\citenamefont {Wen},
  \citenamefont {Yan}, \citenamefont {Yu},\ and\ \citenamefont
  {Yuan}}]{Wen:2024nff}%
  \BibitemOpen
  \bibfield  {author} {\bibinfo {author} {\bibfnamefont {X.-K.}\ \bibnamefont
  {Wen}}, \bibinfo {author} {\bibfnamefont {B.}~\bibnamefont {Yan}}, \bibinfo
  {author} {\bibfnamefont {Z.}~\bibnamefont {Yu}}, \ and\ \bibinfo {author}
  {\bibfnamefont {C.~P.}\ \bibnamefont {Yuan}},\ }\href {\doibase
  10.1103/32nb-d466} {\bibfield  {journal} {\bibinfo  {journal} {Phys. Rev. D}\
  }\textbf {\bibinfo {volume} {112}},\ \bibinfo {pages} {053004} (\bibinfo
  {year} {2025})},\ \Eprint {http://arxiv.org/abs/2411.13845} {arXiv:2411.13845
  [hep-ph]} \BibitemShut {NoStop}%
\bibitem [{\citenamefont {Cheng}\ and\ \citenamefont
  {Yan}(2025)}]{Cheng:2025cuv}%
  \BibitemOpen
  \bibfield  {author} {\bibinfo {author} {\bibfnamefont {K.}~\bibnamefont
  {Cheng}}\ and\ \bibinfo {author} {\bibfnamefont {B.}~\bibnamefont {Yan}},\
  }\href {\doibase 10.1103/gmqz-v4cl} {\bibfield  {journal} {\bibinfo
  {journal} {Phys. Rev. Lett.}\ }\textbf {\bibinfo {volume} {135}},\ \bibinfo
  {pages} {011902} (\bibinfo {year} {2025})},\ \Eprint
  {http://arxiv.org/abs/2501.03321} {arXiv:2501.03321 [hep-ph]} \BibitemShut
  {NoStop}%
\bibitem [{\citenamefont {Cao}\ \emph {et~al.}(2026{\natexlab{b}})\citenamefont
  {Cao}, \citenamefont {Li}, \citenamefont {Wen},\ and\ \citenamefont
  {Yan}}]{Cao:2025qua}%
  \BibitemOpen
  \bibfield  {author} {\bibinfo {author} {\bibfnamefont {Q.-H.}\ \bibnamefont
  {Cao}}, \bibinfo {author} {\bibfnamefont {G.}~\bibnamefont {Li}}, \bibinfo
  {author} {\bibfnamefont {X.-K.}\ \bibnamefont {Wen}}, \ and\ \bibinfo
  {author} {\bibfnamefont {B.}~\bibnamefont {Yan}},\ }\href@noop {} {\bibfield
  {journal} {\bibinfo  {journal} {Rept. Prog. Phys.}\ }\textbf {\bibinfo
  {volume} {89}},\ \bibinfo {pages} {080501} (\bibinfo {year}
  {2026}{\natexlab{b}})},\ \Eprint {http://arxiv.org/abs/2509.18276}
  {arXiv:2509.18276 [hep-ph]} \BibitemShut {NoStop}%
\bibitem [{\citenamefont {Cao}\ \emph {et~al.}(2026{\natexlab{c}})\citenamefont
  {Cao}, \citenamefont {Wen}, \citenamefont {Yan},\ and\ \citenamefont
  {Zhang}}]{Cao:2025wfg}%
  \BibitemOpen
  \bibfield  {author} {\bibinfo {author} {\bibfnamefont {Q.-H.}\ \bibnamefont
  {Cao}}, \bibinfo {author} {\bibfnamefont {X.-K.}\ \bibnamefont {Wen}},
  \bibinfo {author} {\bibfnamefont {B.}~\bibnamefont {Yan}}, \ and\ \bibinfo
  {author} {\bibfnamefont {S.}~\bibnamefont {Zhang}},\ }\href@noop {}
  {\bibfield  {journal} {\bibinfo  {journal} {Phys. Rev. Lett.}\ }\textbf
  {\bibinfo {volume} {136}},\ \bibinfo {pages} {231901} (\bibinfo {year}
  {2026}{\natexlab{c}})},\ \Eprint {http://arxiv.org/abs/2512.16492}
  {arXiv:2512.16492 [hep-ph]} \BibitemShut {NoStop}%
\bibitem [{\citenamefont {He}\ \emph {et~al.}(2026)\citenamefont {He},
  \citenamefont {Li}, \citenamefont {Tian}, \citenamefont {Wen},\ and\
  \citenamefont {Yan}}]{He:2026pxa}%
  \BibitemOpen
  \bibfield  {author} {\bibinfo {author} {\bibfnamefont {Z.-G.}\ \bibnamefont
  {He}}, \bibinfo {author} {\bibfnamefont {G.}~\bibnamefont {Li}}, \bibinfo
  {author} {\bibfnamefont {Y.-J.}\ \bibnamefont {Tian}}, \bibinfo {author}
  {\bibfnamefont {X.-K.}\ \bibnamefont {Wen}}, \ and\ \bibinfo {author}
  {\bibfnamefont {B.}~\bibnamefont {Yan}},\ }\href@noop {} {\  (\bibinfo {year}
  {2026})},\ \Eprint {http://arxiv.org/abs/2603.18874} {arXiv:2603.18874
  [hep-ph]} \BibitemShut {NoStop}%
\bibitem [{\citenamefont {Vossen}\ \emph {et~al.}(2011)\citenamefont {Vossen}
  \emph {et~al.}}]{Belle:2011cur}%
  \BibitemOpen
  \bibfield  {author} {\bibinfo {author} {\bibfnamefont {A.}~\bibnamefont
  {Vossen}} \emph {et~al.} (\bibinfo {collaboration} {Belle}),\ }\href
  {\doibase 10.1103/PhysRevLett.107.072004} {\bibfield  {journal} {\bibinfo
  {journal} {Phys. Rev. Lett.}\ }\textbf {\bibinfo {volume} {107}},\ \bibinfo
  {pages} {072004} (\bibinfo {year} {2011})},\ \Eprint
  {http://arxiv.org/abs/1104.2425} {arXiv:1104.2425 [hep-ex]} \BibitemShut
  {NoStop}%
\bibitem [{\citenamefont {Seidl}\ \emph {et~al.}(2017)\citenamefont {Seidl}
  \emph {et~al.}}]{Belle:2017rwm}%
  \BibitemOpen
  \bibfield  {author} {\bibinfo {author} {\bibfnamefont {R.}~\bibnamefont
  {Seidl}} \emph {et~al.} (\bibinfo {collaboration} {Belle}),\ }\href {\doibase
  10.1103/PhysRevD.96.032005} {\bibfield  {journal} {\bibinfo  {journal} {Phys.
  Rev. D}\ }\textbf {\bibinfo {volume} {96}},\ \bibinfo {pages} {032005}
  (\bibinfo {year} {2017})},\ \Eprint {http://arxiv.org/abs/1706.08348}
  {arXiv:1706.08348 [hep-ex]} \BibitemShut {NoStop}%
\bibitem [{\citenamefont {Airapetian}\ \emph {et~al.}(2008)\citenamefont
  {Airapetian} \emph {et~al.}}]{HERMES:2008mcr}%
  \BibitemOpen
  \bibfield  {author} {\bibinfo {author} {\bibfnamefont {A.}~\bibnamefont
  {Airapetian}} \emph {et~al.} (\bibinfo {collaboration} {HERMES}),\ }\href
  {\doibase 10.1088/1126-6708/2008/06/017} {\bibfield  {journal} {\bibinfo
  {journal} {JHEP}\ }\textbf {\bibinfo {volume} {06}},\ \bibinfo {pages} {017}
  (\bibinfo {year} {2008})},\ \Eprint {http://arxiv.org/abs/0803.2367}
  {arXiv:0803.2367 [hep-ex]} \BibitemShut {NoStop}%
\bibitem [{\citenamefont {Alexeev}\ \emph {et~al.}(2023)\citenamefont {Alexeev}
  \emph {et~al.}}]{COMPASS:2023cgk}%
  \BibitemOpen
  \bibfield  {author} {\bibinfo {author} {\bibfnamefont {G.~D.}\ \bibnamefont
  {Alexeev}} \emph {et~al.} (\bibinfo {collaboration} {COMPASS}),\ }\href
  {\doibase 10.1016/j.physletb.2023.138155} {\bibfield  {journal} {\bibinfo
  {journal} {Phys. Lett. B}\ }\textbf {\bibinfo {volume} {845}},\ \bibinfo
  {pages} {138155} (\bibinfo {year} {2023})},\ \Eprint
  {http://arxiv.org/abs/2301.02013} {arXiv:2301.02013 [hep-ex]} \BibitemShut
  {NoStop}%
\bibitem [{\citenamefont {Adamczyk}\ \emph {et~al.}(2015)\citenamefont
  {Adamczyk} \emph {et~al.}}]{STAR:2015jkc}%
  \BibitemOpen
  \bibfield  {author} {\bibinfo {author} {\bibfnamefont {L.}~\bibnamefont
  {Adamczyk}} \emph {et~al.} (\bibinfo {collaboration} {STAR}),\ }\href
  {\doibase 10.1103/PhysRevLett.115.242501} {\bibfield  {journal} {\bibinfo
  {journal} {Phys. Rev. Lett.}\ }\textbf {\bibinfo {volume} {115}},\ \bibinfo
  {pages} {242501} (\bibinfo {year} {2015})},\ \Eprint
  {http://arxiv.org/abs/1504.00415} {arXiv:1504.00415 [hep-ex]} \BibitemShut
  {NoStop}%
\bibitem [{\citenamefont {Adamczyk}\ \emph {et~al.}(2018)\citenamefont
  {Adamczyk} \emph {et~al.}}]{STAR:2017wsi}%
  \BibitemOpen
  \bibfield  {author} {\bibinfo {author} {\bibfnamefont {L.}~\bibnamefont
  {Adamczyk}} \emph {et~al.} (\bibinfo {collaboration} {STAR}),\ }\href
  {\doibase 10.1016/j.physletb.2018.02.069} {\bibfield  {journal} {\bibinfo
  {journal} {Phys. Lett. B}\ }\textbf {\bibinfo {volume} {780}},\ \bibinfo
  {pages} {332} (\bibinfo {year} {2018})},\ \Eprint
  {http://arxiv.org/abs/1710.10215} {arXiv:1710.10215 [hep-ex]} \BibitemShut
  {NoStop}%
\bibitem [{\citenamefont {Burkert}\ \emph {et~al.}(2020)\citenamefont {Burkert}
  \emph {et~al.}}]{Burkert:2020akg}%
  \BibitemOpen
  \bibfield  {author} {\bibinfo {author} {\bibfnamefont {V.~D.}\ \bibnamefont
  {Burkert}} \emph {et~al.},\ }\href {\doibase 10.1016/j.nima.2020.163419}
  {\bibfield  {journal} {\bibinfo  {journal} {Nucl. Instrum. Meth. A}\ }\textbf
  {\bibinfo {volume} {959}},\ \bibinfo {pages} {163419} (\bibinfo {year}
  {2020})}\BibitemShut {NoStop}%
\bibitem [{\citenamefont {Huang}\ \emph {et~al.}(2014)\citenamefont {Huang}
  \emph {et~al.}}]{SoLID:2014proposal}%
  \BibitemOpen
  \bibfield  {author} {\bibinfo {author} {\bibfnamefont {J.}~\bibnamefont
  {Huang}} \emph {et~al.},\ }\href
  {https://www.jlab.org/exp_prog/proposals/14/E12-10-006A.pdf} {\enquote
  {\bibinfo {title} {Dihadron electroproduction in {DIS} with transversely
  polarized $^3${He} target at 11 and 8.8~{GeV}},}\ } (\bibinfo {year}
  {2014}),\ \bibinfo {note} {{Jefferson} Lab Experiment E12-10-006A, proposal
  to Jefferson Lab PAC 42,
  \url{https://www.jlab.org/exp_prog/proposals/14/E12-10-006A.pdf}}\BibitemShut
  {NoStop}%
\bibitem [{\citenamefont {Arrington}\ \emph {et~al.}(2023)\citenamefont
  {Arrington} \emph {et~al.}}]{Arrington:2022hgr}%
  \BibitemOpen
  \bibfield  {author} {\bibinfo {author} {\bibfnamefont {J.}~\bibnamefont
  {Arrington}} \emph {et~al.},\ }\href {\doibase
  10.1088/1748-0221/18/10/P10042} {\bibfield  {journal} {\bibinfo  {journal}
  {J. Instrum.}\ }\textbf {\bibinfo {volume} {18}},\ \bibinfo {pages} {P10042}
  (\bibinfo {year} {2023})},\ \Eprint {http://arxiv.org/abs/2209.13357}
  {arXiv:2209.13357} \BibitemShut {NoStop}%
\bibitem [{\citenamefont {Accardi}\ \emph {et~al.}(2016)\citenamefont {Accardi}
  \emph {et~al.}}]{Accardi:2012qut}%
  \BibitemOpen
  \bibfield  {author} {\bibinfo {author} {\bibfnamefont {A.}~\bibnamefont
  {Accardi}} \emph {et~al.},\ }\href {\doibase 10.1140/epja/i2016-16268-9}
  {\bibfield  {journal} {\bibinfo  {journal} {Eur. Phys. J. A}\ }\textbf
  {\bibinfo {volume} {52}},\ \bibinfo {pages} {268} (\bibinfo {year} {2016})},\
  \Eprint {http://arxiv.org/abs/1212.1701} {arXiv:1212.1701 [nucl-ex]}
  \BibitemShut {NoStop}%
\bibitem [{\citenamefont {Abdul~Khalek}\ \emph {et~al.}(2022)\citenamefont
  {Abdul~Khalek} \emph {et~al.}}]{AbdulKhalek:2021gbh}%
  \BibitemOpen
  \bibfield  {author} {\bibinfo {author} {\bibfnamefont {R.}~\bibnamefont
  {Abdul~Khalek}} \emph {et~al.},\ }\href {\doibase
  10.1016/j.nuclphysa.2022.122447} {\bibfield  {journal} {\bibinfo  {journal}
  {Nucl. Phys. A}\ }\textbf {\bibinfo {volume} {1026}},\ \bibinfo {pages}
  {122447} (\bibinfo {year} {2022})},\ \Eprint
  {http://arxiv.org/abs/2103.05419} {arXiv:2103.05419 [physics.ins-det]}
  \BibitemShut {NoStop}%
\bibitem [{\citenamefont {Gamberg}\ \emph {et~al.}(2021)\citenamefont
  {Gamberg}, \citenamefont {Kang}, \citenamefont {Pitonyak}, \citenamefont
  {Prokudin}, \citenamefont {Sato},\ and\ \citenamefont
  {Seidl}}]{Gamberg:2021lgx}%
  \BibitemOpen
  \bibfield  {author} {\bibinfo {author} {\bibfnamefont {L.}~\bibnamefont
  {Gamberg}}, \bibinfo {author} {\bibfnamefont {Z.-B.}\ \bibnamefont {Kang}},
  \bibinfo {author} {\bibfnamefont {D.}~\bibnamefont {Pitonyak}}, \bibinfo
  {author} {\bibfnamefont {A.}~\bibnamefont {Prokudin}}, \bibinfo {author}
  {\bibfnamefont {N.}~\bibnamefont {Sato}}, \ and\ \bibinfo {author}
  {\bibfnamefont {R.}~\bibnamefont {Seidl}},\ }\href {\doibase
  10.1016/j.physletb.2021.136255} {\bibfield  {journal} {\bibinfo  {journal}
  {Phys. Lett. B}\ }\textbf {\bibinfo {volume} {816}},\ \bibinfo {pages}
  {136255} (\bibinfo {year} {2021})},\ \Eprint
  {http://arxiv.org/abs/2101.06200} {arXiv:2101.06200 [hep-ph]} \BibitemShut
  {NoStop}%
\bibitem [{\citenamefont {Kovchegov}\ and\ \citenamefont
  {Sievert}(2019)}]{Kovchegov:2018zeq}%
  \BibitemOpen
  \bibfield  {author} {\bibinfo {author} {\bibfnamefont {Y.~V.}\ \bibnamefont
  {Kovchegov}}\ and\ \bibinfo {author} {\bibfnamefont {M.~D.}\ \bibnamefont
  {Sievert}},\ }\href {\doibase 10.1103/PhysRevD.99.054033} {\bibfield
  {journal} {\bibinfo  {journal} {Phys. Rev. D}\ }\textbf {\bibinfo {volume}
  {99}},\ \bibinfo {pages} {054033} (\bibinfo {year} {2019})},\ \Eprint
  {http://arxiv.org/abs/1808.10354} {arXiv:1808.10354 [hep-ph]} \BibitemShut
  {NoStop}%
\bibitem [{\citenamefont {Pobylitsa}(2003)}]{Pobylitsa:2003ty}%
  \BibitemOpen
  \bibfield  {author} {\bibinfo {author} {\bibfnamefont {P.~V.}\ \bibnamefont
  {Pobylitsa}},\ }\href@noop {} {\  (\bibinfo {year} {2003})},\ \Eprint
  {http://arxiv.org/abs/hep-ph/0301236} {arXiv:hep-ph/0301236} \BibitemShut
  {NoStop}%
\bibitem [{\citenamefont {Alexeev}\ \emph {et~al.}(2025)\citenamefont {Alexeev}
  \emph {et~al.}}]{COMPASS:2025dwu}%
  \BibitemOpen
  \bibfield  {author} {\bibinfo {author} {\bibfnamefont {G.~D.}\ \bibnamefont
  {Alexeev}} \emph {et~al.} (\bibinfo {collaboration} {COMPASS}),\ }\href@noop
  {} {\  (\bibinfo {year} {2025})},\ \Eprint {http://arxiv.org/abs/2512.22311}
  {arXiv:2512.22311 [hep-ex]} \BibitemShut {NoStop}%
\bibitem [{\citenamefont {Adolph}\ \emph {et~al.}(2012)\citenamefont {Adolph}
  \emph {et~al.}}]{COMPASS:2012bfl}%
  \BibitemOpen
  \bibfield  {author} {\bibinfo {author} {\bibfnamefont {C.}~\bibnamefont
  {Adolph}} \emph {et~al.} (\bibinfo {collaboration} {COMPASS}),\ }\href
  {\doibase 10.1016/j.physletb.2012.05.015} {\bibfield  {journal} {\bibinfo
  {journal} {Phys. Lett. B}\ }\textbf {\bibinfo {volume} {713}},\ \bibinfo
  {pages} {10} (\bibinfo {year} {2012})},\ \Eprint
  {http://arxiv.org/abs/1202.6150} {arXiv:1202.6150 [hep-ex]} \BibitemShut
  {NoStop}%
\bibitem [{\citenamefont {Adolph}\ \emph {et~al.}(2014)\citenamefont {Adolph}
  \emph {et~al.}}]{COMPASS:2014ysd}%
  \BibitemOpen
  \bibfield  {author} {\bibinfo {author} {\bibfnamefont {C.}~\bibnamefont
  {Adolph}} \emph {et~al.} (\bibinfo {collaboration} {COMPASS}),\ }\href
  {\doibase 10.1016/j.physletb.2014.06.080} {\bibfield  {journal} {\bibinfo
  {journal} {Phys. Lett. B}\ }\textbf {\bibinfo {volume} {736}},\ \bibinfo
  {pages} {124} (\bibinfo {year} {2014})},\ \Eprint
  {http://arxiv.org/abs/1401.7873} {arXiv:1401.7873 [hep-ex]} \BibitemShut
  {NoStop}%
\bibitem [{\citenamefont {Soffer}(1995)}]{Soffer:1994ww}%
  \BibitemOpen
  \bibfield  {author} {\bibinfo {author} {\bibfnamefont {J.}~\bibnamefont
  {Soffer}},\ }\href {\doibase 10.1103/PhysRevLett.74.1292} {\bibfield
  {journal} {\bibinfo  {journal} {Phys. Rev. Lett.}\ }\textbf {\bibinfo
  {volume} {74}},\ \bibinfo {pages} {1292} (\bibinfo {year} {1995})},\ \Eprint
  {http://arxiv.org/abs/hep-ph/9409254} {arXiv:hep-ph/9409254} \BibitemShut
  {NoStop}%
\bibitem [{\citenamefont {Bacchetta}\ and\ \citenamefont
  {Radici}(2004)}]{Bacchetta:2003vn}%
  \BibitemOpen
  \bibfield  {author} {\bibinfo {author} {\bibfnamefont {A.}~\bibnamefont
  {Bacchetta}}\ and\ \bibinfo {author} {\bibfnamefont {M.}~\bibnamefont
  {Radici}},\ }\href {\doibase 10.1103/PhysRevD.69.074026} {\bibfield
  {journal} {\bibinfo  {journal} {Phys. Rev. D}\ }\textbf {\bibinfo {volume}
  {69}},\ \bibinfo {pages} {074026} (\bibinfo {year} {2004})},\ \Eprint
  {http://arxiv.org/abs/hep-ph/0311173} {arXiv:hep-ph/0311173} \BibitemShut
  {NoStop}%
\bibitem [{\citenamefont {Zhou}\ \emph {et~al.}(2021)\citenamefont {Zhou},
  \citenamefont {Cocuzza}, \citenamefont {Delcarro}, \citenamefont
  {Melnitchouk}, \citenamefont {Metz},\ and\ \citenamefont
  {Sato}}]{Zhou:2021llj}%
  \BibitemOpen
  \bibfield  {author} {\bibinfo {author} {\bibfnamefont {Y.}~\bibnamefont
  {Zhou}}, \bibinfo {author} {\bibfnamefont {C.}~\bibnamefont {Cocuzza}},
  \bibinfo {author} {\bibfnamefont {F.}~\bibnamefont {Delcarro}}, \bibinfo
  {author} {\bibfnamefont {W.}~\bibnamefont {Melnitchouk}}, \bibinfo {author}
  {\bibfnamefont {A.}~\bibnamefont {Metz}}, \ and\ \bibinfo {author}
  {\bibfnamefont {N.}~\bibnamefont {Sato}} (\bibinfo {collaboration} {Jefferson
  Lab Angular Momentum (JAM)}),\ }\href {\doibase 10.1103/PhysRevD.104.034028}
  {\bibfield  {journal} {\bibinfo  {journal} {Phys. Rev. D}\ }\textbf {\bibinfo
  {volume} {104}},\ \bibinfo {pages} {034028} (\bibinfo {year} {2021})},\
  \Eprint {http://arxiv.org/abs/2105.04434} {arXiv:2105.04434 [hep-ph]}
  \BibitemShut {NoStop}%
\bibitem [{\citenamefont {Ball}\ \emph {et~al.}(2022)\citenamefont {Ball} \emph
  {et~al.}}]{NNPDF:2021njg}%
  \BibitemOpen
  \bibfield  {author} {\bibinfo {author} {\bibfnamefont {R.~D.}\ \bibnamefont
  {Ball}} \emph {et~al.} (\bibinfo {collaboration} {NNPDF}),\ }\href {\doibase
  10.1140/epjc/s10052-022-10328-7} {\bibfield  {journal} {\bibinfo  {journal}
  {Eur. Phys. J. C}\ }\textbf {\bibinfo {volume} {82}},\ \bibinfo {pages} {428}
  (\bibinfo {year} {2022})},\ \Eprint {http://arxiv.org/abs/2109.02653}
  {arXiv:2109.02653 [hep-ph]} \BibitemShut {NoStop}%
\bibitem [{\citenamefont {Echevarria}\ \emph {et~al.}(2021)\citenamefont
  {Echevarria}, \citenamefont {Kang},\ and\ \citenamefont
  {Terry}}]{Echevarria:2020hpy}%
  \BibitemOpen
  \bibfield  {author} {\bibinfo {author} {\bibfnamefont {M.~G.}\ \bibnamefont
  {Echevarria}}, \bibinfo {author} {\bibfnamefont {Z.-B.}\ \bibnamefont
  {Kang}}, \ and\ \bibinfo {author} {\bibfnamefont {J.}~\bibnamefont {Terry}},\
  }\href {\doibase 10.1007/JHEP01(2021)126} {\bibfield  {journal} {\bibinfo
  {journal} {JHEP}\ }\textbf {\bibinfo {volume} {01}},\ \bibinfo {pages} {126}
  (\bibinfo {year} {2021})},\ \Eprint {http://arxiv.org/abs/2009.10710}
  {arXiv:2009.10710 [hep-ph]} \BibitemShut {NoStop}%
\bibitem [{\citenamefont {Bhattacharya}\ \emph {et~al.}(2022)\citenamefont
  {Bhattacharya}, \citenamefont {Kang}, \citenamefont {Metz}, \citenamefont
  {Penn},\ and\ \citenamefont {Pitonyak}}]{Bhattacharya:2021twu}%
  \BibitemOpen
  \bibfield  {author} {\bibinfo {author} {\bibfnamefont {S.}~\bibnamefont
  {Bhattacharya}}, \bibinfo {author} {\bibfnamefont {Z.-B.}\ \bibnamefont
  {Kang}}, \bibinfo {author} {\bibfnamefont {A.}~\bibnamefont {Metz}}, \bibinfo
  {author} {\bibfnamefont {G.}~\bibnamefont {Penn}}, \ and\ \bibinfo {author}
  {\bibfnamefont {D.}~\bibnamefont {Pitonyak}},\ }\href {\doibase
  10.1103/PhysRevD.105.034007} {\bibfield  {journal} {\bibinfo  {journal}
  {Phys. Rev. D}\ }\textbf {\bibinfo {volume} {105}},\ \bibinfo {pages}
  {034007} (\bibinfo {year} {2022})},\ \Eprint
  {http://arxiv.org/abs/2110.10253} {arXiv:2110.10253 [hep-ph]} \BibitemShut
  {NoStop}%
\end{thebibliography}%
\end{document}